\documentclass[aps,prd,twocolumn,showpacs,groupedaddress,superscriptaddress,nofootinbib,floatfix,preprintnumbers,longbibliography]{revtex4-1}

\usepackage{amssymb,amsmath,graphicx, color}
\usepackage{physics}
\usepackage{bm}
\usepackage[tight]{subfigure}
\usepackage[export]{adjustbox}
\usepackage{braket}
\usepackage{svg}
\usepackage{mathtools}
\usepackage{bbm}
\usepackage{dsfont}

\usepackage{multirow}
\usepackage{rotating,booktabs}
\usepackage[verbose]{placeins}
\usepackage{color}

\newcommand\bea{\begin{eqnarray}}
\newcommand\eea{\end{eqnarray}}

\newcommand\sopp{\hat{\mathcal{S}}_o^{++}}
\newcommand\sopm{\hat{\mathcal{S}}_o^{+-}}
\newcommand\somp{\hat{\mathcal{S}}_o^{-+}}
\newcommand\somm{\hat{\mathcal{S}}_o^{--}}

\newcommand\sipp{\hat{\mathcal{S}}_i^{++}}
\newcommand\sipm{\hat{\mathcal{S}}_i^{+-}}
\newcommand\simp{\hat{\mathcal{S}}_i^{-+}}
\newcommand\simm{\hat{\mathcal{S}}_i^{--}}

\newcommand{\mc}[1]{\mathcal{#1}}
\newcommand{\gket}[1]{|#1\rangle}

\usepackage[colorlinks=true,citecolor=blue,linkcolor=red]{hyperref}
\usepackage{cleveref}
\graphicspath{{figures/}}

\begin{document}

\title{
String-breaking statics and dynamics in a (1+1)D SU(2) lattice gauge theory
}

\author{Navya~Gupta${}^{\diamond}$}
\email{navyag@umd.edu}
\affiliation{
Maryland Center for Fundamental Physics and Department of Physics, 
University of Maryland, College Park, MD 20742, USA}
\affiliation{
Joint Center for Quantum Information and Computer Science, University of Maryland, College Park, Maryland 20742, USA}

\author{Emil~Mathew${}^{\diamond}$}
\email{p20210036@goa.bits-pilani.ac.in}
\affiliation{
Department of Physics, Birla Institute of Technology and Sciences- Pilani, K K Birla Goa Campus, Zuarinagar, Sancole, Goa 403726, India}
\affiliation{Center for Research in Quantum Information and Technology, Birla Institute of Technology and Sciences-Pilani, K K Birla Goa Campus, Zuarinagar, Sancole, Goa 403726, India}

\author{Saurabh~V.~Kadam}
\affiliation{
InQubator for Quantum Simulation (IQuS), Department of Physics, University of Washington, Seattle, WA 98195, USA}

\author{Jesse~R.~Stryker}
\affiliation{
Physics Division, Lawrence Berkeley National Laboratory, Berkeley, CA 94720, USA}

\author{Aniruddha~Bapat}
\affiliation{
Joint Center for Quantum Information and Computer Science, University of Maryland, College Park, Maryland 20742, USA}
\affiliation{
Physics Division, Lawrence Berkeley National Laboratory, Berkeley, CA 94720, USA}

\author{Niklas~Mueller}
\affiliation{Center for Quantum Information and Control, University of New Mexico, Albuquerque, NM 87106, USA}
\affiliation{Department of Physics and Astronomy, University of New Mexico, Albuquerque, NM 87106, USA}

\author{Zohreh~Davoudi${}^{\diamond\diamond}$}
\email{davoudi@umd.edu}
\affiliation{
Maryland Center for Fundamental Physics and Department of Physics, University of Maryland, College Park, MD 20742, USA}
\affiliation{
Joint Center for Quantum Information and Computer Science, University of Maryland, College Park, Maryland 20742, USA}

\author{Indrakshi~Raychowdhury${}^{\diamond\diamond~}$}
\email{indrakshir@goa.bits-pilani.ac.in}
\affiliation{
Department of Physics, Birla Institute of Technology and Sciences- Pilani,
K K Birla Goa Campus, Zuarinagar, Sancole, Goa 403726, India}
\affiliation{Center for Research in Quantum Information and Technology, Birla Institute of Technology and Sciences-Pilani, K K Birla Goa Campus, Zuarinagar, Sancole, Goa 403726, India}

\begingroup
\renewcommand\thefootnote{$\diamond$}
\footnotetext{These authors contributed equally to this work.}
\endgroup

\begingroup
\renewcommand\thefootnote{$\diamond\diamond$}
\footnotetext{These authors jointly supervised this work.}
\endgroup

\preprint{UMD-PP-026-02}
\preprint{IQuS@UW-21-122}

%%%%%%%%%%%%%%%%%%%%%%%%%%%%%%%%%%%%%%%%%%%%%%%%%%%%%%%%%%%%%%%%%%%%%%%%%%%%%%%%%%%%%%%%%%%%%%%%%%%%%%
\begin{abstract}
String breaking is at the core of hadronization models of relevance to particle colliders. Yet, studies of string-breaking dynamics rooted in quantum chromodynamics remain fundamentally challenging. Tensor networks enable sign-problem-free studies of static and dynamical properties of lattice gauge theories. In this work, we develop and apply a tensor-network toolkit based on the loop-string-hadron formulation of an SU(2) lattice gauge theory in 1+1 dimensions with dynamical fermions. We apply this toolkit to study static and dynamical aspects of strings and their breaking in this theory. The simple, gauge-invariant, and local structure of the loop-string-hadron states and constraints removes the need to impose non-Abelian constraints in the algorithm, and allows for a systematic computation of observables at increasingly large bosonic cutoffs, and toward the infinite-volume and continuum limits. Our study of static strings yields a determination of the string tension in the continuum and thermodynamic limits. Our study of dynamical string breaking, performed at a fixed lattice spacing and system size, illuminates underlying processes at play during the quench dynamics of a string. The loop, string, and hadron description offers a systematic and intuitive way to diagnose these processes, including string expansion and contraction, endpoint splitting and particle shower, chain scattering events, and inelastic processes resulting from string dissociation and recombination, and particle production. We relate these processes to several features of the dynamics, such as energy transport, entanglement-entropy production, and correlation spreading. This work opens the way to future tensor-network studies of string breaking and particle production in increasingly complex lattice gauge theories.

\end{abstract}
\maketitle
\tableofcontents

%%%%%%%%%%%%%%%%%%%%%%%%%%%%%%%%%%%%%%%%%%%%%%%%%%%%%%%%%%%%%%%%%%%%%%%%%%%%%%%%%%%%%%%%%%%%%%%%%%%%%%%%%%%%%%%%%%%%%%%%%%%%%%%%%%%%%%%%%%%%%%%%%%%%%%%%%%%%%%%%%%
\section{Introduction
\label{sec:intro}}
\noindent
A primary quest in nuclear and particle physics is to discover new phases of matter, and new particles and interactions, in high-energy particle colliders~\cite{Ludlam:2005cfp,Lisa:2005dd,Lovato:2022vgq,Accardi:2012qut,CidVidal:2018eel,Narain:2022qud,Shiltsev:2019rfl,DUNE:2020fgq}. The final state of the collisions consists of abundant hadrons of various types, created out of color charges, i.e., quarks and gluons, which only exist in isolation immediately after the collision. Quantum chromodynamics (QCD), the gauge-field theory of the strong force, underlies this complex ``hadronization'' process. First-principles simulations of such  dynamics based in QCD, nonetheless, have proven elusive. Yet, successful phenomenological models have enabled predictions of hadron yields in the detectors~\cite{Andersson:1997xwk,Sjostrand:1982fn,Andersson:1983ia,Webber:1983if,Buckley:2011ms,Bierlich:2022pfr,Gieseke:2025mcy,Altmann:2024kwx,Reuter:2025imz,Christiansen:2024bhe}. These models, nonetheless, have a set of parameters that are not determined from the underlying theory, but are fit to experimental output; the models, furthermore, do not universally describe all observables in all types of experiments~\cite{Altmann:2024kwx,Gieseke:2025mcy,Christiansen:2024bhe}.

A primary hadronization process is believed to be the formation and breakdown of strings (i.e., matter charges connected by gluonic flux), which occur at a rate predicted by the Schwinger pair-production mechanism~\cite{Schwinger:1951nm,Gelis:2015kya}, and is modeled stochastically~\cite{Andersson:1983ia,Bierlich:2022pfr, Andersson:1997xwk}. A fully quantum-mechanical simulation of string dynamics in QCD requires real-(Minkowski-)time methods. Conventional Monte-Carlo-based lattice-gauge-theory (LGT) computations face a sign problem when Minkowski-time observables are sought~\cite{Gattringer:2016kco,Alexandru:2020wrj}. As a way forward, studies of simpler gauge theories of non-Abelian color charges and strings can inform our understanding of hadronization processes in real time. Furthermore, computational tools such as tensor networks (TNs) can open the way to first-principles dynamical simulations.
 
Over the last two decades, TNs have emerged as a powerful computational method in quantum many-body physics~\cite{White:1992zz,Fannes:1990ur,Ostlund:1995zz,Verstraete:2004zza,Schollwoeck:2010uqf,Meurice:2020pxc,Banuls:2019rao}. In particular, the application of TN methods to LGTs has seen steady progress over the years. Notably, the first TN simulation of a LGT (the lattice Schwinger model~\cite{Schwinger:1962tp}) was performed in Ref.~\cite{Byrnes:2002nv} using the Density-Matrix-Renormalization-Group (DMRG)~\cite{White:1992zz,Schollwoeck:2010uqf} algorithm. 
Since then, there has been a flurry of numerical strategies and formalism development for implementing TN simulations of LGTs~\cite{Tagliacozzo:2010vk,Rico:2013qya,Buyens:2013yza,Silvi:2014pta,Tagliacozzo:2014bta,Zohar:2014qma,Canals:2024bqp}. These efforts span both Abelian (discrete and continuous) and non-Abelian gauge groups. 

In (1+1)-dimensional LGTs, TNs have been successfully employed to investigate static quantities such as low-lying energy spectra and phase transitions~\cite{Buyens:2015tea,Banuls:2016jws,Banuls:2016gid,Magnifico:2018wek,Magnifico:2019ulp,Papaefstathiou:2021cho,Petrova:2022luq,Funcke:2023lli,Angelides:2023bme,Fujii:2024reh,ArguelloCruz:2024xzi,Popov:2024ysn,Farrell:2023fgd,Davoudi:2024wyv,Rogerson:2026jjt,Grieninger:2025rdi}, finite-temperature phases~\cite{Banuls:2015sta,Banuls:2016lkq,Buyens:2016ecr,Buyens:2016hhu,Banuls:2022iwk,Angelides:2025hjt}, real-time dynamics and out-of-equilibrium phenomena~\cite{Pichler:2015yqa,Buyens:2017crb,Magnifico:2019kyj,Banuls:2022iwk,Florio:2023dke,Barata:2023jgd,Barata:2024bzk,Barata:2024apg,Janik:2025bbz,Farrell:2024fit, Florio:2024aix,Florio:2025hoc}, including particle scattering~\cite{Pichler:2015yqa,Rigobello:2021fxw,Belyansky:2023rgh,Papaefstathiou:2024zsu,Barata:2025rjb,Davoudi:2025rdv,Schuhmacher:2025ehh}, and structure quantities such as parton distribution functions~\cite{Banuls:2025wiq,Kang:2025xpz,Grieninger:2025mbm}. Importantly, string-breaking dynamics has been a primary target of these real-time simulations~\cite{Pichler:2015yqa,Buyens:2015tea,Grieninger:2026bdq}. More recently, there has been growing interest in extending TN techniques to higher-dimensional gauge theories too. TN ansatzes such as Tree TNs~(TTNs)~\cite{Gerster:2014spa,Silvi:2017srb} and Projected Entangled-Pair States (PEPS) \cite{Cirac:2020obd} have facilitated this effort, leading to various implementations of Abelian and non-Abelian LGTs in two or more spatial dimensions~\cite{Zohar:2015eda,Zohar:2017yxl,Emonts:2018puo,Felser:2019xyv,Emonts:2020drm,Magnifico:2020bqt,Zohar:2021wqy,Robaina:2020aqh,Emonts:2022yom,Emonts:2023ttz,Cataldi:2023xki,Roose:2024ruv,Kelman:2024exo,Chandrasekharan:2025smw,Xu:2025ean,Magnifico:2024eiy,Turro:2024pxu,Xu:2025abo,Cao:2026qky,Tian:2025mbv,Cataldi:2025cyo}. Despite this progress, non-Abelian gauge theories continue to pose theoretical and numerical challenges. In part, this challenge arises from the noncommutativity of the Gauss's laws, which impedes deployment of basis states that are simultaneously physical (gauge-invariant) and local. Nevertheless, much progress has been made for non-Abelian gauge theories in (1+1)D, both the SU(2) \cite{Kuhn:2015zqa,Silvi:2016cas,Banuls:2017ena,Sala:2018dui,Turro:2024pxu,Cataldi:2025cyo} and SU(3) \cite{Silvi:2019wnf,Ciavarella:2021lel,Rigobello:2023ype,Hayata:2023pkw} gauge groups.

In this work, we show how a local gauge-invariant description provided by the loop-string-hadron (LSH) formulation~\cite{Raychowdhury:2019iki} can be used to construct gauge-invariant TN ansatzes for non-Abelian LGTs based on matrix product states (MPSes) and matrix product operators (MPOs)~\cite{Mathew:2025fim}. In this formulation, the degrees of freedom are onsite `snapshots' of general nonlocal gauge-invariant observables. These local snapshots, referred to as loops, strings, and hadrons, can be naturally translated to the language of TN states, allowing for a fully gauge-invariant description. In particular, only a set of Abelian Gauss's laws (AGLs) that ensure electric-flux continuity need to be imposed on the states. Using this formulation, which is equivalent to the Kogut-Susskind Hamiltonian formulation in the physical sector of the Hilbert space~\cite{Raychowdhury:2019iki,Davoudi:2020yln}, we determine the static string potential and the string tension in the continuum limit. We further study the real-time dynamics of string breaking in the presence of dynamical fermions. Our results extend the existing literature \cite{Kuhn:2015zqa,Banuls:2017ena,Sala:2018dui} by exploring larger lattice volumes, and controlling Hilbert-space truncation systematics, whilst including fully dynamical fermions and without making an assumption of Gaussianity for the state of the gauge bosons~\cite{Sala:2018dui}. Our work, therefore, provides a more comprehensive study of string-breaking dynamics, leading to new qualitative and quantitative understandings.

Specifically, we create, out of the interacting vacuum, a meson-string excitation, i.e., two dynamical fermions connected by a string of dynamical electric-field flux. We work in the weak-coupling limit with a nontrivial vacuum, and consider two bare fermion masses that yield distinct evolutions. The local gauge-invariant loops, strings, and hadrons offer a systematic and intuitive way to diagnose the underlying processes contributing to the meson-string dynamics. First, fermion-hopping processes lead to the expansion and contraction of the meson string, the string-endpoint splitting, and particle showers. Second, the scattering of excitations in the bulk plays a significant role: These scattering processes can be inelastic, which involve string dissociation and recombination, and particle creation and annihilation. We connect this microscopic picture to several features of the dynamics, such as energy transport, entanglement-entropy generation, and the correlations spreading in two-point electric-energy and two-point matter-density correlators.

The TN construction presented in this work can be carried over directly to the (1+1)D SU(3) LSH formulation developed recently in Ref.~\cite{Kadam:2022ipf}. As such, precision studies of static properties, phase transitions, and string-breaking and scattering dynamics in (1+1)D SU(2) and SU(3) LGTs can be pursued within the LSH TN framework. The LSH formulation, moreover, has been extended to higher dimensions~\cite{Raychowdhury:2019iki,Kadam:2024ifg,Kadam:2025trs}, and is shown to retain its simple structure: the presence of only local gauge-invariant degrees of freedom with AGLs. It is, therefore, well suited to various (2+1)D tensor-network ansatzes, to be explored in future work.

The paper is organized as follows: In Sec.~\ref{sec:formalism}, we introduce our formalism and method, namely the LSH framework and the LSH-based MPS and MPO ansatzes. In Sec.~\ref{sec:static}, we study strings made up of static endpoint charges, probe their ground-state excitations at various lengths, and compute the static potential in the continuum and thermodynamic limits. In Sec.~\ref{sec:dynamic}, we study the breaking of strings with dynamical endpoint charges after a quench, and discuss microscopic, energetic, entropic, and other features of dynamics in depth. A summary of our results, and an outlook for future directions, are presented in Sec.~\ref{sec:conclusions}. The paper includes several appendices; they supplement our discussions and present further details on our formalism, methods, and results.

%%%%%%%%%%%%%%%%%%%%%%%%%%%%%%%%%%%%%%%%%%%%%%%%%%%%%%%%%%%%%%%%%%%%%%%%%%%%%%%%%%%%%%%%%%%%%%%%%%%%%%%%%%%%%%%%%%%%%%%%%%%%%%%%%%%%%%%%%%%%%%%%%%%%%%%%%%%%%%%%%%
\section{
Formalism and tensor-network construction
\label{sec:formalism}}
\noindent
The Hamiltonian formulation of an SU(2) LGT coupled to dynamical fermions was first developed by Kogut and Susskind~\cite{Kogut:1974ag}. The Kogut-Susskind Hamiltonian and its Hilbert space are briefly reviewed in Appendix~\ref{app:KS-formulation}. The loop-string-hadron formalism is a reformulation of the Kogut-Susskind formalism using inherently gauge-invariant degrees of freedom~\cite{Raychowdhury:2019iki}, as is described in Appendix~\ref{app:lsh-formalism}. In this section, we summarize the key ingredients of the LSH formulation, including its degrees of freedom, basis states and operators, as well as the resulting Hamiltonian.
Thereafter, we describe how TN ansatzes can be constructed for this LGT in the LSH formulation.

%%%%%%
%%%%%%
\subsection{Loop-string-hadron formulation}
\label{sec:LSH-formulation}
\begin{figure}[t!]
\includegraphics[scale=0.45]{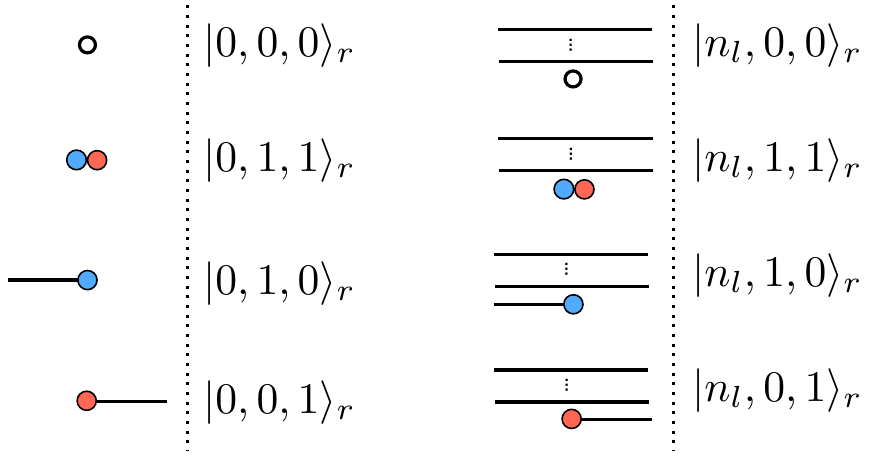}
    \caption{
    The site-local Hilbert space of the LSH formulation is characterized by the following quantum numbers: a loop $n_l$, a fermion sourcing one unit of flux to the left, called an incoming string $n_i$, a fermion sourcing one unit of flux to the right, called an outgoing string $n_o$. Shown are a few examples of the basis states $\ket{n_l,n_i,n_o}_r$ at a given site $r$, along with their pictorial representation. An empty circle at an even (odd) site denotes the absence (presence) of a hadron while a doubly filled site at odd (even) denotes the absence (presence) of a hadron. Open incoming and outgoing strings are the endpoint of nonlocal closed string operators. The states in the left have zero loop quantum number while those in the right can take any positive integer values up to a specified cutoff.}
    \label{fig:lsh-states}
\end{figure}
The LSH formulation of the (1+1)D SU(2) LGT is obtained by first expressing the SU(2) angular-momentum degrees of freedom in terms of Schwinger-boson doublets. Together with the SU(2) fermionic doublets, one can form local SU(2) singlets out of two fermions (forming a `hadron'), two bosons (forming a `loop'), or a fermion and a boson (forming a `string-in' or `string-out'), as outlined in Appendix~\ref{app:lsh-formalism}. Consequently, the states spanning the onsite Hilbert space can be labeled by one bosonic quantum number $n_l\in \mathbb{Z}^+ \cup \{0\}$ and two fermionic quantum numbers $n_i, n_o\in\{0,1\}$. First, the $n_l$ quantum number denotes the total number of loop excitations. Here, a `loop' represents a local snapshot 
of a Wilson loop (when working in more than one spatial dimension), the interior region of a meson string, or some background flux configuration. Second, the state with $n_i=n_o=0$ is the vacuum state of fermions, while that with $n_i=n_o=1$ holds a `hadron' excitation.
Finally, the $(n_i,n_o)=(1,0)$ [$(n_i,n_o)=(0,1)$] state denotes a single fermion at the site, sourcing one unit of flux to the left (right) side of the site. Since these are local snapshots of extended strings starting (ending) at the fermions, we call them string-in (string-out) excitations. Such states $\ket{n_l,n_i,n_o}_r$ at each lattice site $r$ are manifestly gauge invariant and have the non-Abelian Gauss's law constraints intrinsically built into them, as discussed in Appendix~\ref{app:lsh-formalism}. A few representative local basis states of the LSH formulation are shown in Fig.~\ref{fig:lsh-states}.

The Hilbert space of the entire lattice is spanned by tensor-product states of local basis states:
\begin{equation}
    \ket{\Psi} = \bigotimes_{r=1}^{N} \ket{n_l,n_i,n_o}_{r},
    \label{eq:globalPsi}
\end{equation}
where $N$ denotes the number of lattice sites. However, a physical state in this Hilbert space is subject to Abelian constraints, namely AGLs, ensuring that:
\begin{equation}
    \label{eq:AGL-app}
    N_L(r)=N_R(r+1),
\end{equation}
for all $r$, where
\begin{subequations}
\label{eq:NLRasnlnino}
\begin{align}
    &N_L (r) = n_l(r)+n_o(r)(1-n_i(r)),\label{eq:NLasnlnino}\\
    &N_R (r) = n_l(r)+n_i(r)(1-n_o(r)).\label{eq:NRasnlnino}
\end{align}
\end{subequations}
The quantum number $N_L(r)$ ($N_R(r+1)$) is equal to twice the total angular momentum on the left, $j_L(r)$ (right, $j_R(r+1)$), side of the link originating from site $r$; each AGL, thus, ensures the continuity of angular momentum, or ``flux lines,'' across the corresponding link.

There are two types of operators at each site: the diagonal LSH operators $\hat{n}_{l/i/o}$, whose eigenvalues are given by the previously defined quantum numbers $n_{l/i/o}$, and the related $\hat{N}_{L/R}$ operators according to Eq.~\eqref{eq:NLRasnlnino}; as well as ladder operators $\hat{\lambda}^{\pm}_{l,i,o}$ and related string operators $\hat{S}^{\pm}_{i/o}$. These operators and their action on the LSH basis states are presented in Appendix~\ref{app:lsh-formalism}. In terms of these operators, the (scaled, dimensionless) Kogut-Susskind Hamiltonian subject to open boundary conditions\footnote{With open boundary conditions, $N_R(1)$ and $N_L(N)$ are not dynamical degrees of freedom: $N_R(1)$ is fixed by the boundary condition and $N_L(N)$ is fixed by the consecutive action of AGLs.} can be expressed as:
\begin{align}
    \label{eq:H_dimless}
    \hat{\tilde{H}}
    &\coloneq \hat H_E + \hat H_M + \hat H_I \nonumber\\
    &=\sum_{r=1}^{N-1}\hat{h}'_E(r)+\sum_{r=1}^{N}\hat{h}'_M(r)+\sum_{r=1}^{N-1}\hat{h}'_I(r),
\end{align}
where the electric-energy term, $H_E$, mass-energy term, $H_M$, and interaction-energy term, $H_I$, are given by a sum over local Hamiltonian terms:
\begin{widetext}
\begin{subequations}
\label{eq:HEMI_LSH}
\begin{align}
\label{eq:HE_LSH}
    &\hat{h}'_E(r)=\frac{1}{2}\left[\frac{\hat{N}_L(r)}{2}\left( \frac{\hat{N}_L(r)}{2}+1 \right)+\frac{\hat{N}_R(r+1)}{2}\left( \frac{\hat{N}_R(r+1)}{2}+1 \right)\right], \\
    \label{eq:HM_LSH}
    &\hat{h}'_M(r)= \mu(-1)^r\Big[\hat n_i(r)+\hat n_o(r)\Big],\\
    \label{eq:HI_LSH}
    &\hat{h}'_I(r)= x\Bigg\{\frac{1}{\sqrt{\hat{N}_L(r)+1}}\Big[\hat{\mathcal{S}}_o^{++}(r)\hat{\mathcal{S}}_i^{+-}(r+1)+\hat{\mathcal{S}}_o^{+-}(r)\hat{\mathcal{S}}_i^{--}(r+1)\Big]\frac{1}{\sqrt{\hat{N}_R(r+1)+1}}+ \text{H.c}.\Bigg\}.
\end{align}
\end{subequations}
\end{widetext}
Here, $\mu=\frac{2m\sqrt{x}}{g}$ and $x=\frac{1}{a^2g^2}$ are dimensionless couplings, with $a$, $g$, and $m$ being, respectively, the lattice spacing, coupling, and the fermion mass in the original Kogut-Susskind Hamiltonian, see Appendix~\ref{app:KS-formulation}.  Here, $\hat{H}_E$ and $\hat{H}_M$ contain only the LSH number operators [defined in Eq.~\eqref{eq:nlninodefn}], and $\hat{H}_I$ is composed of string-type operators [defined in Eq.~\eqref{eq:Spm-defs}], accompanied by inverse square-root factors of the number operators. Note that due to the AGL in Eq.~\eqref{eq:AGL-app}, contributions from the $N_L$ and $N_R$ operators to the electric Hamiltonian are equal.

The theory exhibits two conserved quantum numbers associated with two global symmetry transformation, see Appendix~\ref{app:KS-formulation}. First, the total number of fermions present,
\begin{eqnarray}
    \hat{Q} \coloneq \sum_{r=1}^{N} \hat{Q}_r= \sum_{r=1}^{N} \Big[\hat{n}_i(r) + \hat{n}_o(r)\Big],
    \label{eq:QinLSH}
\end{eqnarray}
constitutes a conserved charge. The other conserved charge is:
\begin{equation}
    \hat{q} \coloneq \sum_{r=1}^{N} \hat{q}_r= 
    \sum_{r=1}^{N} \Big[\hat{n}_o(r) - \hat{n}_i(r)\Big].
    \label{eq:qinLSH}
\end{equation}
Considering the AGL in Eq.~\eqref{eq:AGL}, along with the relations in Eq.~\eqref{eq:NLRasnlnino}, this charge measures the difference in the electric-flux lines entering the first and leaving the last lattice site: $\hat q = \hat N_L(N)-\hat N_R(1)$.

The locally gauge-invariant basis of LSH offers a convenient way to describe the excitations of the strong-coupling vacuum. The strong-coupling vacuum is the ground state of the Hamiltonian in Eq.~\eqref{eq:H_dimless} in the limit $x=0$. Such a state corresponds to the Dirac-sea configuration with unoccupied even (particle) sites and occupied odd (antiparticle) sites, and no electric-flux excitations. In terms of the LSH local quantum numbers $(n_l,n_i,n_o)$, this state is obtained by setting $(0,0,0)$ at all even (particle) sites and to $(0,1,1)$ at all odd (antiparticle) sites (so it is empty at even sites and holds a hadron at odd sites). This vacuum has, therefore, the global quantum numbers $Q=N$ and $q=0$.

The lowest-lying excitations on top of such a vacuum are the degenerate string-in and string-out states, which carry an energy of $+\mu$ relative to the strong-coupling vacuum. Moreover, a (bare) ``baryon" is described by the hadron (empty) state at an even (odd) site, which carries an energy of $+2\mu$ relative to the strong-coupling vacuum. In other words, an LSH hadron can be interpreted as the vacuum or a baryon at a site depending on its energy.  We can diagnose the presence of these excitations using the operator
\begin{align}
\label{eq:excitations}
    \hat n(r)\coloneq
    \begin{cases}
        \hat n_i(r)+\hat n_o(r), & r\text{ is even;} \\
        2 - \hat n_i(r) - \hat n_o(r), & r\text{ is odd.}
    \end{cases}
\end{align}
We will refer to the expectation value of $\hat n(r)$ as the (matter) excitation number. Thus, the vacuum would carry an excitation number of zero, while the string-in/out and baryonic excitations carry excitation numbers +1 and +2, respectively. To exclusively count one of the three types of excitations, we further split $\hat n$ into three components:
\begin{align}
    \hat n(r)&\coloneq \hat{\tilde{n}}_{i}(r)+\hat{\tilde{n}}_{o}(r)+2\hat{n}_{b}(r),
\end{align}
where string-in ($\hat{\tilde{n}}_i$), string-out ($\hat{\tilde{n}}_o$), and baryon ($\hat n_{b}$) number operators are defined by
\begin{subequations}
\label{eq:exitation-defs}
\begin{align}
    &\hat{\tilde{n}}_i\coloneq\hat n_i(1-\hat n_o), \\
    &\hat{\tilde{n}}_o\coloneq\hat n_o(1-\hat n_i), \\
    &\hat n_{b}(r)\coloneq
    \begin{cases}
        \hat n_i(r)\hat n_o(r), & r\text{ is even;} \\
        \left(1-\hat n_i(r)\right)\left(1-\hat n_o(r)\right), & r\text{ is odd.}
    \end{cases}
\end{align}
\end{subequations}
Probing the expectation values of these operators, together with that of the loop number operator $\hat n_l$, allows us to characterize the local particle and flux contents. Note that the right/incoming (left/outgoing) flux quantum number operator $\hat N_R$ ($\hat N_L$) is simply the sum of the loop number operator and the string-in (string-out) number operator, see Eqs.~\eqref{eq:NLasnlnino} and \eqref{eq:NRasnlnino}. In the following, we shall use a symmetrized definition of flux:
\begin{align}
    \hat{\tilde{N}}(r) \coloneq
    \frac{\hat N_R(r)+\hat N_L(r)}{2}.
\end{align}

Two remarks are in order. First, all notions of baryons and mesons in this work refer to bare excitations, i.e., the simplest excitations sharing the quantum numbers of a baryon and a meson. These, therefore, do not represent asymptotic, fully-dressed hadronic states. Second, in general, there may be several ways to associate strings and loops across lattice sites to given extended (bare) mesonic states. Different choices are not distinguishable by the LSH quantum numbers and are merely different graphical representations of the same state: they are consistent with the AGL and yield the same energy. This is another reason why it is more meaningful to track the sitewise string-in/out and the loop numbers separately, and not attempt to measure the number of, e.g., extended strings. More details and examples are provided in Appendix~\ref{app:local-vs-extended}.\\

Last but not least, with OBCs and $N_R(1)=0$, the largest value the $N_{L/R}$ quantum numbers take in the $q=0$ sector is $N/2$. For large system sizes, the Hilbert-space size poses challenges to computations. Therefore, in practice, one needs to impose a lower cutoff on the local bosonic Hilbert space. Here, we assume the electric flux has a physical upper bound of $2j_{\rm max}$, i.e., $N_{L/R}(r) \leq 2j_{\rm max} < \frac{N}{2}$. Following Eq.~(\ref{eq:NLasnlnino}), these cutoffs translate to a cutoff on the onsite $n_l$ value: when $n_i(r)=n_o(r)$, the maximum allowed value of $n_l(r)$ is $2j_{\rm max}$, while when $n_i(r) \neq n_o(r)$, the maximum $n_l(r)$ value is $2j_{\rm max}-1$. This truncation yields a local Hilbert-space dimension $d=8j_{\rm max}+2$.

%%%%%%
%%%%%%
\subsection{LSH-based MPS and MPO ansatzes
\label{sec:MPS-MPO}}
\begin{figure*}[t!]
    \centering
    \includegraphics[scale=1]{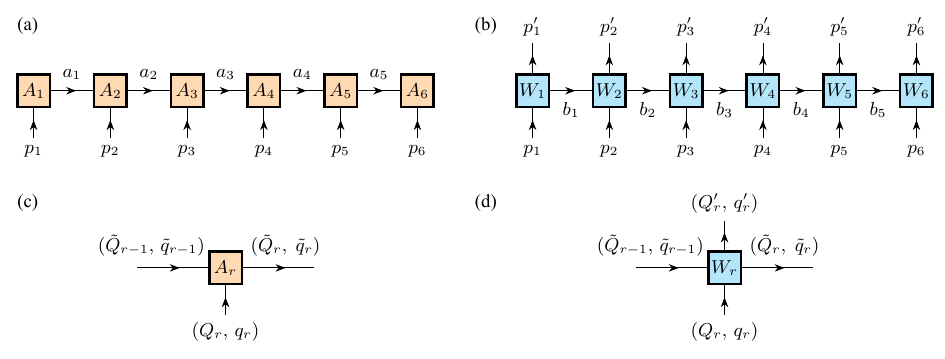}
    \caption{(a) a 6-site MPS representation composed of tensors with a physical leg, $p_r$, and two virtual bonds, $a_{r-1}$ and $a_r$, at site $r$. (b) a 6-site MPO representation consisting of tensors with two physical legs $p_r,p'_r$ and two virtual bonds $b_{r-1},b_r$
    (c) Each local MPS tensor has a block-diagonal structure imposed by the super-selection rules between the local charge $(Q_r,q_r)$ associated with the physical legs, and charges $(\tilde Q_{r-1},\tilde q_{r-1})$ and $(\tilde Q_r,\tilde q_r)$ associated with the left and right virtual bonds, respectively. The rule is given by $\tilde{Q}_r=Q_r+\tilde{Q}_{r-1}$ and $\tilde{q}_r=q_r+\tilde{q}_{r-1}$.
    (d) The local MPO tensors inherit an analogous block-diagonal structure imposed by the super-selection rule $\tilde{Q}_r = \tilde{Q}_{r-1}+(Q_r-Q'_r)$ and $\tilde{q}_r = \tilde{q}_{r-1}+(q_r-q'_r)$.
    }
    \label{fig:tensor_QN}
\end{figure*}

Consider a generic MPS of the form
\begin{equation}
    \ket{\Psi[A]} = \sum_{p_1,\ldots,p_N}A^{a_1}_{p_1}A^{a_1,a_2}_{p_2}\ldots A^{a_{N-1}}_{p_N}\ket{p_1,p_2,\ldots,p_N}.
\end{equation}
Here, the index $p_r$ labels the physical quantum numbers at the staggered site $r$. For an MPS representing LSH states, the physical indices correspond to three quantum numbers $n_l,n_i$, and $n_o$. Explicitly, $\ket{p_1,p_2,\ldots,p_N} = \bigotimes_{r=1}^N \ket{p_r}$ where 
$\gket{p_r} \coloneq \gket{n_l, n_i, n_o}_r$. A representative six-site MPS is depicted in Fig.~\ref{fig:tensor_QN}(a).
The onsite tensors $A$ are complex-valued matrices, whose matrix elements are labeled by the virtual indices $a_{r-1}, a_r$. If the index $a_r$ associated with the bond connecting sites $r$ and $r+1$ were allowed to take all $d^{\mathrm{min}(r,N-r)}$ values for all sites $r\in\{1,\cdots,N-1\}$, then the MPS would be an exact representation of all states in the Hilbert space. This dimension of the virtual bond space is referred to as the bond dimension. It turns out that low-lying energy eigenstates of gapped Hamiltonians can be well approximated using a bond dimension no larger than some fixed number $D \ll d^{N/2}$ throughout the MPS. This allows a representation whose dimension is only polynomial in $N$. The truncation of the MPS to the required bond dimension is achieved by discarding states at all bonds with the lowest associated Schmidt values. Thus, the MPS truncation error can be characterized by the sum of the squares of these discarded Schmidt values $\lambda_{a_r}$, i.e., by $\sum_{a_r>D} \lambda_{a_r}^2$, at each bond. We will denote the maximum value of this quantity by $\varepsilon$. 
The tensors $A$ with a maximum possible bond dimension of $D$ constitute the variational degrees of freedom, which are eventually optimized using a DMRG algorithm. \\

The global symmetries of the LSH Hamiltonian can be incorporated into the MPS ansatz using standard techniques~\cite{Singh:2009cd}. An MPS with definite conserved charge values $(Q,q)$ can be assigned local charges $(Q_r,q_r)$ at site $r$ and charges $(\tilde{Q}_r,\tilde{q}_r)$ to the virtual bond to the right of site $r$. The charges $(Q_r,q_r)$ are simply the eigenvalues of the local charge operators $\hat Q_r$ and $\hat q_r$ defined in Eqs~\eqref{eq:QinLSH} and \eqref{eq:qinLSH}, respectively.  The physical charges flow inward while the virtual charges flow from the left to the right.
Each bulk tensor is block-sparse, where the blocks are defined with respect to the local conservation (``fusing") rules 
\begin{subequations}
\begin{align}
&\tilde{Q}_r=Q_r+\tilde{Q}_{r-1},\\
&\tilde{q}_r=q_r+\tilde{q}_{r-1},
\end{align}
\end{subequations}
as depicted in Fig.~\ref {fig:tensor_QN}(c). At the leftmost tensor, the conservation rule takes the form:
\begin{subequations}
\begin{align}
&\tilde{Q}_0\coloneq\tilde{Q}_1-Q_1=0,\\
&\tilde{q}_0\coloneq\tilde{q}_1-q_1=0,
\end{align}
\end{subequations}
while at the rightmost tensor, they take the form:
\begin{subequations}
\begin{align}
&\tilde Q_N \coloneq \tilde{Q}_{N-1}+Q_N=\sum_r Q_r=Q,\\
&\tilde q_N \coloneq \tilde{q}_{N-1}+q_N=\sum_r q_r=0.
\end{align}
\end{subequations}
The imposition of these local conservation  rules in each individual tensor results in an MPS representation that is block diagonal and respects global symmetries. \\

One can further define a MPO representation of an arbitrary operator $\hat{O}$ as follows:
\begin{align}
    \hat{O} = \sum_{    p_1,\ldots,p_N}\sum_{p_1',\ldots,p_N'}&W^{a_1}_{p_1,p'_1}W^{a_1,a_2}_{p_2,p'_2}\ldots W^{a_{N-1}}_{p_N,p'_N} \\ \nonumber
    &\ket{p_1,\ldots,p_N}\bra{p_1',\ldots,p_N'}.
\end{align}
Here, sitewise rank-4 $W^{a_{r-1},a_r}_{p_r,p'_r}$ (and rank-3 $W^{a_1}_{p_1,p'_1}$ and $W^{a_{N-1}}_{p_N,p'_N}$) matrices carry two physical legs instead of one. A generic quantum-many-body Hamiltonian with local or nearest-neighbor interactions admits a compact definition for the local tensors $W_{p_r,p'_r}$. The LSH Hamiltonian, as defined in Sec.~\ref{sec:LSH-formulation}, consists of 1-body and 2-body operators, and hence can be represented efficiently. Concretely, the Hamiltonian can be written as
\begin{equation}
    \hat{H} = v_L \hat{F}_1 \hat{F}_2 \ldots \hat{F}_N v_R^T,
\end{equation}
where $v_L=(1,0,0,0,0,0)$, $v_R=(0,0,0,0,0,1)$, and
\begin{align}
    \hat{F}_r \coloneq \begin{pmatrix}
        \hat{I}_r  & \hat{O}^1_r & \hat{O}^2_r & \hat{O}^3_r & \hat{O}^4_r & \hat{D}_r\\
        0 & 0 & 0 & 0 & 0 & \hat{O}^5_r\\
        0 & 0 & 0 & 0 & 0 & \hat{O}^6_r\\
        0 & 0 & 0 & 0 & 0 & \hat{O}^7_r\\
        0 & 0 & 0 & 0 & 0 & \hat{O}^8_r\\
        0 & 0 & 0 & 0 & 0 & \hat{I}_r\\
    \end{pmatrix}.
\end{align}
Here, 
\begin{align}
    &\hat{O}^1_r \coloneq \frac{1}{\sqrt{\hat{N}_L(r)+1}}\hat{\mathcal{S}}^{++}_o(r),
    ~~ \hat{O}_r^2 \coloneq \frac{1}{\sqrt{\hat{N}_L(r)+1}}\hat{\mathcal{S}}^{+-}_o(r),
    \nonumber\\
    &\hat{O}^3_r \coloneq \frac{1}{\sqrt{\hat{N}_L(r)+1}}\hat{\mathcal{S}}^{--}_o(r),
    ~~ \hat{O}^4_r \coloneq \frac{1}{\sqrt{\hat{N}_L(r)+1}}\hat{\mathcal{S}}^{-+}_o(r),
    \nonumber\\
    &\hat{O}^5_r \coloneq \hat{\mathcal{S}}^{+-}_i(r)\frac{1}{\sqrt{\hat{N}_R(r)+1}},
    ~~ \hat{O}^6_r \coloneq \hat{\mathcal{S}}^{--}_i(r)\frac{1}{\sqrt{\hat{N}_R(r)+1}},
    \nonumber\\
    &\hat{O}^7_r \coloneq \hat{\mathcal{S}}^{-+}_i(r)\frac{1}{\sqrt{\hat{N}_R(r)+1}},
    ~~ \hat{O}^8_r \coloneq \hat{\mathcal{S}}^{++}_i(r)\frac{1}{\sqrt{\hat{N}_R(r)+1}},
    \nonumber\\
    &\hat{D}_r \coloneq 
    \frac{\hat{N}_{L/R}^2(r)}{4}(1-\delta_{r,N})+\frac{\hat{N}_{L/R}(r)}{2}(1-\delta_{r,N}) \nonumber\\
    & \hspace{2.5 cm}+ \mu(-1)^r\hat{n}_i(r)+\mu(-1)^r\hat{n}_o(r).
\end{align}
Each of the operators in the above matrix is defined at a single site and, hence, carries the additional $(p_r,p'_r)$ physical indices. This will result in a block-diagonal structure on the MPOs. \\

The operators considered in this work conserve the global symmetries, and this structure can be incorporated into an MPO in direct analogy with the MPS. The two physical legs of the MPO tensor at site $r$ (labeled by $p_r$ and $p'_r$) are, respectively, assigned the local charges $(Q_r,q_r)$ and $(Q'_r,q'_r)$, while the virtual bond index to the right of site $r$ carries a charge $(\tilde Q_r,\tilde q_r)$. The index $p'_r$ labels the ``incoming" state while the index $p_r$ labels the ``outgoing" state. This means that the bulk MPO tensors are block-sparse, with the blocks defined by the conservation (fusing) rules
\begin{subequations}
\begin{align}
&\tilde Q_r=\tilde Q_{r-1}+\bigl(Q_r-Q'_r\bigr),\\
&\tilde q_r=\tilde q_{r-1}+\bigl(q_r-q'_r\bigr),
\end{align}
\end{subequations}
which are depicted in Fig.~\ref{fig:tensor_QN}(d). For charge-preserving operators, the MPO carries zero total charge, implemented via the boundary conditions:
\begin{subequations}
\begin{align}
&\tilde Q_0\coloneq \tilde Q_1 - (Q_1-Q'_1)=0,\\
&\tilde q_0\coloneq \tilde q_1 - (q_1-q'_1)=0,
\end{align}
\end{subequations}
at the left end, and the boundary conditions 
\begin{subequations}
\begin{align}
&\tilde Q_N \coloneq \tilde Q_{N-1} + (Q_N-Q'_N)=0,\\
&\tilde q_N \coloneq \tilde q_{N-1} + (q_N-q'_N)=0,
\end{align}
\end{subequations}
at the right end. These conditions are equivalent to the conditions  $\sum_r (Q_r-Q'_r)=0$ and $\sum_r (q_r-q'_r)=0$.\\

Finally, a crucial component of the LSH framework is the AGL, which needs to be preserved throughout the computations. One way to impose the AGL at each link is via a penalty term, as described in Sec.~\ref{sec:static}. However, for the (1+1)D LSH framework, it has been established~\cite{Mathew:2022nep, Mathew:2024bed} that conservation of all the AGLs follows directly from conservation of only the global symmetries. As described above, the MPS and MPOs in this work preserve global symmetries by construction and, therefore, no additional effort is needed to preserve AGLs. 

We emphasize that the primary advantage of the LSH formulation is that it eliminates the need to impose the SU(2) symmetry locally: the degrees of freedom are gauge-invariant by construction. Since many of the existing TN packages incorporate global Abelian symmetries in the MPS and MPO constructions, one can readily use these packages to perform computations within the LSH framework. We rely upon \texttt{ITensors.jl} package~\cite{itensor} to evaluate properties of extended static and dynamical strings in the following sections.

%%%%%%%%%%%%%%%%%%%%%%%%%%%%%%%%%%%%%%%%%%%%%%%%%%%%%%%%%%%%%%%%%%%%%%%%%%%%%%%%%%%%%%%%%%%%%%%%%%%%%%%%%%%%%%%%%%%%%%%%%%%%%%%%%%%%%%%%%%%%%%%%%%%%%%%%%%%%%%%%%%
\section{String-breaking statics
\label{sec:static}}
\noindent
The potential between two static charges in a confining theory probes confinement, and identifies the energy scales relevant for the string-breaking statics and dynamics. The ground- and low-lying excited-states energies, as well as the static potential, have been studied in the (1+1)D SU(2) LGT using TN methods in both truncated Kogut-Susskind~\cite{Kuhn:2015zqa} and locally gauge-integrated bases~\cite{Banuls:2017ena}, as well as with Gaussian variational states~\cite{Sala:2018dui}. In this section, using the LSH-based MPS and MPO constructions introduced in the previous section, we compute the ground-state energy of the (1+1)D SU(2) LGT in the presence of a pair of static charges, in order to obtain the static potential and string tension in the continuum and infinite-volume limits. We present the computation of the ground-state energy in the absence of static charges in Appendix~\ref{app:ground-state}. These computations provide a good benchmark for the LSH-based TN, and set the stage for the dynamical computations in the next section.\\

\subsection{Computational setup for static potential}
We introduce a pair of static charges at sites $r_1$ and $r_2$ and compute the system's ground-state energy. These static charges are introduced as follows. We first initialize an MPS which obeys a modified AGL:
\begin{align}
\label{eq:penalty}
    \hat{N}_L(r)-\hat{N}_R(r+1)=
    \hat{\mathcal{Q}}(r,r+1).
\end{align}
Here, 
\begin{align}
    \hat{\mathcal{Q}}(r_1,r_1+1)=-1, \ \hat{\mathcal{Q}}(r_2-1,r_2)=1,
    \label{eq:static_config}
\end{align}
with $r_1<r_2-1$, and $\mathcal{Q}(r,r+1)=0$ for $r\neq r_1,r_2-1$. States in this AGL sector can be created by applying the operator $\hat{\mathcal{L}}^{++}(r_1+1)\ldots\hat{\mathcal{L}}^{++}(r_2-1)$ to any state with \emph{zero} static charge carrying the global quantum numbers $(Q,q)=(N,0)$. We then add the penalty term:
\begin{align}
\label{eq:penalty}
    \hat{H}_\text{P} \coloneq \Lambda_\text{P}\sum_r \Big[\hat{N}_L(r)-\hat{N}_R(r+1)-
    \hat{\mathcal{Q}}(r,r+1)\Big]^2
\end{align}
to the LSH Hamiltonian and perform the DMRG to evolve the initial MPS toward the ground state. The coefficient of the penalty term, $\Lambda_\text{P}$, is taken to be large compared to the single-site energy; it is observed that the computations are effectively restricted to the desired static-charge sector with this choice. Further details on this construction, and how it relates to the notion of static charges in the Kogut-Susskind formulation are presented in Appendix~\ref{app:static_charges}. The positions of the static charges, $r_1$ and $r_2$, are chosen to lie symmetrically about the midpoint of the lattice. Given the staggering of the Hamiltonian, even and odd choices for $r_1$ and $r_2$ lead to different kinds of strings. In particular, we choose $r_1$ to be an even (particle) site and $r_2$ to be an odd (antiparticle) site. \\

Each computation is specified by a set of values $(\frac{m}{g},N,x,j_{\rm max},D,gl)$, corresponding to lattice spacing $ga=\frac{1}{\sqrt{x}}$, physical volume $gL=\frac{N}{\sqrt{x}}$, and physical static-string length $gl \coloneq ga(r_2-r_1)$ (in units of inverse coupling, $g^{-1}$).\footnote{With $gl=0$ denoting no static charges, hence no string.} Recall that $N$ is the number of staggered lattice sites, $x$ is the dimensionless hopping parameter, $\frac{m}{g}$ is the fermion mass in units of coupling, $j_{\rm max}$ is the electric-flux cutoff, and $D$ is the bond dimension. The continuum value at a fixed value for $\frac{m}{g}$ is estimated by taking the successive limits: 1) $D \rightarrow \infty$, 2) $j_{\rm max}\rightarrow \frac{N}{2}$, 3) $N \rightarrow \infty$, and finally, 4) $ga \rightarrow 0$ (or $x \to \infty$). For the rest of this section, we fix $\frac{m}{g}=0.5$ but report the value of the string tension at two other values of the bare mass.

\subsection{Static potential and string tension}
The static potential is defined as the difference between the ground-state energy in the presence and in the absence of static charges:
\begin{align}
\label{eq:static-potential}
    V(gl)\coloneq 
    E(gl)-
    E(gl=0).
\end{align}
Here, it is assumed that the extrapolation to zero MPS truncation error has already been performed, and the remaining arguments ($\frac{m}{g},N,x,j_{\rm max}$) (which are held fixed) are suppressed for brevity. Appendix~\ref{app:ground-state} provides a description of the extrapolation in the MPS truncation error. \\

\begin{figure*}[t!]
    \centering
    \includegraphics[width=\linewidth]{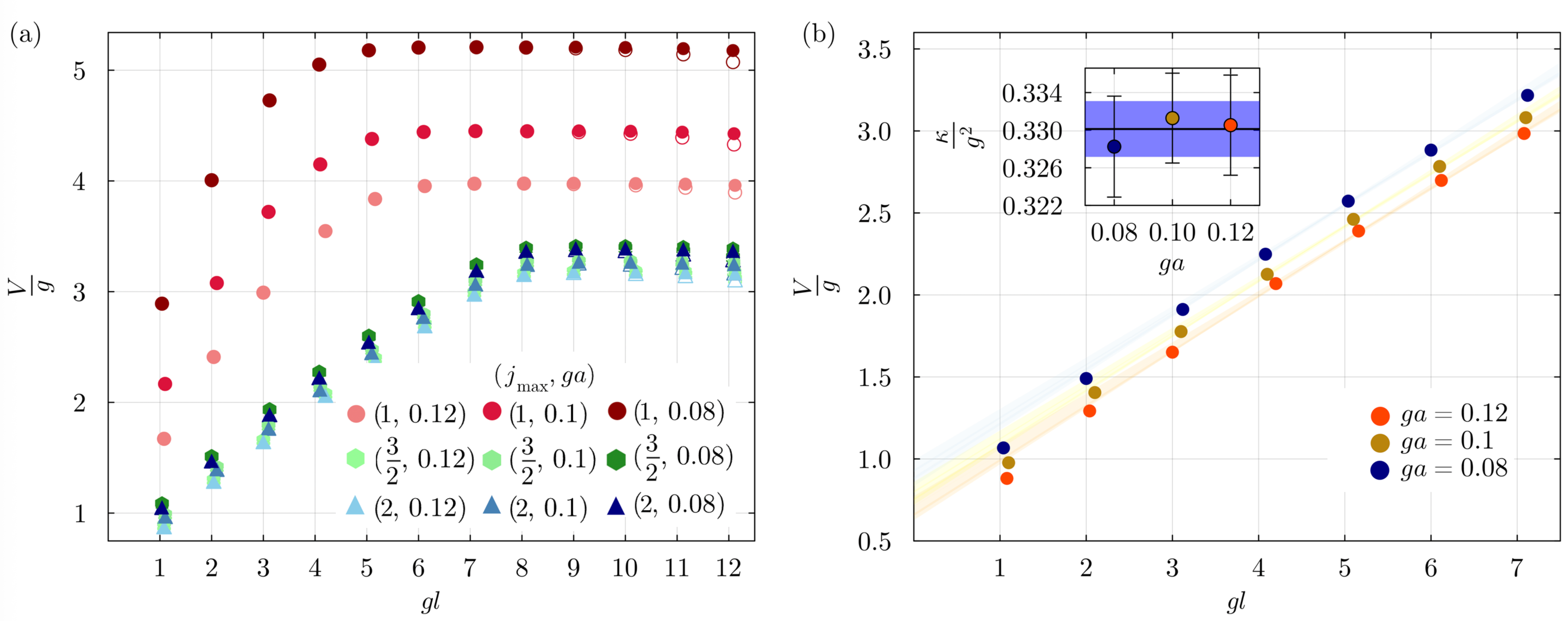}
    \caption{(a) Static potential as a function of $gl$ for cutoff values $j_{\rm max}\in\{1,\frac{3}{2},2\}$ and lattice-spacing values $ga\in\{0.08,0.1,0.12\}$. Solid markers correspond to system volume $gL=16$ while hollow markers correspond to system volume $gL=14$. An MPS-error truncation extrapolation is already performed but the resulting error bars are too small to be visible. Values are computed at a fixed $\frac{m}{g}=0.5$. (b) Static potential plotted against the distance, $gl$, between static charges in the thermodynamic limit at a fixed $\frac{m}{g}=0.5$. The lines show fits to all contiguous regions in $gl$ with 3 to 5 points in the linear part of the potential between $gl\approx 3$ and $gl\approx 7$. Inset displays the weighted average of the slopes of all these lines used to estimate the string tension at various values of $ga$. The continuum estimate, in turn, is the weighted average of these string-tension estimates at various values of $ga$, and yields the value $\frac{\kappa}{g^2}=0.330(3)$.}
    \label{fig:static-potential}
\end{figure*}

We separately compute, using DMRG, $E(gl)$ and $E(0)$, and hence the difference in Eq.~\eqref{eq:static-potential}. Figure~\ref{fig:static-potential}(a) displays $V(gl)/g$ as a function of $gl$ for $gL \in \{14,16\}$, $ga \in \{0.12,0.10,0.08\}$, and $ j_{\rm max} \in \{1,\frac{3}{2},2\}$ at fixed $\frac{m}{g}=0.5$. Here, a physical volume of $gL=16$ corresponds to taking $N=136,160$, and $200$ for $ga=0.12,0.1$, and $0.08$, respectively, while a physical volume of $gL=14$ corresponds to taking $N=120,140$, and $176$ for $ga=0.12,0.1$, and $0.08$, respectively.

In each case, the potential first rises roughly linearly with $gl$ before it flattens into a plateau beyond some critical charge separation. This behavior is a signature of confinement: It indicates that at short separations, the ground state consists of a continuous flux tube connecting the static charges. As this separation increases, the energy of the flux tube increases until a critical separation is reached. At this critical distance, this unbroken-string configuration becomes comparable in energy to a broken-string configuration. The broken-string configuration consists of dynamical charges that form smaller flux tubes with the static charges, leading to a screening of the electric field in the central region. Therefore, the ground-state energy becomes independent of the separation between the static charges beyond this critical separation. However, due to the finite total volume of the system, boundary effects cause this plateau to slope downwards for the smaller volume $gL=14$. Increasing the system volume to $gL=16$ alleviates this effect. For smaller separations, the difference between the value of the potential at the two volumes considered is negligible. \\

There is a substantial difference in the behavior of the static potential for the lowest cutoff considered, i.e., $j_{\rm max}=1$, and for the two higher values $j_{\rm max} \in \{\frac{3}{2},2\}$. For $j_{\rm max}=1$, both the magnitude of the static potential and the slope of the linear part have a pronounced dependence on $ga$. In contrast, the results for $j_{\rm max}\in \{\frac{3}{2},2\}$ are nearly coincident and exhibit only mild lattice-spacing dependence. This observation suggests that the energies are nearly converged in flux-cutoff values $j_{\rm max}\in \{\frac{3}{2},2\}$. For all flux-cutoff values, the critical-separation value appears insensitive to lattice spacing. However, the critical separation for $j_{\rm max}=1$ is shorter than that for $j_{\rm max}\in\{\frac{3}{2},2\}$. \\

The slope of the linear part of the static potential is referred to as the string tension. It provides a natural mass scale for theory, and is a diagnostic for confinement. Using the data for energies at various values of $(gL, ga, j_{\rm max})$, the continuum string tension can be estimated. The bond-dimension and flux-cutoff extrapolations proceed as described in Appendix~\ref{app:ground-state}. Since the static potential is the difference between the ground-state energy in the presence and absence of a defect (the two static charges) in the bulk of the lattice, it is not expected to depend on $gL$ for sufficiently large $gL$ values. As observed earlier in Fig.~\ref{fig:static-potential}(a), this is indeed the case for $gl\leq 7$ at the minimum and maximum considered system volumes $gL=14$ and $gL=16$, respectively (we additionally consider the volume $gL=15$). However, when $gl> 7$, the static potential becomes sensitive to $gL$. This is because the string length becomes comparable to the system volume and finite-size effects become important. In fact, the string-breaking transition occurs at $gl\sim 7$. Since we are interested in the linear regime of the potential, we will estimate the thermodynamic limit of the static potential in this regime by averaging the values at different volumes, weighted by the bond-dimension and flux-cutoff extrapolation uncertainties.  

The resulting plot of the linear part of the static potential in the thermodynamic limit is shown in Fig.~\ref{fig:static-potential}(b) for various values of $ga$. The instantaneous slope,
\begin{equation}
    \kappa(gl+\frac{1}{2})\coloneq \frac{V(gl+ga)-V(gl)}{ga},
\end{equation}
is relatively stable between $gl\approx3$ and $gl\approx 7$, from which one can estimate the string tension. We perform linear fits for all possible contiguous blocks of string lengths $gl\approx3$ to $gl\approx7$ (which amounts to performing fits involving 3 to 5 points) for each lattice spacing. Thereafter, the string tension is computed by taking the mean of the slopes corresponding to these linear fits (weighted by the uncertainty associated with the corresponding slope estimate). The results from this computation are shown in the inset plot in Fig.~\ref{fig:static-potential}(b). As the value of the string tension does not change significantly with lattice spacing within the uncertainties, the continuum string tension is estimated by taking the weighted mean of the values at different lattice spacings, yielding a value of
\begin{align}
\frac{\kappa}{g^2}=0.330(3)
\end{align}
at $\frac{m}{g}=0.5$. For comparison, the static potential was computed in Ref.~\cite{Sala:2018dui} using a Gaussian variational ansatz and in the gauge-integrated basis, which requires no truncation of the flux quantum number, at the larger values of $ag \in \{0.707,1.01\}$, and for the masses $\frac{m}{g} \in \{0.707, 0.354\}$. While no explicit numerical values were reported for the string tension, the value for $\frac{\kappa}{g^2}$ can be estimated to be around $0.3-0.4$ based on the plots presented in that reference. 

In the next section, we will study the dynamics of string breaking for a lighter and a heavier fermion mass compared to the one considered in this section, $\frac{m}{g} \in \{0.2,2.0\}$, and a fixed value of $(N,x,j_{\rm max})=(128,16,\frac{5}{2})$. For the lighter mass, the instantaneous slope of the static potential starts at 0.5 for the smallest possible string length of $gl=0.75$ and drops continuously until it hits zero at around $gl=10$. The continuous drop of the potential implies that there is no linear regime from which one could estimate a string tension. Thus, we just note that the instantaneous slope at the string-length of $gl=4$ (which will be the length of the dynamical string whose evolution we study in the next section) turns out to be $\approx 0.293$. In contrast, for the heavier mass, the instantaneous slope of the static potential stabilizes at $\approx 0.373$ for $gl\geq 2$, and stays constant till the largest considered string length of $gl=10$. Therefore, with all other parameters fixed, the string tension increases with $\frac{m}{g}$. In fact, the heavy-mass string tension is close to that of the pure Yang-Mills theory value $\frac{\kappa}{g^2}=\frac{3}{8}=0.375$~\cite{Reinhardt:2011fq}. This is expected since the fermions approach their static limit.

\subsection{A microscopic look into unbroken and broken strings}
\begin{figure*}[t!]
    \centering
\includegraphics[width=\linewidth]{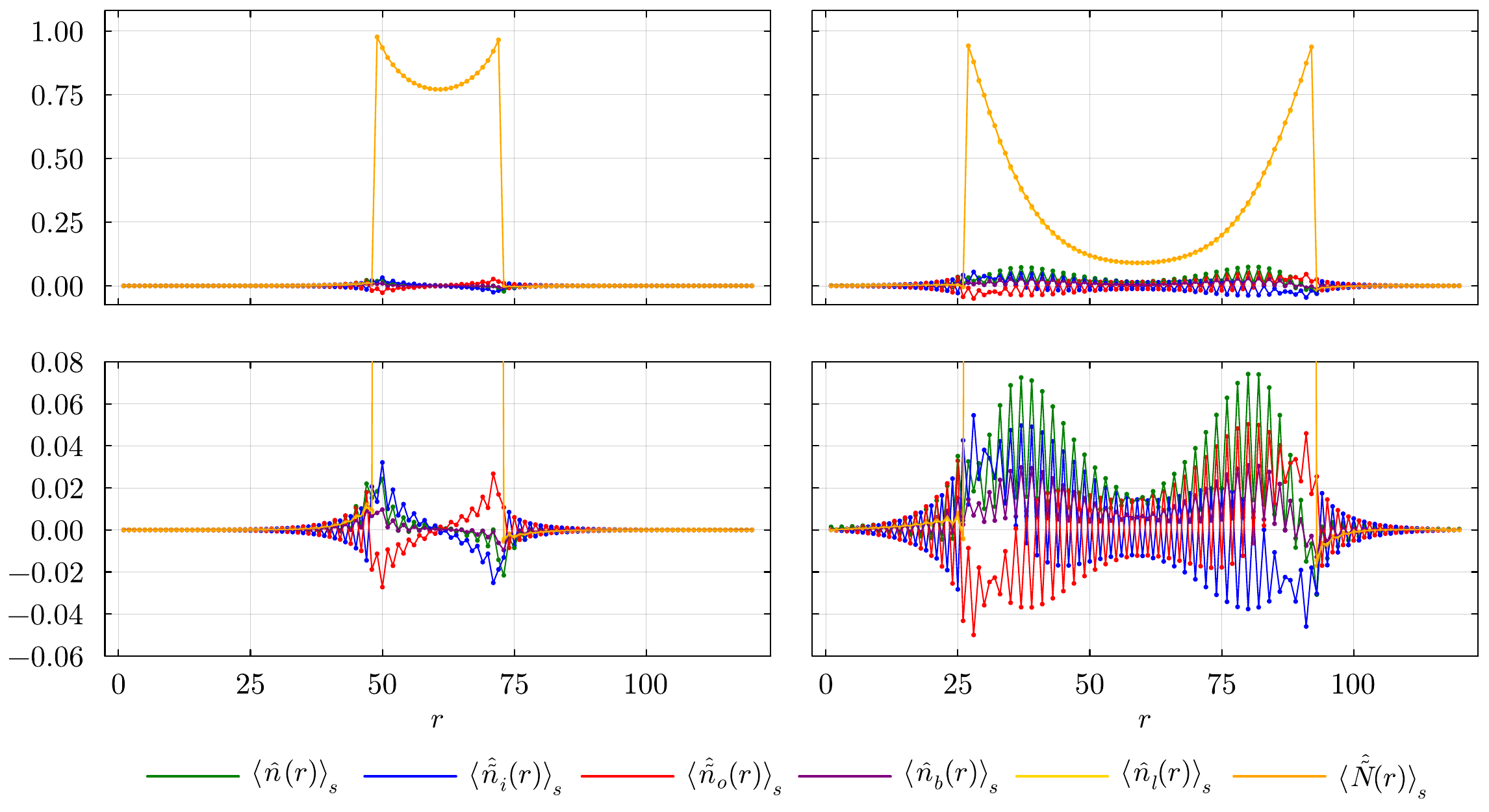}
    \caption{Values of the vacuum-subtracted total excitation, string-in, string-out, baryon, loop, and symmetrized flux expectation values ($\langle\hat{n}(r)\rangle_s$, $\langle\hat{\tilde{n}}_{i}(r)\rangle_s$, $\langle\hat{\tilde{n}}_{o}(r)\rangle_s$, $\langle\hat{n}_b(r)\rangle_s$, $\langle\hat n_l(r)\rangle_s$, and $\langle\hat{\tilde{N}}(r)\rangle_s$, respectively) for the ground state in the presence of static charges separated by a distance $gl=3$ (left) and $gl \approx 8$ (right) for $(\frac{m}{g},N,ga,j_{\rm max},D)=(0.5,128,0.12,2,200)$. Static charges are inserted at sites $\{r_1, r_2\}=\{48, 73\}$ and $\{r_1, r_2\}=\{26, 93\}$ for $gl=3$ and $gl \approx 8$, respectively. The bottom row denotes the same plots as in the top row but with a smaller range along the y-axis to reveal finer features of the matter excitations.}
    \label{fig:static_excitation}
\end{figure*}
How do the unbroken- and broken-string ground states differ microscopically? To address this question, we probe the particle and flux contents of the states by computing the expectation values of total matter excitation $\hat{n}(r)$, string-in $\hat{\tilde{n}}_{i}(r)$, string-out $\hat{\tilde{n}}_{o}(r)$, baryon $ \hat{n}_b(r)$, loop $\hat n_l(r)$, and symmetric flux $\hat{\tilde{N}}(r)$ at $\frac{m}{g}=0.5$. These quantities were introduced at the end of Sec.~\ref{sec:LSH-formulation}. In the strong-coupling limit, all the ground-state excitation numbers are trivially zero in the absence of static charges. The introduction of two static charges, one at $r_1$ and one at $r_2>r_1+1$ increases the loop quantum number by one for sites $r_1+1$ through $r_2-1$, while leaving the matter quantum numbers unchanged throughout. 

In general, ground-state excitation expectation values in the absence of static charges assume a constant nonvanishing value in the bulk of the lattice (as the Hamiltonian is translationally invariant in the thermodynamic limit), but fluctuate closer to the edges due to the open boundaries (see Fig.~\ref{fig:vacuum} for examples at two other mass values). In particular, for the parameter values, $(\frac{m}{g},N,ga,j_{\rm max},D)=(0.5,128,0.12,2,200)$, these expectation values exhibit a roughly constant value between the lattice sites $20\leq r \leq 108$. For some bulk point $r_0$, these expectation values are given by $\langle\hat{n}(r_0)\rangle=0.832$, $\langle\hat{\tilde{n}}_{i}(r_0)\rangle=\langle\hat{\tilde{n}}_{o}(r_0)\rangle=0.242$, $\langle \hat{n}_b(r_0)\rangle=0.173$, $\langle \hat{n}_l(r_0)\rangle=0.151$, and $\langle\hat{\tilde{N}}(r_0)\rangle=0.394$. In the presence of static charges, the creation of matter excitations modifies the above excitation profile.

Consider two static-charge separations $gl=3$ and $gl=8.04$, which correspond to inserting static charges at $\{r_1, r_2\}=\{48, 73\}$ and $\{r_1, r_2\}=\{26, 93\}$, respectively. Figure~\ref{fig:static_excitation} displays the change in the excitation expectation values due to the introduction of static charges relative to the ground state without static charges (and hence the notation $\langle \cdot \rangle_s$ to denote this subtraction). For both string lengths, one observes an alternating pattern of string-in and string-out excitations between adjacent sites in the exterior of the string close to the string ends, with diminishing amplitude away from the string ends. These point to a small production of pairs of fermions adjacent to each other (i.e., production of minimal-length meson strings). This production is in addition to a small net baryon production. The behavior in the interior of the string, on the other hand is different. The string-in excitations dominate over the string-out excitations in the left side of the string close to the string ends while the pattern reverses in the right side. These string-in and -out states consume some of the flux sourced by the static charges at the left and right end of the static string, respectively. The hybridization of a static charge with several nearby dynamical charges, which yields meson strings of various lengths, is responsible for the flux depletion. For an interior region away from the string ends, the alternating pattern of string-in and string-out excitations between adjacent sites is restored. The pattern correspond to the creation of various minimal-length mesons, which do not contribute to the loop excitation.
The excitation pattern is similar in both string-length cases but the magnitude of excitation numbers are larger in the $gl \approx 8$ case, causing more pronounced effects. In particular, the flux in the middle is nearly diminished for the longer string, hence the notion of a broken string.

%%%%%%%%%%%%%%%%%%%%%%%%%%%%%%%%%%%%%%%%%%%%%%%%%%%%%%%%%%%%%%%%%%%%%%%%%%%%%%%%%%%%%%%%%%%%%%%%%%%%%%%%%%%%%%%%%%%%%%%%%%%%%%%%%%%%%%%%%%%%%%%%%%%%%%%%%%%%%%%%%%
\section{String-breaking dynamics
\label{sec:dynamic}}
\noindent As the static-potential study confirmed, the (1+1)D SU(2) model exhibits confinement of color charges: the ground-state string consisting of static color charges breaks once the charges' separation reaches a critical value. How does string breaking occur in nonstatic, far-from-equilibrium conditions, such as in a quench? In this section, we set up a dynamical string-breaking process and study the resulting outcome in detail. Specifically, in place of static charges, we apply to the interacting vacuum a pair of dynamical color charges with an electric flux tube connecting them. Such a state is not an eigenstate of the Hamiltonian; hence it evolves nontrivially under Hamiltonian dynamics. With the aid of the quantities the TN method gives access to, and via the clear lens of the LSH excitations, we illuminate the complex picture of dynamical string breaking. We first present details of the set up; then focus on the underlying microscopic dynamics by contrasting the interacting vacuum, the initial meson-string state, and the time-evolved states; finally, we support the microscopic picture via the time- and site-resolved evolution of energy, entanglement entropy, and correlations.

%%%
%%%
\subsection{Computational setup for quench dynamics\label{sec:string-setup}}
Consider the meson-string operator
\begin{align}
    \label{eq:String_operator_LSH}
    \hat{S}_{r,\Delta r} =&\sum_{\sigma_1,\sigma_2,
    \ldots,\sigma_{\Delta r}=\pm}\frac{1}{\sqrt{\hat{N}_L(r)+1}}\hat{\mathcal{S}}_o^{+,\sigma_1}(r)
    \nonumber\\
    &\hat{\overline{\mathcal{L}}}^{\sigma_1,\sigma_2}(r+1)
    \ldots 
    \hat{\overline{\mathcal{L}}}^{\sigma_{\Delta r-1},\sigma_{\Delta r}}(r+\Delta r-1)
    \nonumber\\
    &\hat{\mathcal{S}}_i^{\sigma_{\Delta r},-}(r+\Delta r)\frac{1}{\sqrt{\hat{N}_R(r+\Delta r)+1}},
\end{align}
expressed in terms of the LSH operators. The operators $\mathcal{\overline{L}}$ are redefinitions of the original loop operators $\mathcal{L}$ with some extra normalization factors. The definitions of these operators, and of the string-in/out operators $\mathcal{S}_{i/o}$, can be found in Appendix~\ref{app:lsh-formalism}. This meson-string operator in Eq.~\eqref{eq:String_operator_LSH} creates a fermion excitation at position $r$ and annihilates a fermion excitation at position $r+\Delta r$; the loop operators connecting the two fermionic operators ensure gauge invariance, see Appendix~\ref{app:meson-string}. Since the ultimate interest is in dynamical strings with Dirac fermions at the endpoints, we further introduce the Dirac meson-string operator
\begin{align}
    \label{eq:string_op_dirac}
    \hat S^{\rm Dirac}_{r,\Delta r}\coloneq \frac{1}{ga}\Big[\hat{S}_{r,\Delta r}-\hat{S}_{r+1,\Delta r}\Big],
\end{align}
with odd $r$ and even $\Delta r$. These choices for $r$ and $\Delta r$ ensure that $\hat S^{\rm Dirac}_{r,\Delta r}$ gets contributions from two possible Dirac-fermion bilinear structures, as shown in Appendix~\ref{app:meson-string}. For brevity, in what follows, we use the terms Dirac meson-string operator and meson-string operator interchangeably.\\

We apply the Dirac meson-string operator to the interacting vacuum $\gket{\Omega}$, i.e., the ground state of the Hamiltonian in Eq.~\eqref{eq:H_dimless} in the zero static-charge sector, to form the initial state for real-time evolution:
\begin{align}
    \gket{\phi(t=0)}\eqcolon\gket{\phi(0)} =\frac{1}{\mathcal{N}_{0}}\hat S_{r,\Delta}^{\rm Dirac}\gket{\Omega},
\end{align}
where $\mathcal{N}_0 \coloneq \norm{\hat S_{r,\Delta}^{\rm Dirac}\gket{\Omega}}$. This state is then subjected to the unitary evolution
\begin{align}
    \gket{\phi(t)}=e^{-i{\tilde{t}} {\hat{\tilde{H}}}}\gket{\phi(0)},
\end{align}
where ${\hat{\tilde{H}}}$ is the dimensionless Hamiltonian in Eq.~\eqref{eq:H_dimless}, and $\tilde t$ is the dimensionless time $\tilde t \coloneq \frac{ag^2}{2}t=\frac{1}{2}x^2\big(\frac{t}{a}\big)$. Here, we use ${\hat{\tilde{H}}}$ for our computations and quote energies in the corresponding dimensionless unit. Nonetheless, we opt to quote the results as a function of time $gt$ (and not $\tilde t$). The initial mesonic-string state enjoys an excess energy compared with the interacting vacuum, due to the additional dynamical charges and the flux lines connecting them. This excess energy is transferred to various excitations throughout the time dynamics. These dynamics are governed by the fermion mass, hopping strength, and the length of the string. We probe, in particular, the fermion-mass dependence, and analyze the evolutions in detail in this section.

DMRG is used to approximate the ground state $\gket{\Omega}$ up to a truncation error $\varepsilon=10^{-8}$ on an MPS. Thereafter, an MPO representation of $\hat S_{r,\Delta r}^{\rm Dirac}$ is constructed and applied to the ground-state MPS. This raises the bond dimension of the state, but we truncated it back to a lower-bond-dimension representation up to a truncation error $\varepsilon=10^{-8}$. This truncation allows us to start the evolution algorithm with a sufficiently low bond dimension. This state is evolved in time using a single-site time-dependent variational-principle (TDVP) algorithm for the MPS ansatzes~\cite{Paeckel:2019yjf,Vanderstraeten:2019voi} coupled with a Krylov-basis-expansion protocol~\cite{Yang:2020ykn} to dynamically expand bond dimensions to no more than $D_\text{max} = 600$ during the evolution. While we do not perform extrapolations in the bond dimension explicitly, we demonstrate that the results are sufficiently converged at the chosen values of $D_\text{max}=600$. The effects of bond-dimension truncations are discussed in more detail in the end of this section and later in an appendix. 

All computations in this section are carried out with these sets of parameters: $j_{\rm max}=\frac{5}{2}$, $N=128$, $x=16$, and two different values of bare fermion mass, $\frac{m}{g}=0.2, 2.0$. Evolutions are performed in steps of size $g\,dt = 0.08$ for a total evolution time $gt=8$ for the lighter mass and $gt=16$ for the heavier mass (so as to keep the bond-dimension truncation error under control). We do not perform an extrapolation in the flux cutoff either, but we demonstrate that the cutoff value $j_{\rm max}=5/2$ is large enough to ensure convergence at all times. The explicit extrapolations of the dynamical observables considered in this section to the thermodynamic and continuum limits are computationally demanding, and are left to future work. Nonetheless, we choose the system size and lattice spacing sufficiently close to the thermodynamic limit and away from the strong-coupling limit to allow for nontrivial dynamics.

%%%%%%
%%%%%%
\subsection{A microscopic look into the dynamics
\label{sec:microscopic}}
One can characterize the interacting-vacuum, initial meson-string, and evolved meson-string states using elementary onsite gauge-invariant excitations introduced in Sec.~\ref{sec:LSH-formulation}, namely the total matter excitation, $\hat{n}(r)$, string-in, $\hat{\tilde{n}}_{i}(r)$, string-out, $\hat{\tilde{n}}_{o}(r)$, baryon, $\hat{n}_{b}(r)$, loop, $\hat n_l(r)$, and symmetrized flux, $\hat{\tilde{N}}(r)$, expectation values. Aided by this analysis, we illuminate dynamical processes that contribute to the evolution and breaking of the meson string.\\

\begin{figure*}[t!]
    \centering
\includegraphics[scale=0.505]{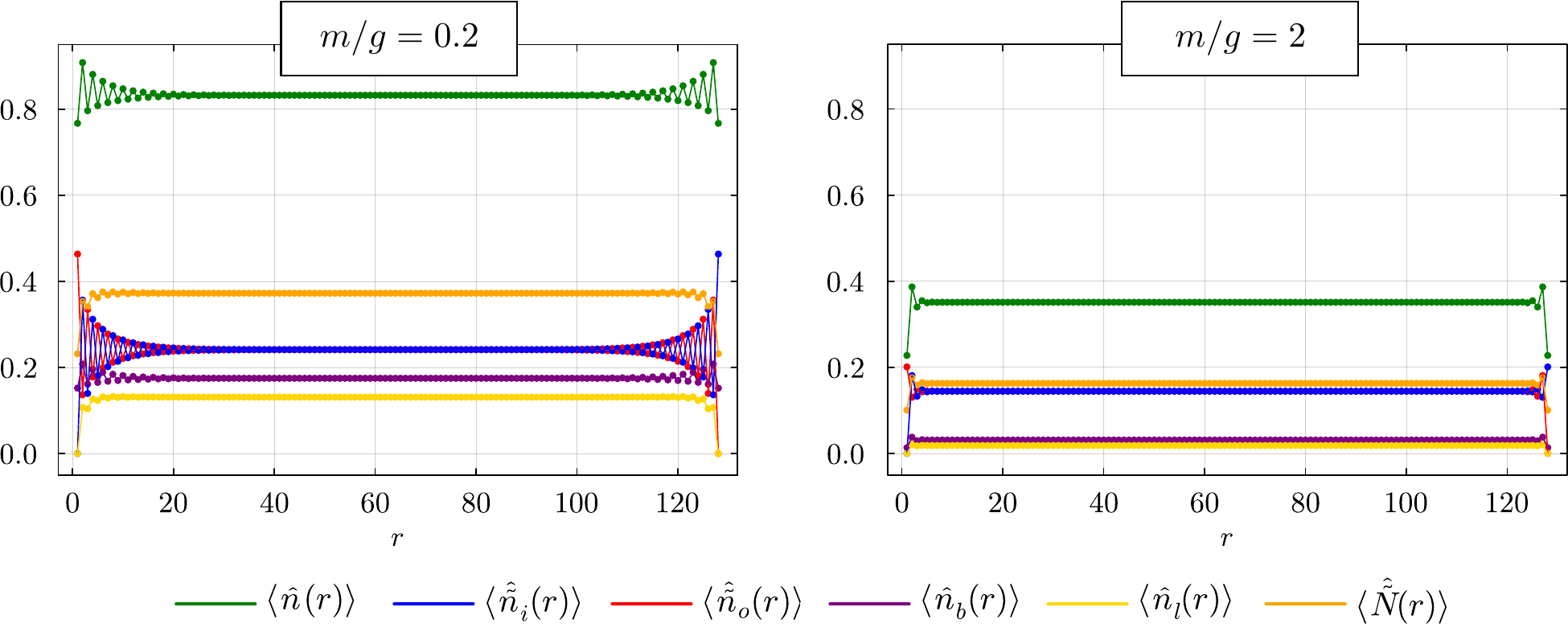}
    \caption{Values of the total excitation, string-in, string-out, baryon, loop, and symmetrized flux expectation values ($\langle\hat{n}(r)\rangle$, $\langle\hat{\tilde{n}}_{i}(r)\rangle$, $\langle\hat{\tilde{n}}_{o}(r)\rangle$, $\hat{n}_b(r)$, $\langle\hat n_l(r)\rangle$, and $\langle\hat{\tilde{N}}(r)\rangle$, respectively) for the interacting vacuum for two different bare fermion masses. The other parameters are $N=128$, $x=16$, and $j_{\rm max}=\frac{5}{2}$.}
    \label{fig:vacuum}
\end{figure*}

\subsubsection{Interacting-vacuum state}
We first compute the expectation values of the above quantum numbers throughout the spatial lattice for the interacting ground state for the parameter sets specified earlier ($N=128$, $x=16$, and $j_{\rm max}=\frac{5}{2}$ for two different values of bare fermion mass $\frac{m}{g}=0.2, 2.0$). These ground states will be the starting point for studying string-breaking dynamics.

Figure~\ref{fig:vacuum} displays $\langle\hat{n}(r)\rangle$, $\langle\hat{\tilde{n}}_{i}(r)\rangle$, $\langle\hat{\tilde{n}}_{o}(r)\rangle$, $\langle\hat{n}_b(r)\rangle$, $\langle\hat n_l(r)\rangle$, and $\langle\hat{\tilde{N}}(r)\rangle$ as a function of $r$ for both these masses. As expected, there are artifacts from the open boundary conditions in both cases, and they are more pronounced for the lighter mass. However, the large size of the lattice ensures that there is a translationally invariant bulk. Within this bulk, the total excitation number $\braket{\hat{n}}$ for the light mass (0.832) is much greater than that of the heavy mass (0.351), which is expected because the greater value of mass penalizes excitations. The inversion symmetry and 2-site translational invariance in the bulk of the lattice imply that the bulk string-in $\langle\hat{\tilde{n}}_{i}\rangle$ and string-out $\langle\hat{\tilde{n}}_{o}\rangle$ numbers are equal to each other. They are given by 0.241 and 0.144 for the light and heavy masses, respectively. This means that the baryon numbers $\braket{\hat{n}_b(r)}$ for the light and heavy masses are given by 0.175 and 0.031, respectively. Thus, string-in/out excitations constitute 58$\%$ and 82$\%$ of the total excitations for the light and heavy masses, respectively. The lower concentration of baryons for the heavier mass is due to the higher energy cost of baryon creation. 
The loop-number expectation value, $\langle\hat n_l(r)\rangle$, is 0.131 and 0.018 for the light and heavy masses, respectively.

The strong-coupling vacuum is characterized by $\langle\hat{n}(r)\rangle=\langle\hat{\tilde{n}}_{i}(r)\rangle=\langle\hat{\tilde{n}}_{o}(r)\rangle=\langle\hat{n}_b(r)\rangle=\langle\hat n_l(r)\rangle=\langle\hat{\tilde{N}}(r)\rangle=0$. Thus, the local excitation numbers for the heavier-mass vacuum are closer to those for the strong-coupling vacuum, owing to their smaller magnitude compared to those for the lighter mass. 
In fact, the overlap between the heavier-mass vacuum and the strong-coupling vacuum is $3.16 \times 10^{-6}$ while that between the lighter-mass and the strong-coupling vacuum is zero within numerical precision. Ultimately, these differences in the local excitation profile of the interacting vacua will impact the local excitation profile of the initial meson-string states and their subsequent evolution.

\subsubsection{Initial meson-string state}
\begin{figure*}[t!]
    \centering
    \includegraphics[width=\linewidth]{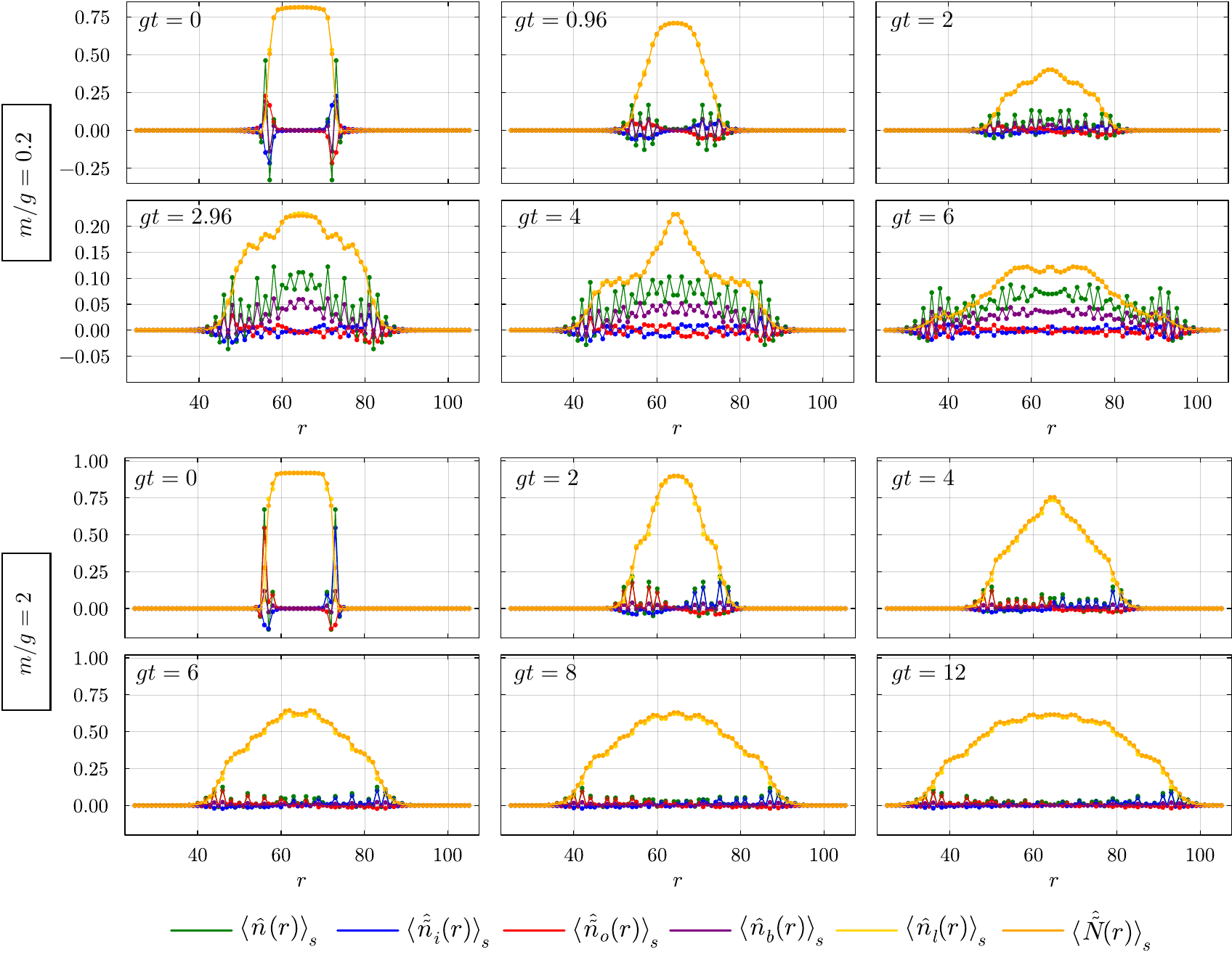}
    \caption{Values of the vacuum-subtracted total excitation, string-in, string-out, baryon, loop, and symmetrized flux expectation values ($\langle\hat{n}(r)\rangle_s$, $\langle\hat{\tilde{n}}_{i}(r)\rangle_s$, $\langle\hat{\tilde{n}}_{o}(r)\rangle_s$, $\langle\hat{n}_b(r)\rangle_s$, $\langle\hat n_l(r)\rangle_s$, and $\langle\hat{\tilde{N}}(r)\rangle_s$, respectively) in the initial meson-string states and a few subsequent times for two different bare fermion masses. The other parameters are $N=128$, $x=16$, and $j_{\rm max}=\frac{5}{2}$.}
    \label{fig:excitation_time_slices}
\end{figure*}
\begin{figure*}[t!]
    \centering
    \includegraphics[scale=0.585]{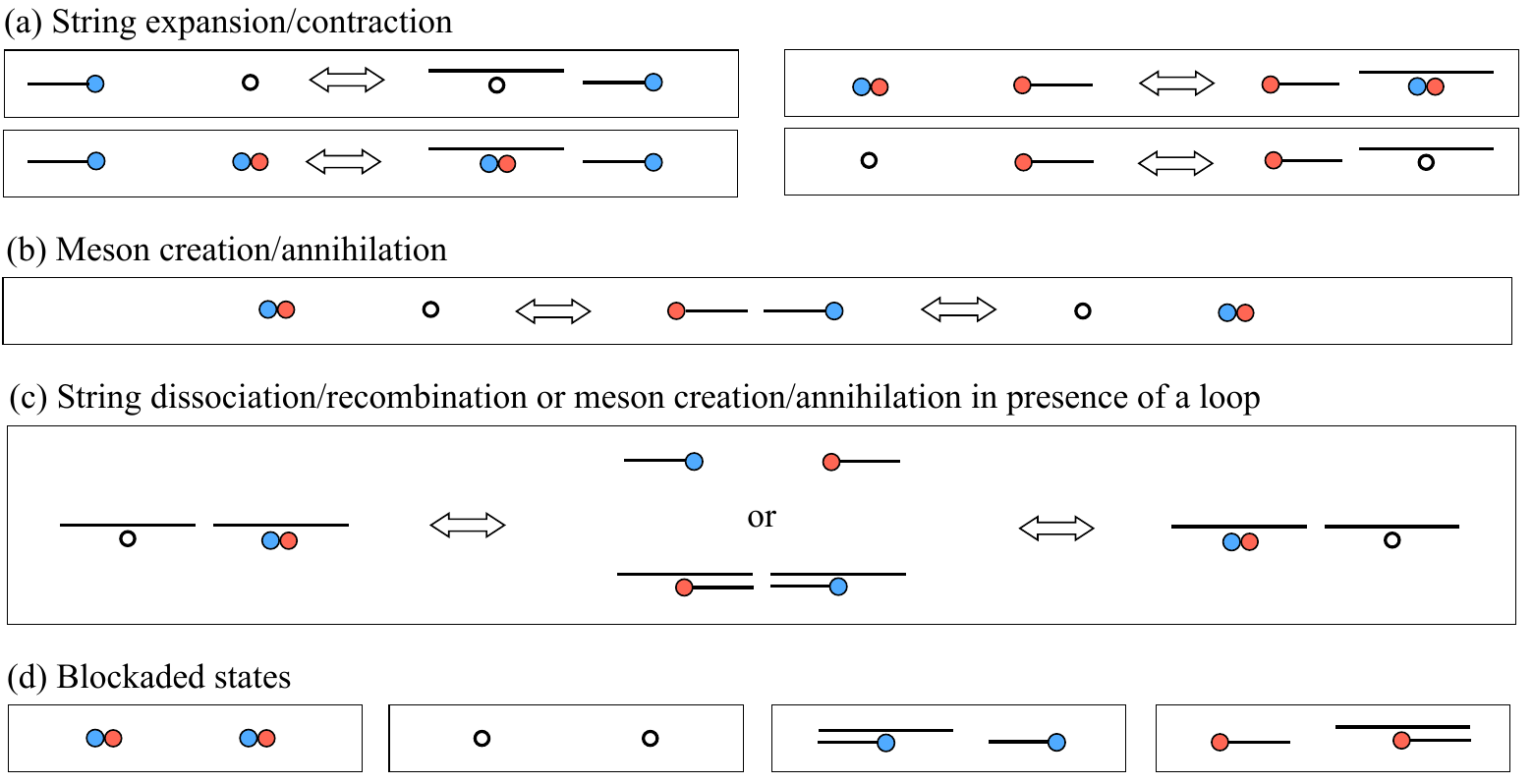}
    \caption{All possible transitions between the states of two adjacent sites induced by a single application of the hopping Hamiltonian. Transitions shown in (a) and (c) can also happen in the presence of some background flux. The translation between the pictorially represented states and the LSH quantum numbers is given in Fig.~\ref{fig:lsh-states}.  The interpretation of the empty/hadron state as the vacuum or baryonic excitation depends upon the parity of its site.}
    \label{fig:single_hop_transitions}
\end{figure*}
Recall that the initial state is created by applying
\begin{align}
    \hat S^{\rm Dirac}_{r,\Delta r}\coloneq \frac{1}{ga}\Big[\hat{S}_{r,\Delta r}-\hat{S}_{r+1,\Delta r}\Big]
\end{align}
to the interacting ground states $\ket{\Omega}$, with an odd $r$ and even $\Delta r$, as discussed in Sec.~\ref{sec:string-setup}. We choose $r=55$ and $\Delta r=16$; thus the Dirac meson string is well within the translationally-invariant bulk of the ground state for either mass. Each operator $\hat{S}_{r,\Delta r}$ is composed of a string-in and a string-out operator connected by a series of loop operators, according to Eq.~\eqref{eq:String_operator_LSH}. The discussion in the previous section illustrates that a given site in the translationally-invariant bulk of the ground state predominantly consists of low-loop, low-flux states. The action of various pieces of $\hat{S}_{r,\Delta r}$ on such states is shown in Fig.~\ref{fig:string-action} in Appendix~\ref{app:lsh-formalism}. In particular, notice that $\hat S^\text{Dirac}_{r,\Delta r}$ with even $\Delta r$ annihilates the strong-coupling vacuum.

The change caused by the application of the Dirac-string operator in excitation, string-in/out, baryon, loop, and symmetric flux quantum numbers relative to the ground state is plotted in Fig.~\ref{fig:excitation_time_slices} (panels marked by $gt=0$ for each mass). First, one observes that $\hat S^\text{Dirac}_{r,\Delta r}$ with even $\Delta r$ acts nontrivially on the interacting vacua. The norm of the Dirac meson-string operator acting on the interacting vacuum is $12.76$ for the lighter mass and $10.04$ for the heavier mass. These highlight the nontrivial composition of the interacting vacua compared with the strong-coupling vacuum. Second, it is observed that the disturbance caused by the string operator leaks out of the domain of support of $S^{\rm Dirac}_{r,\Delta r}$ itself.
However, the finite-range correlations of the ground state contain this leakage to a small distance.

The nearly uniform energy density in the middle of the meson string is $\approx 0.286 g^2$ for the lighter mass and $\approx 0.335 g^2$ for the heavier mass, which are surprisingly close to the corresponding string-tension values obtained from a static meson string in the previous section. This similarity may be due to the fact that instantaneously, the dynamical charges can be viewed as nearly static ones in the background of the interacting ground state. Nonetheless, these states, i.e., the dynamical string in the background of the interacting vacuum and the ground state in the presence of two static charges, are intrinsically different: one is a nonstationary state while the other is stationary.  In fact, the excitation profile of the two types of states exhibit interesting differences (compare the $gt=0$ panels of Figs.~\ref{fig:excitation_time_slices} with left panel of Fig.~\ref{fig:static_excitation}): in the former case, the loop excitations deplete near the dynamical edges and remain undepleted in the interior, since the dynamical string-in and -out sources at the meson-string endpoints diffuse; in the latter case, the loop excitations are peaked near the static edges and depleted in the interior, since the immovable static charges constantly source one unit of flux near the meson-string endpoints.
 
To summarize, the changes in the matter quantum numbers $\langle\hat{\tilde{n}}_{i/o}(r)\rangle$ and $\langle\hat{n}_b(r)\rangle$ are restricted to a small domain near the two endpoints $r\sim 55,71$ of the meson string. The outer sides of these two lumps of matter disturbances are surrounded by the equilibrium ground state, while the interior of the meson string is a nonequilibrium region characterized by a background flux applied to the ground state. Since the energy injected by the Dirac meson string is much higher at its ends than in its interior, we expect the dynamics to be primarily driven by the motion of these endpoints. We diagnose such dynamics in more detail next.

%%%
\subsubsection{Time-evolved broken-string state}
\begin{figure*}[t!]
    \centering
    \includegraphics[scale=0.525]{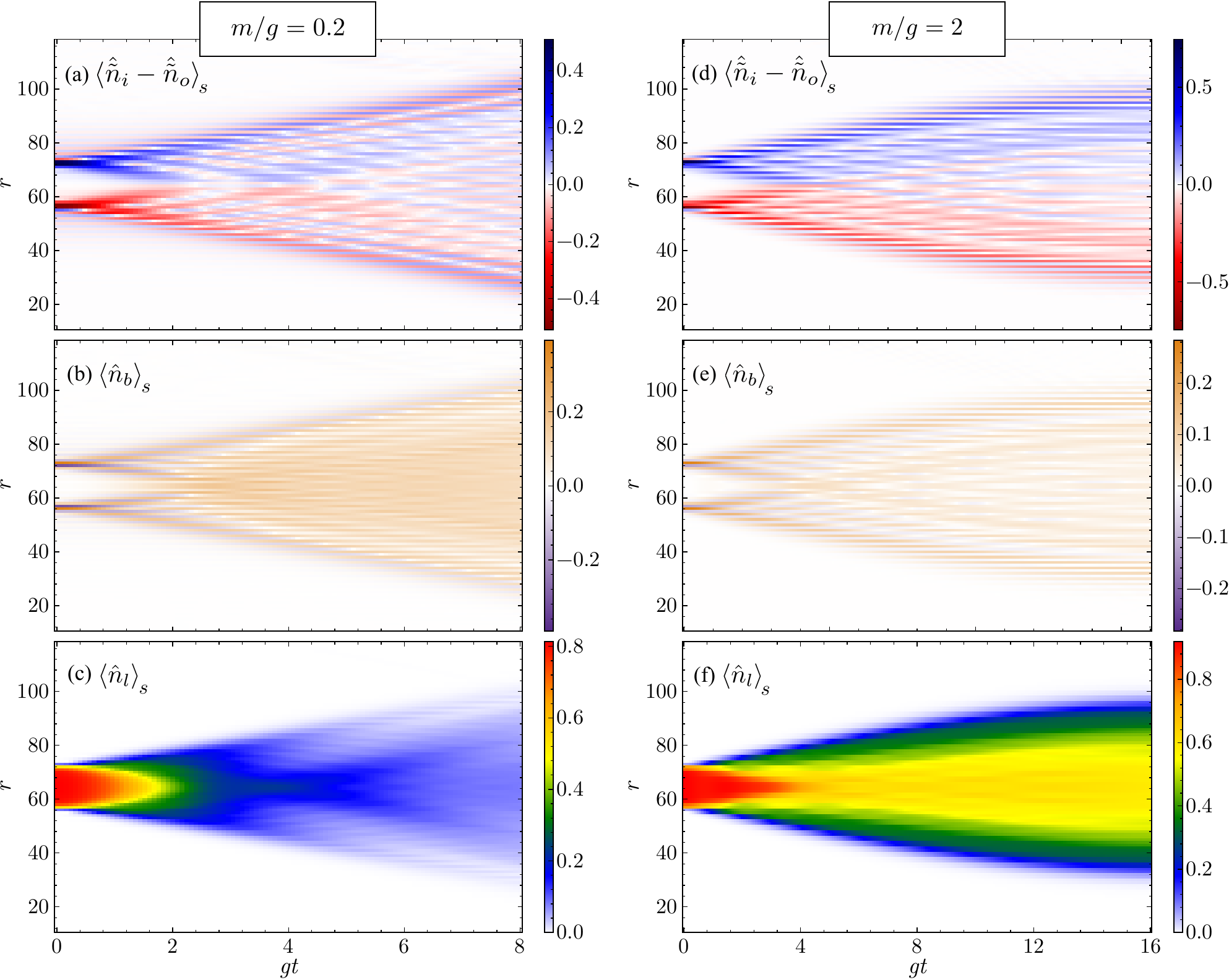}
    \caption{
    Evolution of the expectation values of the difference in the string-in and string-out excitations, $\langle\hat{\tilde{n}}_{i}-\hat{\tilde{n}}_{o}\rangle_s$, baryons, $\langle\hat n_b\rangle_s$, and loop excitations, $\langle \hat n_l\rangle_s$ as a function of time for two values of bare fermion mass. The other parameters are $N=128$, $x=16$, and $j_{\rm max}=\frac{5}{2}$.}
    \label{fig:excitation_dynamics}
\end{figure*}
During real-time evolution, transitions in the LSH quantum numbers are driven by the hopping Hamiltonian. Recall that the hopping Hamiltonian connecting sites $r$ and $r+1$ is a sum of four terms: $\sopp(r)\sipm(r+1)$, $\sopm(r)\simm(r+1)$, $\somm(r)\simp(r+1)$, and $\somp(r)\sipp(r+1)$ (up to some factors that are diagonal in the LSH basis). All possible transitions between the LSH quantum numbers at two adjacent sites induced by a single application of the hopping Hamiltonian are depicted in Fig.~\ref{fig:single_hop_transitions}. As shown in Fig.~\ref{fig:single_hop_transitions}(a), when exactly one of the two sites hosts either a string-in or a string-out state and the other site is empty or hosts a hadron, the hopping term causes the string associated with the string-in or string-out state to expand or contract, and each process can yield creation or annihilation of a bare baryon (depending on the sites' parity). Such string contraction and expansion dynamics can also happen in the presence of a background flux, i.e., an arbitrary number of additional loops, running through the two sites. Figure~\ref{fig:single_hop_transitions}(b) shows the transition between a length-one string into an empty site and a hadron (hence vacuum or two baryons depending on the sites' parity). This transition is responsible for bare-meson creation and annihilation. This transition can also happen in the presence of a background flux, as shown in Fig.~\ref{fig:single_hop_transitions}(c). But in the presence of this background flux, the configuration which consists of one hadron and an empty site can also transition into a configuration with two separate strings while depleting the flux hosted between the two sites. This process (named string dissociation) and its reverse (named string recombination) are also shown in Fig.~\ref{fig:single_hop_transitions}(c). Finally, Fig.~\ref{fig:single_hop_transitions}(d) shows the configurations which are annihilated by the hopping Hamiltonian.\\

The electric and mass Hamiltonians cannot induce transitions between LSH basis states. However, they determine the cost associated with the different transitions caused by the hopping Hamiltonian. All the hopping transitions between adjacent sites are associated with either the annihilation or creation of a single unit of flux between the two sites. In the strong-coupling limit (small $x$), this leads to an electrical-energy cost of $\mp \frac{3}{4}$. Likewise, since the hopping Hamiltonian is particle-number conserving, all transitions results in the annihilation of a fermion at one site, and the creation of a fermion at the other. Due to the staggering in the mass term, depending upon the parity of the sites at which fermion annihilation/creation occurs, a matter-energy cost of $\mp 2\mu$ is paid in the strong-coupling limit. In general, both the mass $\mu$ and the coupling (i.e., hopping coefficient), $x$, together determine which processes are more favored energetically. The other factor determining the prevalence of these hopping transitions is the state itself, as it specifies the local densities of empty, string-in, string-out, hadron, and loops available to undergo various transitions. \\

Figure~\ref{fig:excitation_time_slices} displays the expectation values of total matter excitation, $\langle\hat{n}(r)\rangle_s$, string-in, $\langle\hat{\tilde{n}}_{i}(r)\rangle_s$, string-out, $\langle\hat{\tilde{n}}_{o}(r)\rangle_s$, baryon, $\langle\hat{n}_b(r)\rangle_s$, loop, $\langle\hat n_l(r)\rangle_s$, and symmetrized flux, $\langle\hat{\tilde{N}}(r)\rangle_s$, at select time slices during the meson-string evolution for the two different bare fermion masses. The $\langle \cdot \rangle_s$ notation denotes that the corresponding expectation values in the interacting vacuum are subtracted out. The large, nearly uniform loop/flux values in the original meson string decrease as the string evolves, and this decrease occurs in a nonuniform manner. The loop/flux excitations further expand beyond the extent of the original string, and this expansion occurs at a faster rate for the lighter mass than for the heavier mass. For example, at $gt=6$, while the peak values of loop/flux drop by almost a factor of seven for the lighter mass, they only decrease by less than a factor of two for the heavier mass. The enhanced initial values of the various matter excitations at the string endpoints also decline, and such excitations start to populate both the interior and the exterior of the initial meson string.

To enable a more fine-grained analysis of the evolution, and to reveal the underlying string-evolution mechanisms at play during various stages of the dynamics, we plot in Fig.~\ref{fig:excitation_dynamics} $\langle\hat{\tilde{n}}_{i}-\hat{\tilde{n}}_{o}\rangle_s$, $\langle\hat n_b\rangle_s$, and $\langle \hat n_l\rangle_s$ as a function of time for the two bare masses. For both mass values, the initial dynamics is dominated by the string expansion/contraction processes described in Fig.~\ref{fig:single_hop_transitions}(a). The inward motion (string contractions) of the string-out and string-in states at the two ends of the meson string depletes the flux in the interior of the string starting at the edges. Similarly, the outward motion (string expansions) of these states distribute flux to the outward region of the string. As seen in Fig.~\ref{fig:single_hop_transitions}(a), the hopping of string-in/out states is also accompanied by the transport of baryons, which is likewise visible in the dynamics. This process marks the conversion of the potential energy of the string to the kinetic energy carried out by the fermions. Due to the left and right hopping of the fermions continuously throughout the evolution, each particle trajectory repeatedly splits into two, creating particle ``showers.'' Each subsequent ``shower" carries less mean matter excitations than its parent, but continues moving with nearly the same speed. As the mean matter excitations on the original meson-string state decrease with these repeated splittings, there is also a depletion of flux in the interior region of the string. This interior flux-depletion effect is minimal in the heavier-mass dynamics, but is much more pronounced in the lighter-mass dynamics, due to the lower barrier to hopping for the latter.

The outward-moving string-in/string-out and baryonic wavefronts continue moving uninterruptedly toward the boundaries, while the inward-moving ones ultimately collide. When the first such collision occurs for the heavier mass (around $gt = 4$), the net string-in/out difference, $\langle \tilde{n}_i-\tilde{n}_o\rangle$, suggests that the string-in/out wavefronts move past each other without much interference. The local baryon density gets attenuated as the baryonic disturbances arriving from the endpoints move past each other, but the baryons still continue tracking the motion of the string endpoints. Since the flux in the interior region ceases to deplete after the collision of these wavefronts, the predominant mechanism after collision continues to be string expansion/contraction. Moreover, the slight attenuation in baryon density and concurrent small enhancement of flux in the interior indicates that process  Fig.~\ref{fig:single_hop_transitions}(b), which converts baryons into mesons, may also be at play. 

In contrast, when the string-in and -out wavefronts collide for the lighter mass (around $gt=2$), the net string-in/out difference, $\langle \tilde{n}_i-\tilde{n}_o\rangle$, vanishes. In fact, we find that $\langle \tilde{n}_i\rangle$ and $\langle \tilde{n}_o \rangle$ vanish separately. At the same time, the collision of the baryonic wavefronts is accompanied by the creation of more baryons. Together with the fact that the flux in the center of the lattice continues depleting rapidly in the wake of this collision, this is the signature of the string-dissociation process shown in Fig.~\ref{fig:single_hop_transitions}(c). Subsequently, meson creation in presence of any number of loops, i.e., Fig.~\ref{fig:single_hop_transitions}(b) and (c), creates net string-in/out excitations that also move and split. These dynamics get further complicated by the proliferation of showers of excitations and their subsequent scattering throughout the bulk.

To better understand the difference between the heavy- and light-mass dynamics after the collision of the interior beams, one can examine the energetic cost of the underlying processes. First, consider a transition from a length-one string with no loop to a dissociated string (to vacuum or baryons). These processes involve a cost of $-\frac{3}{4}$ for flux annihilation and $\pm 2\mu$ for vacuum or baryon generation. The baryon generation is penalized more for the heavy mass than for the light mass, hindering one possible channel for flux depletion. Second, consider a transition from a length-one string state with a loop to a dissociated string. This process requires two applications of the hopping Hamiltonian. Due to the staggering of the lattice, both these applications are associated with opposite matter energy costs $\pm 2\mu$. However, both result in a reduction of the electric energy by $-\frac{3}{4}$. For the heavier-mass case, therefore, even if one of the hops is energetically favored due to a dominant $-2\mu$ cost, the other hop gets penalized due to the $+2\mu$ cost, and there are not enough net savings from flux depletion. Such arguments can be repeated for other string states with any number of flux, which are the primary states at the point of collision of the string-in and string-out wavefronts (the high-flux states are suppressed in the dynamics, as will be demonstrated in Sec.~\ref{sec:entropy}). This discussion explains why string dissociation is not as prevalent for the heavier mass, and supports the lack of substantial electric-field screening in the interior of the lattice. For the lighter mass, in contrast, these matter-energy costs are not as important relative to the gains from flux reduction and the hopping energy.\footnote{While the hopping energy is a significant contribution to the energy at the chosen value of $x$, the relative significance of transitions between light and heavy mass remains qualitatively the same.}

For the lighter mass, a third phase of evolution is seen, most notably in the local loop-excitation evolutions. Not only the loop-excitation depletion for the lighter mass features a temporally and spatially nonuniform pattern, it signals a secondary phase (around $gt=4$) in which the outer loop excitations nearly diminish while an inner cone shapes up. An inspection of the difference in the string-in and string-out excitations reveals that this moment coincides with the secondary showers from the inner rays colliding immediately after the initial scattering has resulted in string dissociation. So while the outer showers continue to move uninterruptedly to deplete fully the local loop/flux excitations, the inner showers engage in nontrivial scattering processes, which redistributes the loop/flux excitations in the bulk. In contrast, for the heavier mass, loop excitations stabilize in the bulk, and the outer loop excitation never fully deplete. Secondary scatterings occur here too but they do not have sufficient energy to lead to significant inelastic channels. This picture is supported by the proliferation of baryons in the bulk for the lighter mass resulting from inelastic production channels. In contrast, for the heavier mass, the evolution of baryons is consistent with the dominantly undisturbed passing of the baryonic tails, although nonvanishing but small amount of baryons are still generated after these collision events. 

\begin{figure*}[t!]
    \centering
    \includegraphics[scale=0.645]{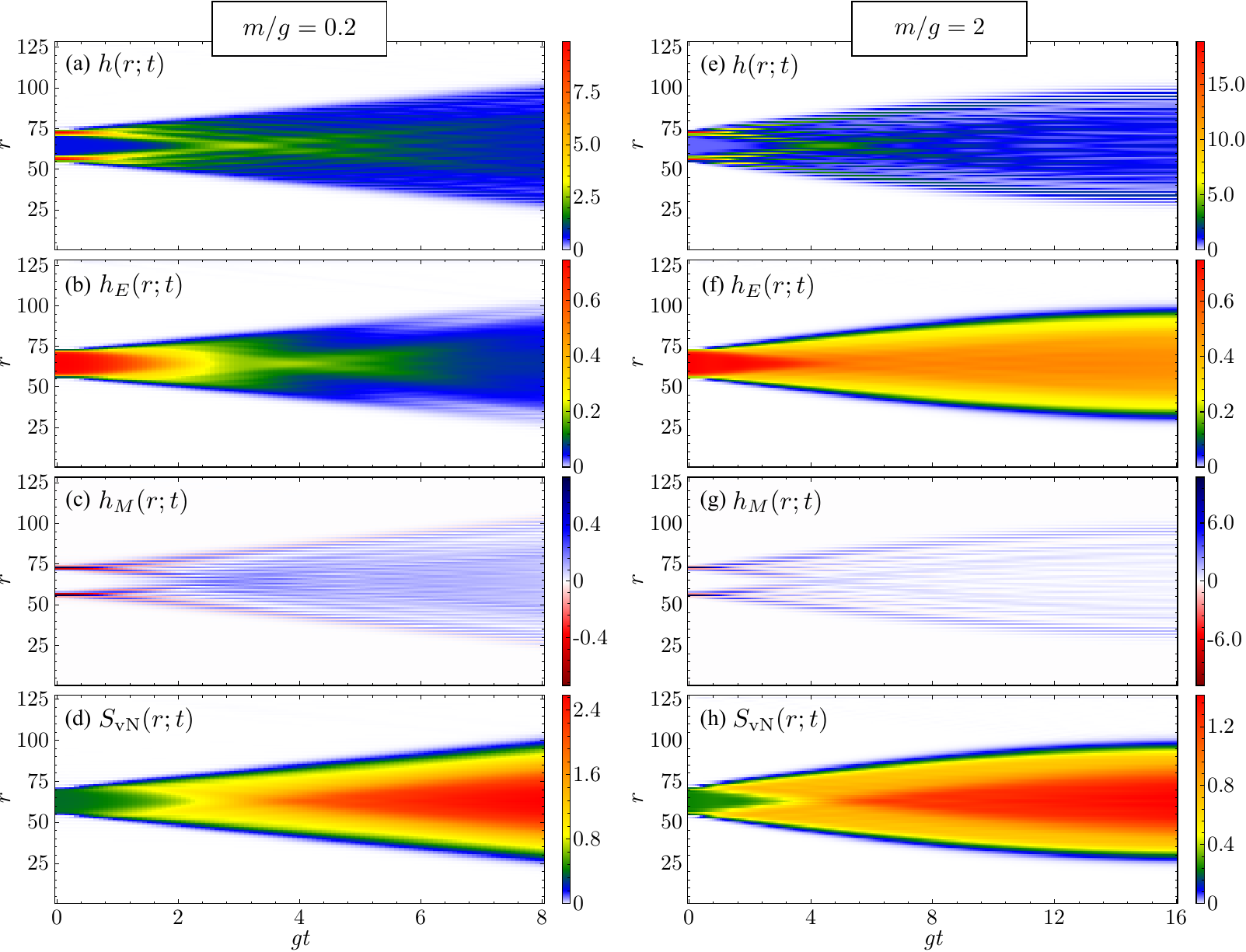}
    \caption{Evolution of the vacuum-subtracted expectation values of:  (a),(e) energy density, (b),(f) electric-energy density, (c),(g) mass-energy density, as well as the value of bipartite entanglement entropy for a cut at position $r$, (d),(h), for two values of bare fermion mass. The other parameters are $N=128$, $x=16$, and $j_{\rm max}=\frac{5}{2}$.
    }
    \label{fig:energy-entropy-2D-plots}
\end{figure*}
%

%%%%%%%%%%%%%%%%%%%%%%%%%%%%%%%%%%%%%%%%%%%%%%%%%%%%%%%%%%%%%%%%%%%%%%%%%%%%%%%%%%%%%%%%%%%%%%%%%%%%%%%%%%%%%%%%%%%%%%%%%%%%%%%%%%%%%%%%%%%%%%%%%%%%%%%%%%%%%%%%%%
\subsection{Energy, entropy, and correlation evolutions
\label{sec:global}}
The fine-grained picture of matter and flux excitations throughout the meson-string evolution developed in the previous section can assist interpreting evolution of various energy densities, bipartite entanglement entropies, as well as the spreading of correlations throughout the evolution.

\subsubsection{Energy diffusion\label{sec:energy-entropy}}
\begin{figure*}[t!]
    \centering
    \includegraphics[scale=0.645]{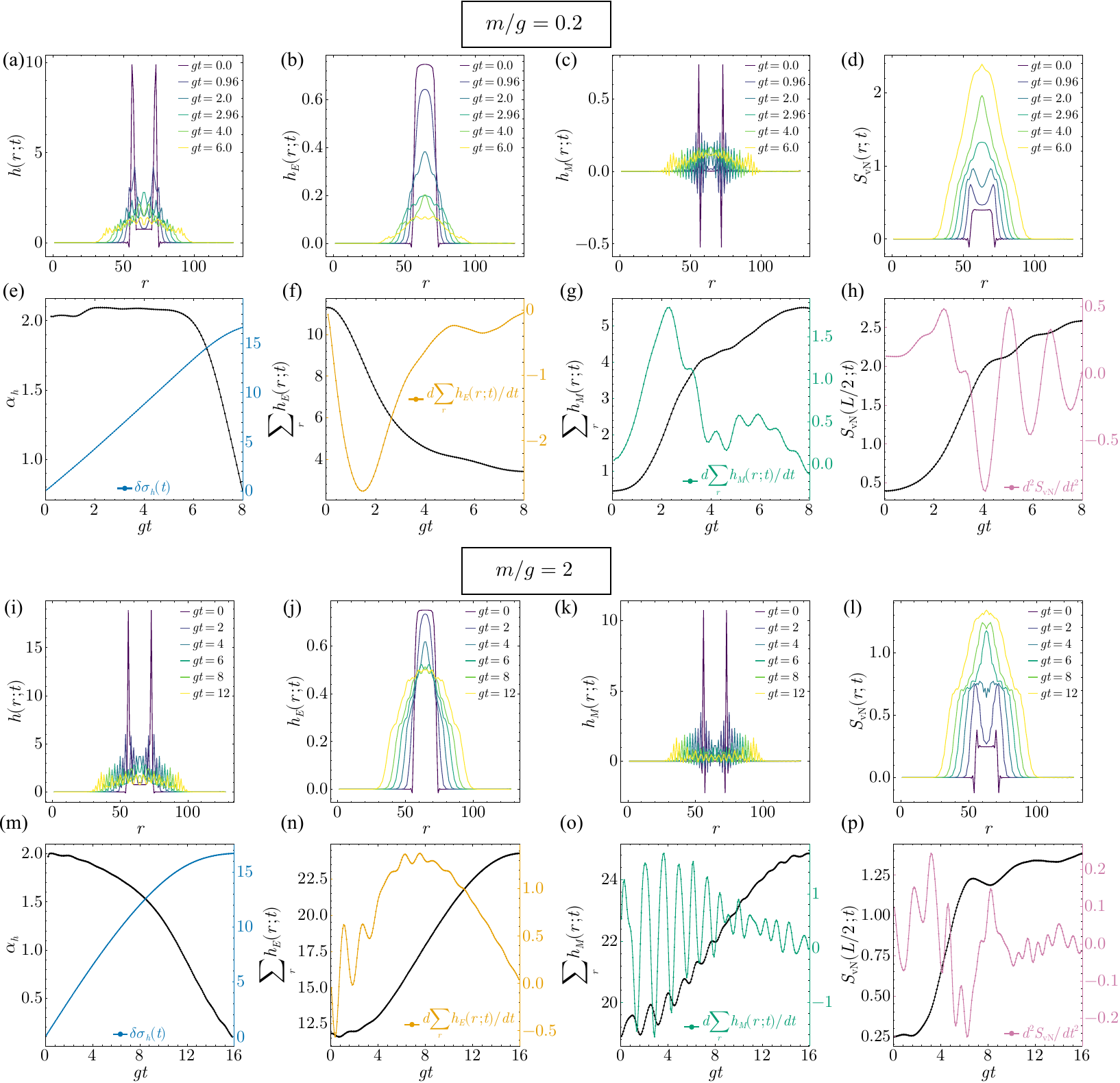}
    \caption{(a),(i) Vacuum-subtracted expectation values of energy density, (b),(j) electric-energy density, (c),(k) mass-energy density, and (d),(l) bipartite entanglement entropy with a cut at position $r$, for each bare-mass value at select times during the meson-string evolution. (e),(m) The logarithmic time derivative of the spatial variance of the energy-density distribution defined in Eq.~\eqref{eq:delta-sigma-E} as a function of lattice site for each mass. (f),(n) Total vacuum-subtracted electric energy, (g),(o) mass energy, as well as (h),(p) vacuum-subtracted half-cut bipartite entanglement entropy as a function of evolution time. The first derivative of the total vacuum-subtracted electric and mass energies, and the second derivative of the vacuum-subtracted half-cut entanglement entropy are also shown.
    }
    \label{fig:energy-entropy-1D-plots}
\end{figure*}

Consider the vacuum-subtracted energy density:
\begin{align} 
    h(r;t)\coloneq \langle \phi(t)| \hat h(r)|\phi(t) \rangle-\langle \Omega | \hat h(r) | \Omega \rangle.
\end{align}
Here, the left-right symmetrized onsite energy density is defined as:
\begin{align} 
    \hat h(r) \coloneq \hat h_E(r)+\hat h_M(r)+\hat h_I(r),
\end{align}
with
\begin{subequations}
\label{eq:hr-defs}
\begin{align} 
    &\hat h_E(r) \coloneq (1-\delta_{r,1}
    )\frac{\hat h_E'(r-1)+\hat h_E'(r)}{2}
    +\delta_{r,1}\frac{\hat h_E'(1)}{2},
    \label{eq:hEr-defs}
    \\
    &\hat h_M(r)\coloneq \hat h'_M(r),
    \label{eq:hMr-defs}
    \\
    &\hat h_I(r) \coloneq (1-\delta_{r,1}
    )\frac{\hat h_I'(r-1)+\hat h_I'(r)}{2}
    +\delta_{r,1}\frac{\hat h_I'(1)}{2}.
    \label{eq:hIr-defs}
\end{align}
\end{subequations}
$\hat h'_E(r)$, $\hat h'_M(r)$, and $\hat h'_I(r)$ are defined in Eqs.~\eqref{eq:HEMI_LSH} [with $\hat h'_E(N)=\hat h'_I(N)=0$]. Figures~\ref{fig:energy-entropy-2D-plots}(a) and (e) display this energy density for the lighter and heavier masses, respectively (with the corresponding 3D plots presented in Figs.~\ref{fig:energy-entropy-3D-plots} of Appendix~\ref{app:3D-plots}). The same quantities at select times are also plotted in Figs.~\ref{fig:energy-entropy-1D-plots} (a) and (i). The plots demonstrate that the application of the meson-string operator to the interacting vacuum creates excess energy density along the length of the string. This excess is bimodal; there are two peaks around the meson string's endpoints, and a smaller uniform excess energy density in the interior region of the mesonic string. Over time, these two lumps of excess energy density spread both to the exterior and interior regions of the meson string. The peak energy at the edges is $\approx 9.9$ for the lighter mass and $\approx 18.9$ for the heavier mass. As a comparison, the first excitation energy gap per lattice site is $\approx 5.5$ for the lighter mass and $\approx 34.7$ for the heavier mass. While energetic considerations favor producing of the lowest-lying hadronic excitation for the lighter mass, identifying the asymptotic hadron content of the state requires computing overlap onto such a hadronic states at long times, which is beyond the scope of this work. One can nonetheless, observe the ejection of string-in, string-out, and bare-baryon excitations from the edges, as discussed in the previous section.

While the interior of the dynamical meson string does not constitute a uniform background electric field to induce the Schwinger pair-production mechanism, we can still assume an instantaneous electric field at $gt=0$, which as evident from Fig.~\ref{fig:energy-entropy-1D-plots}(b) and (j), is nearly uniform in the middle of the meson string. The Schwinger-mechanism-induced pair-production rate from a semiclassical approximation reads $\frac{\bar{h}_E}{\pi}e^{-\frac{\pi m^2}{\bar{h}_E}}$~\cite{Schwinger:1951nm,Casher:1978wy}. Here, $\bar{h}_E$ is the electric energy per unit length stored in the interior of the meson string, which is $\approx 0.286g^2$ for the lighter mass and $\approx 0.335g^2$ for the heavier mass. The Schwinger pair-production rate per unit length is, thus, $\approx 0.059 g^2$ for the lighter mass and $\approx 5.6 \times 10^{-18} g^2$ for the heavier mass. To obtain the total rate, nonetheless, one needs to introduce a space-time dependent rate and integrate it over space and time. Since the flux profile appears nearly constant in the interior of the meson string for $gt \lesssim 0.5$ for the lighter mass, and for $gt \lesssim 1$ for the heavier mass, one can obtain an integrated pair-production rate of $\approx 0.1$ for the lighter mass and $\approx 2\times 10^{-17}$ for the heavier mass. Therefore, the total production rate at early stages of the evolution is low for the light mass and is significantly suppressed for the heavy mass. Ultimately, the pair production dynamics can be fully tracked in our real-time, quantum approach, removing the need for such semiclassical modelings.

Next, we analyze the spreading rate of the energy density. To diagnose the energy-transport properties, we consider the quantity~\cite{karrasch2014real}:
\begin{align}
\delta \sigma_{h}(t) \coloneq \sqrt{\sigma_{h}^2(t)-\sigma_{h}^2(0)},
\end{align}
where
\begin{align}
\sigma_{h}^2(t)\coloneq \frac{1}{\mathcal{N}}\sum_{r=r_0}^{N-r_0}(r-r_c)^2h(r;t)
\end{align}
is the spatial variance of the excess energy-density ``distribution".\footnote{The excess energy density assumes small negative values for a small number of lattice sites at some times. These negative regions are, nonetheless,  small enough that one can still approximate this excess energy as a distribution.} Here, $\mathcal{N}\coloneq \sum_{r=r_0}^{N-r_0}h(r;t)$ is a normalization constant, $r_c=64$ is the center of the lattice, and we set $r_0=30$ (to exclude the exterior regions where boundary effects set in). For ballistic energy transport, $\delta \sigma_{h}(t)=V_{h} gt$ for a constant velocity $V_{h}$,  whereas for diffusive dynamics, $\delta \sigma_{h}(t)=\sqrt{2D_{h}t}$ for a constant diffusion coefficient $D_{h}$. We plot in Fig.~\ref{fig:energy-entropy-1D-plots} (e) and (m) the value of logarithmic derivative of $\delta \sigma_{h}^2(t)$,
\begin{align}
\alpha_{h} \coloneq \frac{d\mathrm{ln} \ \delta \sigma_{h}^2(t)}{d(\ln(gt))},
\label{eq:delta-sigma-E}
\end{align}
as a function of time:. For ballistic (diffusive) transport, this value is equal to two (one). As seen in the plots, the lighter-mass case exhibits a window in time, $2 \lesssim gt \lesssim 4.96$,  in which transport appears to be ballistic while neither ballistic nor diffusive transport is observed for the heavier mass. For the lighter mass, the energy-transport velocity is $V_h \approx 2.28$ lattice sites per $gt$, which can be compared with the energy-front velocity $v \approx 3.12$ lattice sites per $gt$. The former velocity is computed by performing a linear fit of $\delta\sigma_h(t)$ in the above-mentioned window of time which corresponds to the plateau in the logarithmic derivative. The latter velocity is derived by tracking the expansion of the outer boundary of nonvanishing energy density as a function of time: At each time slice, the $h(r;t)$ values are inspected to find the lattice site at which the local energy density exceeds by $5\%$ compared to its value at the adjacent site in the exterior region. Any site which satisfies this criterion is taken as the wavefront site. Once these sites and their corresponding time values are found, we perform a least-square linear fit to extract the velocity. Uncertainties are expected on these extractions, and the values quoted should be taken as an estimate. A possible explanation for the lower transport velocity compared with the front velocity is that the transport velocity is informed by the energy distribution in the bulk, which is filled with a nontrivial medium, while the front velocity involves propagation of energy outward in the vacuum. For the heavier mass, the front expansion is not linear and we refrain from extracting the front velocity.

Next, consider the diagonal contributions to the energy density, i.e., the electric and matter-mass energy densities,
\begin{align}
    h_{E/M}(r;t)\coloneq \langle \phi(t)| \hat h_{E/M}(r)|\phi(t)\rangle-\langle \Omega|\hat h_{E/M}(r)|\Omega\rangle,
\end{align}
where $h_{E}(r)$ and $h_{M}(r)$ are defined in Eqs.~\eqref{eq:hEr-defs} and \eqref{eq:hMr-defs}, respectively. Figures~\ref{fig:energy-entropy-2D-plots}(b), (c), (f), and (g) display the electric- and mass-energy densities for each bare mass. These quantities at select times are depicted in Figs.~\ref{fig:energy-entropy-1D-plots} (b), (c), (j), and (k), along with the total values over the entire lattice as a function of time in Figs.~\ref{fig:energy-entropy-1D-plots} (f), (g), (n), and (o). The plots reveal distinct behaviors for the two masses. For the lighter mass, the electric-energy density gets depleted in the interior region of the meson string and spreads to the outer regions of the lattice. This behavior is accompanied by the spreading of the initial mass-energy density both toward and away from the meson string from both endpoints. The mass-energy density in the interior of the string is seen to build up. The lattice-summed quantities point to an overall decrease in the electric energy and increase in the mass energy. The total electric energy nears zero at $gt=8$ while the mass energy nears $\approx 3.4$, indicating significant electric-field screening and particle production. In contrast, for the heavier mass, the electric-energy density initially depletes in the region of the initial meson string, but eventually this depletion stops, and the electric energy continues to slowly spread to the external parts of the lattice. Much less mass energy is produced in the bulk region of the initial meson string compared to the lighter-mass case. The lattice-summed electric energy shows an overall increase to an asymptotic value $\approx 25$ at $gt=16$ while the mass energy only slightly increases and stabilizes to the value $\approx 25$. \\ 

\subsubsection{Entanglement-entropy production
\label{sec:entropy}}
\begin{figure*}[t!]
    \centering
        \includegraphics[scale=0.55]{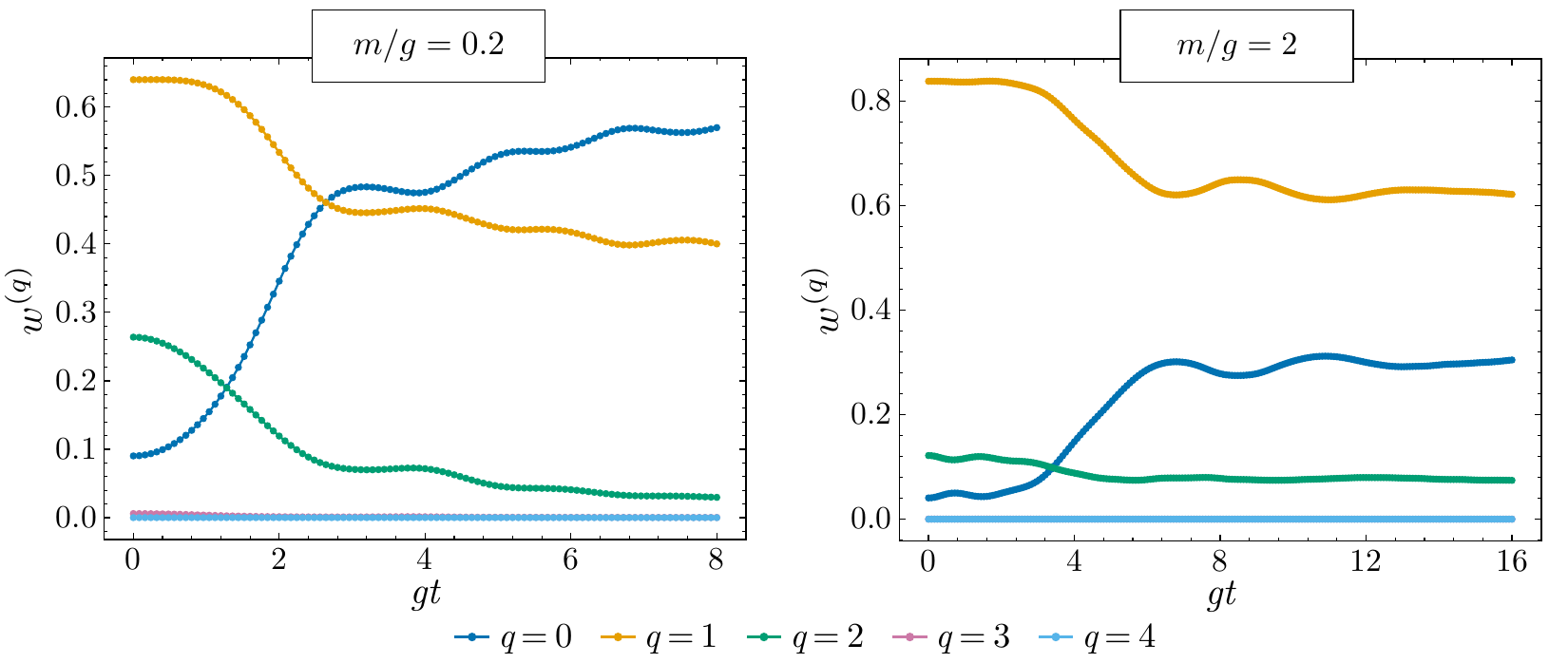}
    \caption{Evolution of symmetry-resolved Schmidt weights $w^{(q)}$ defined in Eq.~\eqref{eq:w-q} with the half-cut reduced density matrices $\rho_\phi^{(q)}$, associated with state $\ket{\phi(t)}$ and given values of the net flux $q$ exiting the subsystem.}
    \label{fig:symmetry-resolved}
\end{figure*}
Consider the vacuum-subtracted bipartite entanglement entropy:
\begin{equation}
        S_{\rm vN}(r;t)\coloneq S_{\rm vN}\left(\hat \rho_{\phi(t)}(r)\right)-S_{\rm vN}\left(\hat\rho_{\Omega}(r)\right).
        \label{eq:vac-SvN}
\end{equation}
Here, $\hat\rho_{\phi}(r)$ is the reduced density matrix of the pure state $\ket{\phi}$ defined on the subsystem spanning sites $\{1,\cdots,r\}$, i.e., $\hat\rho_{\phi}(r) \coloneq \mathrm{Tr}_{\{r+1,\cdots,N\}}(\hat{\rho})$ where $\hat{\rho}$ is the full-system state and $\mathrm{Tr}_{\{r+1,\cdots,N\}}$ denotes partial trace of the state at sites $\{r+1,\cdots,N\}$. The von Neumann entropy is defined as
\begin{align}
    \label{eq:S_ent}
        S_\text{vN}(\hat \rho) &\coloneq-\mathrm{Tr}(\hat \rho\,\mathrm{log}\hat\rho)=-\sum_i p_i \log(p_i),
\end{align}
where $0 \leq p_i \leq 1$ denote the nonvanishing eigenvalues (i.e., singular values) of the Hermitian, positive semidefinite matrix $\hat \rho$, with $\sum_i p_i=1$. 
In the tensor-network language, bipartite entanglement entropies can be computed by transforming the MPS into a mixed canonical form at the bipartition point. The Schmidt coefficients $\lambda_i = \sqrt{p_i}$ can then be extracted from the bond tensor at this bipartition point, and the von Neumann entropy can be computed subsequently.\footnote{We consider only the full entanglement entropy and do not separate the `classical' and `representation' contributions from the `distillable' contribution~\cite{Soni:2015yga,VanAcoleyen:2015ccp,Lin:2018bud,Feldman:2024qif}.}

Figures~\ref{fig:energy-entropy-2D-plots}(d) and (h) display the vacuum-subtracted entanglement entropy for each mass as a function of time for bipartite cuts at site $r$ (with the corresponding 3D plots shown in Fig. \ref{fig:energy-entropy-3D-plots} in Appendix~\ref{app:3D-plots}); Figs.~\ref{fig:energy-entropy-1D-plots}(d) and (l) depict the same quantity at select times during the evolution; and Figs.~\ref{fig:energy-entropy-1D-plots}(h) and (p) display the half-cut values as a function of time. The plots reveal that initially, a significant uniform entanglement entropy is observed across cuts within the original meson-string region but not for cuts made outside. For both masses, the bipartite entanglement entropy is rapidly produced around the endpoints of the string initially. However, entanglement-entropy production eventually increases in the center, and finally it appears to saturate. The build-up of entanglement at initial times at the endpoints is a consequence of endpoint expansion and contraction of the string, leading to splitting and particle-shower events that expand the entanglement to nearby regions. The build-up of entanglement in the middle region in intermediate times coincides with the first main scattering event, which necessarily entangles the left and right regions post collision. 

Overall, entanglement-entropy production is larger for the light mass. For example, the vacuum-subtracted half-cut entanglement entropy at $gt=8$ is nearly twice larger for the lighter mass than the heavier mass. This difference is consistent with the larger particle production for the lighter mass compared with the heavy mass. Beyond the correlation between the overall entanglement-entropy and particle production, the rate of matter production appears correlated with the rate of entanglement-entropy production, as observed from the first derivative of the total mass energy (a measure of particle-production rate) and the second derivative of the half-cut entanglement entropy (a measure of entropy-production acceleration). These quantities are displayed in Figs.~\ref{fig:energy-entropy-1D-plots}(g) and (h), respectively, for the light mass, and in Figs.~\ref{fig:energy-entropy-1D-plots}(o) and (p), respectively, for the heavy mass. One can define an associated front velocity for the lighter mass, defined by the expansion speed of the boundary of the region with nonvanishing entanglement entropy, similar to what was done for the energy-front velocity. We obtain a velocity of $\approx 3.58$ lattice sites per $gt$, which is close to the energy-front velocity. \\

Another related quantity that can signify the breaking of the original meson string is the reduced state's symmetry-resolved Schmidt weight: Consider $\rho_{\phi}(\frac{N}{2})$, i.e., the reduced state of the system spanning lattice sites $\{1,\cdots,\frac{N}{2}\}$. This state can be decomposed into
\begin{align}
\rho_{\phi}=\underset{q}{\oplus}\,\rho_{\phi}^{(q)},
\end{align}
where each block $\rho_{\phi}^{(q)}$ corresponds to a given global charge $q$, regardless of the global charge $Q$. We define the Schmidt weight $w^{(q)}$ associated with the symmetry sector $q$ as 
\begin{align}
w^{(q)} \coloneq \sum_i p_i^{(q)},
\label{eq:w-q}
\end{align}
where $p_i^{(q)}$ are eigenvalues of $\rho^{(q)}$. Note that $\sum_{q=0}^{\frac{N}{2}}w^{(q)}=1$.
We compute the symmetry-resolved Schmidt weights $w^{(q)}$ as a function of time during the meson-string evolution. These weights are plotted in Fig.~\ref{fig:symmetry-resolved} for each mass for the lowest $q$ values (weights not shown are consistent with zero). The dominant weight arises from the $q=1$ sector initially, with small but non-negligible contributions from the $q=0$ and $q=2$ sectors. This feature is expected as cutting the system in the middle of the original string creates a subsystem with a nonvanishing incoming-outgoing flux difference. Importantly, over time the contributions from the $q=1$ and $q=2$ sectors decrease while that from the $q=0$ sector increases. The implication is that while the probability of having a nonvanishing net flux out of the half-system remains significant, the likelihood of having separated mesonic pairs on one or the other side of the lattice increases, signaling string breaking. Interestingly, for the lighter mass, the $q=0$ contribution eventually surpasses the $q=1$ contribution, around the time the main scattering event occurs in the dynamics (i.e., $gt \approx 2$). For the heavier mass, on the other hand, the $q=1$ sector remains the dominant contribution for the full simulation-time duration, consistent with only partial electric-field screening.

\subsubsection{Two-point correlation growth
\label{sec:correlations}}
\begin{figure*}[t!]
    \centering  
    \includegraphics[scale=0.59]{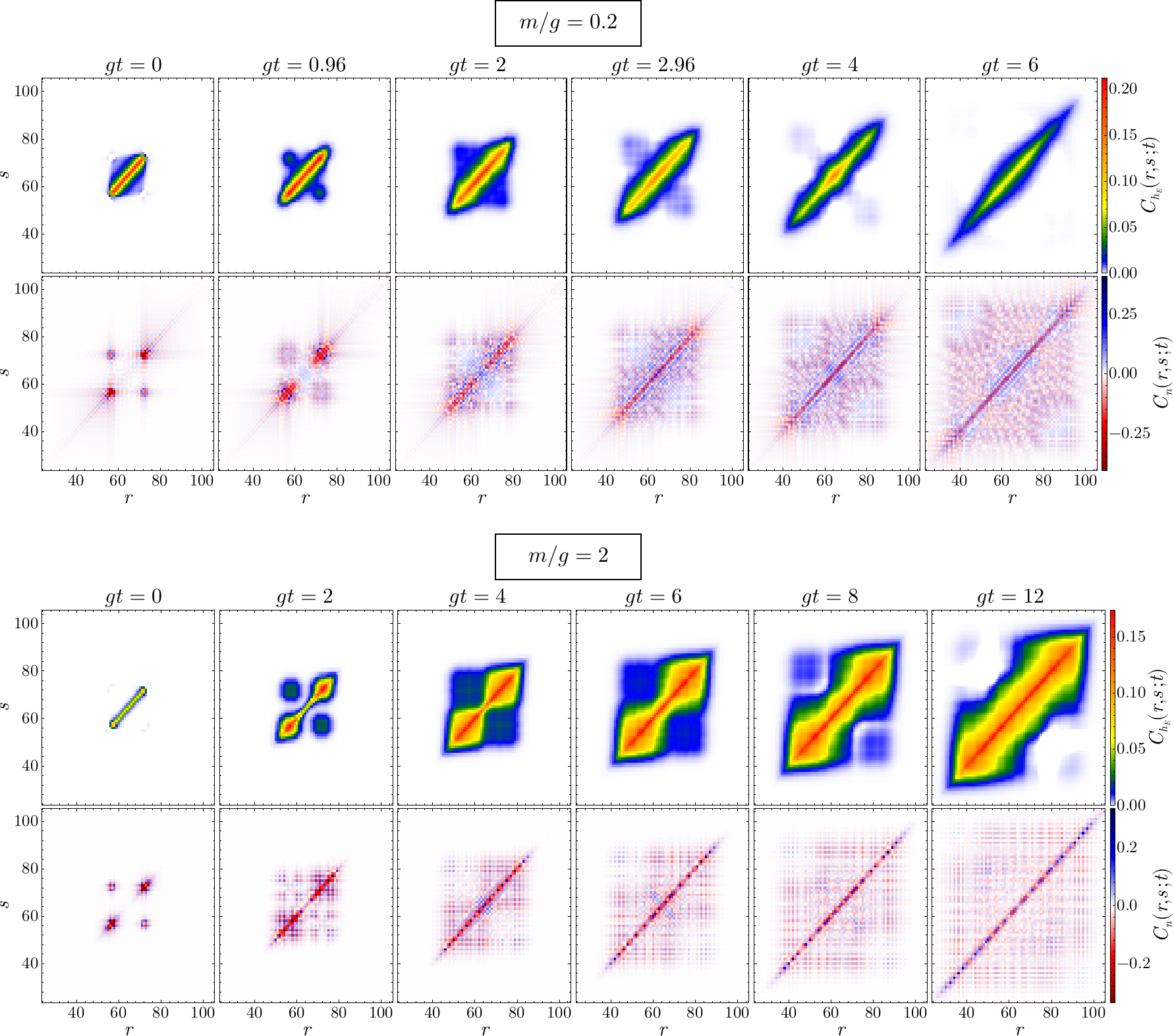}
    \caption{ 
    The vacuum-subtracted, connected correlation-function values between the local electric Hamiltonian (matter-excitation operator) at point $r$ and that at point $s$ are displayed in the top (bottom) panel and for each bare fermion mass, at select times during the meson-string evolution. These correlation functions are defined in Eq.~\eqref{eq:HEn_corr}. Note that only a subset of the full lattice is shown for clarity.}
    \label{fig:correlators}
\end{figure*}
Consider the vacuum-subtracted connected two-point correlation functions of the electric Hamiltonian and of the matter-excitation density, respectively:
\begin{equation}
    \label{eq:HEn_corr}
    C_{h_E/n}(r,s;t)\coloneq C_{h_E/n;\phi(t)}(r,s)-C_{h_E/n;\Omega}(r,s),
\end{equation}
where 
\begin{align}
    C_{h_E;\phi}(r,s)\coloneq \langle \phi| \hat h_E(r)\hat h_E(s)|\phi\rangle -\langle \phi| \hat h_E(r)|\phi\rangle\langle \phi|\hat h_E(s)|\phi\rangle,
\end{align}
and
\begin{align}
    C_{n;\phi}(r,s)\coloneq \langle \phi| \hat n(r)\hat n(s)|\phi\rangle -\langle \phi| \hat n(r)|\phi\rangle\langle \phi|\hat n(s)|\phi\rangle.
\end{align}
The local electric Hamiltonian and the local matter-excitation operator are defined in Eqs.~\eqref{eq:hEr-defs}~and~\eqref{eq:excitations}, respectively. Figures~\ref{fig:correlators} displays these correlators as a function of $(r,s)$ at select times during the evolution.

Initially, the electric-energy correlator $C_{h_E}(r,s;t)$ is concentrated within the domain of the original meson string and is strongest near the diagonal $r=s$. The strength as well as the range of these correlations are greater for the lighter mass as compared to the heavier mass. These correlations are consistent with local AGLs and flux continuity. As the meson string evolves, the excess electric-energy correlations spread outward from the support of the initial meson string, producing diagonal and off-diagonal correlations. Diagonal and near-diagonal (short-range correlations) are consistent with the meson-string expansion/contraction due to hopping processes. Off-diagonal (long-range) correlations result from persisting connected fluxes. For the lighter mass, the correlation spreading remains concentrated mainly near the diagonal, indicating that the state does not retain a long, continuous flux tube over large separations. Instead, screening efficiently suppresses the off-diagonal correlations as the original flux tube is depleted and reorganized. In contrast, for the heavier mass, $C_{h_E}(r,s;t)$ develops substantial off-diagonal support in addition to its diagonal broadening. This reflects the persistence of a stretched, still-connected flux tube whose two ends remain correlated over long distances, even as the string expands into the bulk. In fixed-$r$ slices, this long-range endpoint-to-endpoint correlation appears as a broad plateau around the opposite end rather than a sharp peak; the endpoint energy fronts are no longer sharply localized but are spread over an extended region. 

Initially, the matter-excitation correlator $C_n(r,s;t)$ primarily tracks the correlated motion of the matter-carrying ends of the meson string: the correlations are concentrated along the diagonal at the two ends of the string and along the off-diagonal across the two ends. The nonlocal correlation is a manifestation of the electric flux flowing between the two matter-filled ends and the presence of the AGLs. As the meson string evolves, the  matter-excitation correlations spread outward from their initially endpoint-localized support. Note that excitations at nearest-neighbor sites are all anti-correlated, consistent with the site-by-site oscillatory behavior of the local excitations in Fig.~\ref{fig:excitation_time_slices} on a staggered lattice. Both short- and long-range correlations develop during the dynamics. Shorter-range correlations result from endpoint-splitting processes (hopping dynamics) as well as pair creation and string dissociation/recombination. Longer-range correlations develop as the excitations move throughout the bulk while retaining the correlation they acquired during the local processes mentioned above. For the lighter mass, greater production of matter excitations leads to rapid development of a dense correlation structure throughout the bulk, redistributed across the interior. 
In particular, at around $gt=2$ and $gt=4$ when the first two primary scattering of excitations happen, significant correlations develop in the middle of the lattice. In contrast, for the heavier mass, particle production is suppressed, and the primary processes driving the dynamics are continuous splitting of excitation trajectories and propagation of these excitations into the bulk. As a result, for the heavier mass, the correlation spreading is more orderly and no significant correlation is generated in the middle of the lattice at the primary collision point around $gt=4$.

\vspace{1 cm}

We end this section with a discussion of the numerical accuracy of our dynamical computations. The above results are obtained by evolving the initial state for a fixed flux truncation $j_{\rm max}$, at a fixed maximum allowed bond dimension $D_\text{max}$, and using the 1-site TDVP algorithm~\cite{Paeckel:2019yjf,Vanderstraeten:2019voi}. We analyze the effects of $j_\text{max}$ truncation and $D_\text{max}$ restriction (with 1-site and 2-site algorithms) in Appendices~\ref{app:truncation-error} and \ref{app:bond-dimension}, respectively. 

First, as we demonstrate in Appendix~\ref{app:truncation-error}, Hilbert-space truncation has a pronounced impact when increasing the cutoff from $j_\text{max} = \frac{1}{2}$ to $j_\text{max}=\frac{3}{2}$ while the impact is less severe when increasing the cutoff to $j_\text{max}>\frac{3}{2}$. In fact, dynamics is distinctly different for the minimal cutoff value $j_\text{max} = \frac{1}{2}$ (as used in early studies) and the cutoff value $j_\text{max} = \frac{5}{2}$ used in this work.

Second, in Appendix~\ref{app:bond-dimension}, we repeat the computations for different choices of bond dimension for the lighter-mass case. Simulating long-time evolution is challenging with TNs due to the linear growth of entanglement entropy~\cite{BenediktKloss:2018}. Working in the massive regime, nonetheless, substantially reduces bond-dimension requirements. At early times, the dynamics are accurately captured within the maximum bond dimension used in our simulations. The light-mass results start to accumulate errors due to the finite bond-dimensions of the simulations. We opt to present results for which the relative errors for local observables are at most 10$\%$, amounting to a maximum simulation time of $gt=8$. The heavier-mass results are better converged for the same bond dimensions, allowing the results to be reliable for a maximum simulation time of $gt=16$.

%%%%%%%%%%%%%%%%%%%%%%%%%%%%%%%%%%%%%%%%%%%%%%%%%%%%%%%%%%%%%%%%%%%%%%%%%%%%%%%%%%%%%%%%%%%%%%%%%%%%%%%%%%%%%%%%%%%%%%%%%%%%%%%%%%%%%%%%%%%%%%%%%%%%%%%%%%%%%%%%%%
\section{Conclusions and outlook
\label{sec:conclusions}}
\noindent String-breaking dynamics is at the core of quantum-chromodynamics phenomenology; yet first-principles simulations of string-breaking phenomena in real time are elusive. Tensor networks provide a computational tool to probe such phenomena, albeit in simpler, lower-dimensional gauge theories to date. Aiming to ultimately make such methods amenable to complex, higher-dimensional models, we present in this work TN computations rooted in the loop-string-hadron formulation of non-Abelian lattice gauge theories. The LSH formulation is equivalent to the Kogut-Susskind lattice formulation; is fully local; consists of only gauge-invariant degrees of freedom; requires imposition of only Abelian Gauss's laws; and can be applied for both SU(2) and SU(3) gauge theories in any spatial dimension~\cite{Raychowdhury:2019iki,Kadam:2022ipf,Kadam:2025trs,Kadam:2024ifg}. Focusing on an SU(2) LGT coupled to staggered fermions in (1+1)D, we develop LSH-based matrix-product-state and matrix-product-operator ansatzes, and use the standard TN toolbox, to compute static and dynamic properties of mesonic strings, i.e., static or dynamical color charges connected by color electric flux.

Our study of static strings yields a determination of the string tension in the continuum and thermodynamic limits. For a fermion mass $m=0.5g$, the continuum string tension is $\kappa=0.330(3)g^2$, where $g$ is the gauge-matter coupling. Our study of dynamical string breaking illuminates underlying microscopic processes contributing to the evolution and breaking of a string in quench dynamics, including string expansion and contraction, endpoint splitting and particle shower, chain scattering events, including inelastic ones involving string dissociation and recombination, and bare-baryon and bare-meson production. We relate these underlying processes to several features of dynamics, such as energy transport, entanglement-entropy generation, and correlations spreading. These analyses reveal sharp distinction in string-breaking dynamics between small and large bare fermion masses. While the lighter fermion mass allows for inelastic chain scattering events, abundant particle production, full screening of the original electric flux, ballistic transport of energy, and linear energy- and entropy-front cones, the heavier mass yields suppressed inelastic events, reduced particle production, partial electric-flux screening, nonballistic and nondiffusive energy transport, and nonlinear energy- and entropy-front propagations.

This work, therefore, offers a comprehensive study of string breaking in a (1+1)D SU(2) LGT. It goes beyond preceding studies~\cite{Kuhn:2015zqa,Sala:2018dui,Spitz:2018eps} by extending the system size and local Hilbert-space dimension, and by working closer to the continuum limit, where the interacting vacuum and string states exhibit nontrivial matter and flux excitations. The local LSH gauge-invariant degrees of freedom offer an intuitive and systematic way to characterize the gauge-invariant, phenomenologically motivated subprocesses contributing to the string-breaking dynamics. The LSH formulation also allowed for a more efficient encoding of the link degrees of freedom without sacrificing locality. Hence, it enables efficient computations even with sizable values for flux cutoffs. For example, with a cutoff of $j_{\rm max}=\frac{5}{2}$ that is used in our dynamical study, one needs to encode a 91-dimensional link Hilbert space using the original Kogut-Susskind formulation. This cutoff value translates to a maximum loop quantum number $n_{l, {\rm max}}=5$ in the LSH formulation, demanding only a 6-dimensional Hilbert space for the local bosons. Importantly, we demonstrate that small cutoff values yield qualitatively different real-time phenomena; hence, sizable electric-flux cutoffs may be essential in capturing the accurate dynamics.

Our work can be expanded in several ways. First, performing thermodynamic and continuum limits of dynamical quantities is essential but requires extensive computations. Second, our extensive study of dynamical meson strings and their fate in the quench dynamics reveals close connections between quench and scattering processes. A controlled scattering simulation can allow for better characterization of the scattering outcome and associated probabilities, and relations to the initial conditions. While relatively complete scattering studies using TNs exist for Ising field theories~\cite{Milsted:2020jmf,Jha:2024jan}, Abelian gauge theories~\cite{Pichler:2015yqa,Rigobello:2021fxw,Belyansky:2023rgh,Papaefstathiou:2024zsu}, and quantum link models~\cite{Rico:2013qya} in (1+1)D, non-Abelian studies are still rare~\cite{Barata:2025rjb}. Third, while we have analyzed a wealth of observables to characterize the state at various stages of dynamics, such descriptions can be eventually supplemented by tailored, and more quantitative, studies. For example, various current-current correlators can be computed in the time-evolved state to deduce transport properties of the medium; or the overlap between the time-evolved state and various asymptotic hadrons can be computed to quantify hadronization rate and particle yields.

Last but not least, while the Gauss's law constraints can be solved by fully integrating out the gauge degrees of freedom in (1+1)D~\cite{Hamer:1997dx,Davoudi:2020yln}, or by introducing nonlocal gauge-invariant degrees of freedom in higher dimensions~\cite{Mathur:2016cko,DAndrea:2023qnr,Grabowska:2024emw}, the loss of locality poses computational challenges for the TN method. Recently, Ref.~\cite{Dempsey:2025wia} introduced an alternative approach that encodes gauge fields in the virtual bonds of an explicitly gauge-invariant MPS, restoring locality and translation invariance; however, extensions to (2+1)D ansatzes require contending with orthogonality of gauge-field state vectors in tensor contractions. The ultimate power of the LSH formulation is in simplifying the structure of states and constraints in higher dimensions, providing a more straightforward path to higher-dimensional TN constructions. Given the rich string and scattering phenomenology in (2+1)D systems reported in recent studies~\cite{Cochran:2024rwe,Tian:2025mbv,Cao:2026qky,Cataldi:2025cyo,Pavesic:2025nwm}, development and application of LSH-based tensor-network tools, using MPS, PEPS, or TTNs, presents exciting opportunities for future investigations.

Ultimately, TN studies of confining non-Abelian theories can inform hadronization and fragmentation models in nuclear and high-energy physics~\cite{Andersson:1983ia,Webber:1983if,Buckley:2011ms,Altmann:2024kwx,Gieseke:2025mcy}, and begin to test their assumptions and limitations. Along the way, they with provide benchmarking tools for quantum-simulation experiments~\cite{De:2024smi,Luo:2025qlg,Gonzalez-Cuadra:2024xul,Liu:2024lut,Ilcic:2026cac,Halimeh:2025vvp} and quantum-computing implementations~\cite{Cochran:2024rwe,Ciavarella:2024lsp,Crippa:2024hso,Alexandrou:2025vaj,Halimeh:2025vvp} of the string-breaking physics.

%%%%%%%%%%%%%%%%%%%%%%%%%%%%%%%%%%%%%%%%%%%%%%%%%%%%%%%%%%%%%%%%%%%%%%%%%%%%%%%%%%%%%%%%%%%%%%%%%%%%%%%%%%%%%%%%%%%%%%%%%%%%%%%%%%%%%%%%%%%%%%%%%%%%%%%%%%%%%%%%%%
\section*{Acknowledgments}
\noindent N.G. and E.M. would like to thank, respectively, David Huse and Lukas Devos for valuable discussions. Z.D. is grateful to Ignacio Cirac and the Max Planck Institute for Quantum Optics (MPQ), Garching, Germany, for their hospitality during the completion of this work. The visit to MPQ was made possible by a Humboldt Research Fellowship by the Alexander von Humboldt Foundation. I.R. acknowledges helpful discussions at the QC4HEP working-group~\cite{di2024quantum} meetings. 

N.G. and Z.D. were supported by the U.S. National Science Foundation’s Quantum Leap Challenge Institute (award no. OMA-2120757); the U.S. Department of Energy (DOE), Office of Science, Office of Nuclear Physics (award no. DE-SC0026067); and Maryland Center for Fundamental Physics, Department of Physics, and College of Computer, Mathematical, and Natural Sciences at the University of Maryland.
I.R. was supported by the OPERA award (FR/SCM/11-Dec-2020/PHY) from BITS-Pilani; the Start-up Research Grant (SRG/2022/000972) from ANRF, India; and the cross-discipline research fund (C1/23/185) from BITS-Pilani. 
The work by E.M. was  supported by fellowship support from Birla Institute of Technology and Science (BITS)-Pilani and the International Travel Support (ITS/2024/002694) from ANRF, India.
S.K. was supported by the U.S. DOE, Office of Science, Office of Nuclear Physics, IQuS under Award Number DOE (NP) Award DE-SC0020970 via the program on Quantum Horizons: QIS Research and Innovation for Nuclear Science; the U.S. Department of Energy QuantISED program through the theory consortium ``Intersections of QIS and Theoretical Particle Physics'' at Fermilab (Fermilab subcontract no. 666484); and in part through the Department of Physics and the College of Arts and Sciences at the University of Washington. 
The work by J.R.S. and A.B. was supported by the U.S. DOE, Office of Science (award no. DE-AC02-05CH11231), partially through the Quantum-Information-Science-Enabled Discovery (QuantISED) program for High Energy Physics (KA2401032).
A.B. acknowledges further support from the DOE, Office of Science, Office of Advanced Scientific Computing Research, Quantum Testbed Pathfinder program (awards no.~DE-SC0019040 and DE-SC0024220). 
N.M. acknowledges support from the U.S. DOE, Office of Science, National Quantum Information Science Research Centers, Quantum Systems Accelerator (award no. DE-SCL0000121).

We acknowledge computing resources from the Zaratan HPC cluster at the University of Maryland; an ANRF-funded computing facility at BITS Pilani, Goa  and the Sharanga HPC cluster at the BITS-Pilani, Hyderabad; and the Hyak supercomputer system at the University of Washington.

\appendix
%%%%%%
%%%%%%
\section*{Appendices}
\subsection{Overview of the Kogut-Susskind formulation}
\label{app:KS-formulation}
\noindent Consider an SU(2) LGT on a one-dimensional lattice with an even number of sites $N$, lattice spacing $a$, and open boundary conditions. The Kogut-Susskind Hamiltonian is formulated in a temporal gauge with continuous time. A staggered fermionic matter field in the fundamental representation inhabits the lattice site $r$,\footnote{With the physical position $ra$.} and it is denoted by
\begin{align}
    \hat\psi(r)\coloneq\begin{pmatrix}\hat\psi_1(r)\\ \hat\psi_2(r)\end{pmatrix}.
    \label{eq:psi}
\end{align}
The gauge holonomy or the link operator $\hat{U}(r,r+1) \eqcolon\hat{U}(r)$ is an SU(2) matrix (with operator-valued matrix elements) situated on the link connecting the lattice sites at $r$ and $r+1$. The electric-field operators, i.e., the canonical conjugate variables of $\hat{U}(r)$, are located at the left and right ends of the link, and they are denoted by $\hat E^{\rm a}_L(r)$ and $\hat E^{\rm a}_R(r+1)$, respectively.\footnote{Note that the $L/R$ labels are inherited from the sides of a \emph{link} joining sites $r$ on the left and $r+1$ on the right. There is a potential for confusion because in relation to a \emph{site} $r$, the $L$-side ($R$-side) quantities reside to the \emph{right} (\emph{left}) of the site.}
Here, ${\rm a}=1,2,3$ denotes the adjoint (i.e., color) index. The electric-field operators satisfy
\begin{subequations}
\label{eq:EUcomm}
\begin{align}
    & [\hat E^{\rm a}_L(r),\hat U(r')] =-T^{\rm a}\hat U(r)\delta_{rr'}, \\
   & [\hat E^{\rm a}_R(r+1),\hat U(r')] =\hat U(r)T^{\rm a}\delta_{rr'},
\end{align}
\end{subequations}
where $T^{\rm a}=\frac{\sigma^{\rm a}}{2}$ with $\sigma^{\rm a}$ denoting the Pauli matrices.
Furthermore, the left and right electric fields form independent copies of the SU(2) algebra,
\begin{subequations}
\label{eq:ERELcomm}
\begin{align}
    &[\hat{E}_{L/R}^{\rm a}(r),\hat{E}_{L/R}^{\rm b}(r')]=i\epsilon^{\rm abc}\delta_{rr'} \hat{E}_{L/R}^{\rm c}(r),
    \\
   & {[\hat{E}_L^{\rm a}(r),\hat{E}_R^{\rm b}(r')]}=0.
\end{align}
\end{subequations}
However, Eq.~\eqref{eq:EUcomm} relates $\hat{E}_L^{\rm a}(r)$ and $\hat{E}_R^{\rm b}(r+1)$ through a parallel transport, and imposes a constraint on their Casimirs as
\begin{align}
    E^2(r)\coloneq E_L^{\rm a}(r)E_L^{\rm a}(r)=E_R^{\rm a}(r+1)E_R^{\rm a}(r+1).
    \label{eq:ELER}
\end{align}

The local Hilbert space at each lattice site is spanned by the irreducible-representation (irrep) basis states of the SU(2) groups generated by Eq.~\eqref{eq:ERELcomm}, as well as the staggered fermionic occupation states associated with the $\hat{\psi}_1$ and $\hat{\psi}_2$ operators.
The left (right) irrep states at site $r$ are labeled by the total angular-momentum quantum number $j_L(r)$ $(j_R(r))$ and azimuthal angular-number quantum number $m_L(r)$ $(m_R(r))$. The global Hilbert space is spanned by states that are tensor products of local states satisfying Eq.~\eqref{eq:ELER}, i.e., 
\begin{equation}
    j_L(r) = j_R(r+1),
    \label{eq:jLjRequality}
\end{equation}
at all $r$. However, a physical state in this Hilbert space must be gauge invariant under all local gauge transformations. The generators of gauge transformations at site $r$ are the Gauss's law operators
\begin{align}
    \hat G^{\rm a}(r)\coloneq \hat E^{\rm a}_R(r)+\hat E^{\rm a}_L(r)+\hat\psi^{\dagger}(r)T^{\rm a}\hat \psi(r),
    \label{eq:Ga}
\end{align}
and the states annihilated by $\hat G^{\rm a}(r)$ at all sites are the physical states.

The Kogut-Susskind Hamiltonian can be written as
\begin{eqnarray}
\label{eq:H_KS}
    \hat H&\coloneq& \frac{g^2a}{2} \sum_{r=1}^{N-1} E^2(r)\nonumber \\
    && + m\sum_{r=1}^{N}(-1)^r\hat{\psi}^{\dagger}(r)\hat{\psi}(r)\nonumber \\
    && + \frac{1}{2a}\sum_{r=1}^{N-1}\left[\hat{\psi}^{\dagger}(r) \hat{U}(r)\hat{\psi}(r+1)+{\rm H.c.} \right],
    \label{eq:HKS}
\end{eqnarray}
where the sum over appropriate color indices is implied through matrix multiplication. The first term denotes the electric-field energy stemming from the SU(2) gauge field with $g$ being the bare gauge coupling. The second term indicates matter rest-mass energy with $m$ being the bare fermion mass. The staggering phase implies that sites labeled by an even (odd) $r$ are particle (antiparticle) sites. Finally, the last term is the interaction energy between the matter and the gauge degrees of freedom, where the fermionic fields at both ends of a link interact with the corresponding link operator. This term is interchangeably called the (fermion) hopping Hamiltonian.

The Hamiltonian in Eq.~\eqref{eq:H_KS} commutes with the Gauss's law operators in Eq.~\eqref{eq:Ga}. Thus, the unitary time evolution under this Hamiltonian maps a physical state to another physical state; i.e., dynamics of an initial physical state is restricted within the subspace of all physical states. In addition to the local gauge symmetry, this Hamiltonian is invariant under the global $U(1)$ transformation $\hat \psi(r)\rightarrow e^{i\alpha}\hat\psi(r)$. The conserved charge associated with this symmetry is the total fermion-number operator $\hat Q \coloneq \sum_r \hat\psi^\dagger(r)\hat\psi(r)$.
Furthermore, with open boundary conditions in (1+1)D, one can always transform to a gauge in which $\hat U(r)=\hat{\mathds{1}}$ for all $r$. Then, the Hamiltonian commutes with $\hat q^a \coloneq\sum_r\hat\psi^\dagger(r)T^a\hat\psi(r)$, hence supplying another global conserved charge. Applying Gauss's laws, it is easy to see that this charge operator equals $E^a_L(N)-E^a_R(1)$. Consequently, the difference between incoming $j_R(1)$ and outgoing $j_L(N)$ total angular-momentum quantum numbers is conserved. We denote this conserved quantum number by $q/2$.

Finally, it is useful to define a dimensionless Hamiltonian by rescaling Eq.~\eqref{eq:HKS} with $2/(ag^2)$ such that
\begin{equation}
    \label{eq:H_dimless-app}
    \hat{\tilde{H}}\coloneq \frac{2 \hat{H}}{ag^2} = \hat H_E + \hat H_M + \hat H_I. 
\end{equation}
The electric-energy term, $H_E$, mass-energy term, $H_M$, and interaction-energy term, $H_I$, are given by
\begin{subequations}
\label{eq:HEMI_defn_KS}
\begin{align}
    & \hat H_E \coloneq \sum_{r=1}^{N-1} 
    \hat E^2(r),
    \label{eq:HE_defn_KS}\\
    & \hat H_M \coloneq \mu \sum_{r=1}^{N}(-1)^r\hat{\psi}^{\dagger}(r)\hat{\psi}(r),\label{eq:HM_defn_KS} \\
    & \hat H_I \coloneq x \sum_{r=1}^{N-1}\left[\hat{\psi}^{\dagger}(r) \hat{U}(r)\hat{\psi}(r+1)+{\rm H.c.} \right],
    \label{eq:HI_defn_KS}
\end{align}
\end{subequations}
where $\mu=\frac{2m\sqrt{x}}{g}$ and $x=\frac{1}{a^2g^2}$ are dimensionless couplings. In the next appendix, we briefly summarize the LSH reformulation of this Hamiltonian in the physical sector of the theory.

\subsection{Loop-string-hadron dictionary
\label{app:lsh-formalism}}
\begin{figure*}[t!]
    \centering
    \includegraphics[scale=0.675]{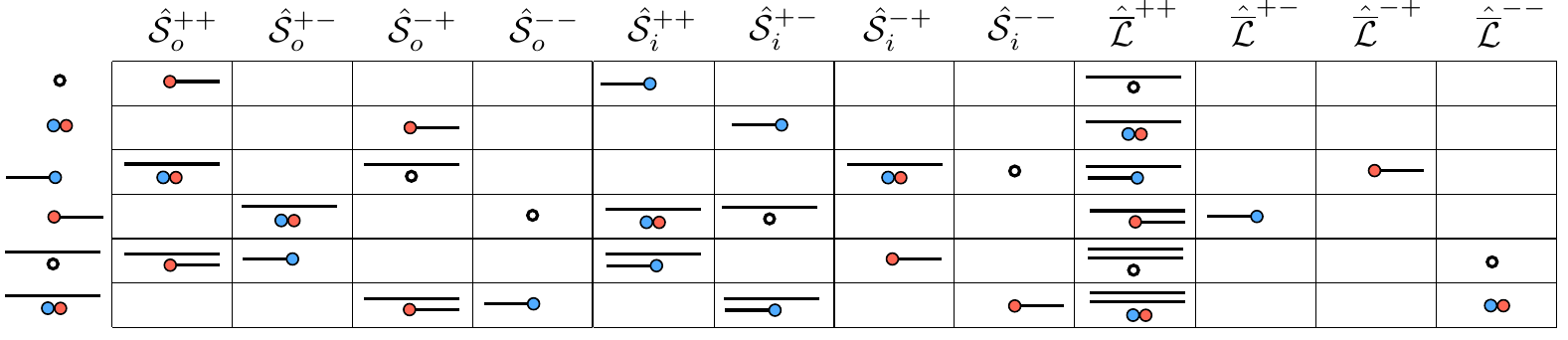}
    \caption{The action of the local LSH string and loop operators (appearing in the hopping Hamiltonian and constituting the mesonic string operator) on a few low-flux onsite LSH basis states. The translation between the pictorially represented states and the LSH quantum numbers is given in Fig.~\ref{fig:lsh-states} of the main text.
    An empty entry in the table implies that the operator annihilates the state.}
    \label{fig:string-action}
\end{figure*}
\noindent The LSH reformulation is built upon the prepotential framework~\cite{Mathur:2004kr,Mathur:2007nu,Mathur:2010wc,Anishetty:2009ai,Anishetty:2009nh,Anishetty:2014tta,Raychowdhury:2013rwa,Raychowdhury:2014eta,Raychowdhury:2018tfj}.
Prepotentials are the bosonic operators that construct the irrep basis states of a simple unitary group from a vacuum (``empty'') state.
For the SU(2) case, the prepotentials are also known as Schwinger bosons~\cite{biedenharn1965quantum}; they are the ladder operators associated with two quantum harmonic oscillators, in form of a color doublet, with the same frequency.
The irreps of the SU(2) groups generated by $E_L(r)$ and $E_R(r)$ in Eq.~\eqref{eq:ERELcomm} can be constructed from a set of doublets at each site $r$:
\begin{align}
    \hat a_{L/R}(r)\coloneq \begin{pmatrix}
        \hat a_{1,L/R}(r) \\
        \hat a_{2,L/R}(r)
    \end{pmatrix}.
\end{align}
The total angular-momentum quantum number of a state, $j_{L/R}(r)$, is equal to half of the total number of colored oscillators acting on the vacuum state. The latter is given by the eigenvalue ${N}_{L/R} (r)$ of the number operator
\begin{equation}
    \hat{N}_{L/R} (r) \coloneq \hat{a}^\dagger_{L/R}(r)\hat{a}_{L/R}(r).
    \label{eq:NLNRdefn}
\end{equation}
Furthermore, prepotentials allow for the construction of gauge-invariant operators by contracting the color indices of local bosonic $(\hat{a}_{L/R})$ and fermionic $(\hat{\psi})$ field operators.
Consider the following local gauge-singlet operators that are constructed only from creation-type operators. 
The first operator,
\begin{align}
\label{eq:Lpp}
\hat{\mc{L}}^{++}(r)\coloneq\epsilon_{\alpha,\beta}\hat a_{\alpha,R}^{\dagger}(r)\hat a_{\beta,L}^{\dagger}(r),
\end{align}
is called a \emph{loop} operator, since it only involves contraction of bosonic operators.
The next two are called \emph{string-in} and \emph{-out} operators, which are combinations of bosonic and fermionic operators:
\begin{subequations}
\begin{align}
\label{eq:Sppi}
\hat{\mc S}^{++}_i(r)\coloneq\hat{a}_{\alpha,R}^{\dagger}(r)\hat \psi_{\beta}^{\dagger}\hat(r) \epsilon_{\alpha,\beta},\\
\label{eq:Sppo}
\hat{\mc S}^{++}_o(r)\coloneq\epsilon_{\alpha,\beta}\hat \psi_{\alpha}^{\dagger}\hat(r) \hat{a}_{\beta,L}^{\dagger}(r).
\end{align}
\end{subequations}
Finally, the singlet formed by the matter operators constitute the \emph{hadron} operator,
\begin{align}
\label{eq:Hpp}
\hat{\mc H}^{++}\coloneq-\frac{1}{2}\epsilon_{\alpha,\beta}\hat \psi_{\alpha}^{\dagger}\hat \psi_{\beta}^{\dagger}.
\end{align}
Similarly, four annihilation-type operators can be constructed upon Hermitian conjugation of the four creation-type operators above.

Consider a local vacuum state, i.e., the state annihilated by all four local annihilation-type operators introduced above. A complete and orthonormal basis that spans the local Hilbert space of gauge-singlet states can be constructed by exciting such a vacuum state using the creation-type operators~\cite{Raychowdhury:2019iki}. 
As already described in Sec.~\ref{sec:LSH-formulation}, the states in this Hilbert space can be labeled by one bosonic quantum number $n_l\in\{0,1,\cdots\}$ and two fermionic quantum numbers $n_i, n_o\in\{0,1\}$. A local Hilbert space at $r$, spanned by states $\ket{n_l,n_i,n_o}_r$, is manifestly gauge invariant and automatically obeys the Gauss's laws. The number operators corresponding to the LSH quantum numbers are defined as 
\begin{equation}
    \hat{n}_{l/i/o}(r)\ket{n_l, n_i,n_o}_r = {n}_{l/i/o}(r)\ket{n_l, n_i,n_o}_r.
    \label{eq:nlninodefn}
\end{equation}

The Hilbert space of the entire lattice is spanned by states that are tensor product of local basis states:
\begin{equation}
    \ket{\Psi} = \bigotimes_{r=1}^{N} \ket{n_l,n_i,n_o}_{r}.
    \label{eq:globalPsi-app}
\end{equation}
However, a physical state in this Hilbert space must obey Eq.~\eqref{eq:jLjRequality}, which from the discussion around Eq.~\eqref{eq:NLNRdefn}, constrains $N_L$ and $N_R$ values at sites across a link according to
\begin{equation}
    \label{eq:AGL}
    N_L(r)=N_R(r+1).
\end{equation}
Equation~\eqref{eq:AGL} is known as the AGL in the LSH formulation; it is an algebraic equality condition on the number of $L$ and $R$ prepotentials at the neighboring sites. The $N_L$ and $N_R$ values can be expressed in terms of LSH quantum numbers, as given in Eq.~\eqref{eq:NLRasnlnino} of the main text. Thus, the AGL can be stated as a continuity requirement on the flux lines along the link connecting two neighboring sites.

The electric-field operator and the gauge-link operator can be expressed in terms of prepotentials as
\begin{subequations}
\label{eq:E-U}
\begin{align}
    \label{eq:ELER_in_pretoentials}
    &\hat{E}_{L/R}(r) = \hat{a}^\dagger_{L/R}(r)T^a\hat{a}_{L/R}(r),\\
    &\hat{U}(r,r+1) = \hat{U}_L(r)\hat{U}_R(r+1),
    \label{eq:ULUR}
\end{align}
\end{subequations}
where
\begin{subequations}
\label{eq:ULR}
\begin{align}
\label{eq:UL}
    & \hat{U}_L(r) \coloneq \frac{1}{\sqrt{\hat{N}_L(r)+1}}
    \begin{bmatrix}
        \hat{a}^\dagger_{2,L} (r) && \hat{a}_{1,L} (r)\\
        -\hat{a}^\dagger_{1,L} (r) && \hat{a}_{2,L}(r)
    \end{bmatrix},\\
    & \hat{U}_R(r) \coloneq
    \label{eq:UR}
    \begin{bmatrix}
        \hat{a}^\dagger_{1,R}(r) && \hat{a}^\dagger_{2,R}(r)\\
        -\hat{a}_{2,R}(r) && \hat{a}_{1,R}(r)
    \end{bmatrix}\frac{1}{\sqrt{\hat{N}_R(r)+1}}.
\end{align}
\end{subequations}
These relations can be used to rewrite the Kogut-Susskind Hamiltonian terms in terms of prepotential and fermion fields, and ultimately in terms of the gauge-invariant LSH operators derived from them. The result is Eq.~\eqref{eq:H_dimless} of the main text. Here, we provide the explicit definitions of the relevant LSH operators used in the construction of the LSH Hamiltonian, and the string operators introduced in Sec.~\ref{sec:string-setup}. The operators and states are defined at a given site $r$; the site arguments or labels will be dropped for brevity.

We first introduce a set of ladder operators for the LSH fermionic and bosonic degrees of freedom, whose actions on the multi-site LSH basis state $\ket{\Psi}$ in Eq.~\eqref{eq:globalPsi-app} are:
\begin{widetext}
\begin{subequations}
\label{eq:lambda-chi-ops}
\begin{align}
\label{nlp}
&\hat \lambda^+(r)\bigotimes_{s} \ket{n_l,n_i,n_o}_{s} = (1-\delta_{n_l(r),n_{l,\rm max}})\ket{n_l+1,n_i,n_o}_{r}\bigotimes_{s\neq r}^{} \ket{n_l,n_i,n_o}_{s}, \\
\label{nlm}
&\hat \lambda^-(r)\bigotimes_{s} \ket{n_l,n_i,n_o}_{s} =(1-\delta_{n_l(r),0}) \ket{n_l-1,n_i,n_o}_{r}\bigotimes_{s\neq r}^{} \ket{n_l,n_i,n_o}_{s}, \\
\label{nip}
&\hat \chi_i^{+}(r)\bigotimes_{s} \ket{n_l,n_i,n_o}_{s} = \delta_{n_i(r),0}(-1)^{\sum_{s<r}^{} (n_i(s)+n_o(s))}\ket{n_l,n_i+1,n_o}_{r}\bigotimes_{s\neq r}^{} \ket{n_l,n_i,n_o}_{s}, \\
\label{nim}
&\hat \chi_i^{-}(r)\bigotimes_{s} \ket{n_l,n_i,n_o}_{s} = \delta_{n_i(r),1}(-1)^{\sum_{s<r}^{} (n_i(s)+n_o(s))}\ket{n_l,n_i-1,n_o}_{r}\bigotimes_{s\neq r}^{} \ket{n_l,n_i,n_o}_{s}, \\
\label{nop}
&\hat \chi_o^{+}(r)\bigotimes_{s} \ket{n_l,n_i,n_o}_{s} = \delta_{n_o(r),0}(-1)^{n_i(r)+\sum_{s<r}^{} (n_i(s)+n_o(s))}\ket{n_l,n_i,n_o+1}_{r}\bigotimes_{s\neq r}^{} \ket{n_l,n_i,n_o}_{s}, \\
\label{nom}
&\hat \chi_o^{-}(r)\bigotimes_{s} \ket{n_l,n_i,n_o}_{s} = \delta_{n_o(r),1}(-1)^{n_i(r)+\sum_{s<r}^{} (n_i(s)+n_o(s))}\ket{n_l,n_i,n_o-1}_{r}\bigotimes_{s\neq r}^{} \ket{n_l,n_i,n_o}_{s}.\end{align}
\end{subequations}
\end{widetext}
To write the equations for the fermionic operators, an ordering for the fermionic states is adopted where the states are firstly ordered by site number, and secondly ordered with the $n_i$ state preceding the $n_o$ state at each site. The ladder operators above are relevant to the construction of the string operators that appear in the interaction Hamiltonian, defined in Eq.~\eqref{eq:HI_LSH}:
\begin{subequations}
\label{eq:Spm-defs}
\begin{align}
&\hat{\mathcal{S}}_o^{++}= \hat \chi_o^+ (\lambda^+)^{\hat n_i}\sqrt{\hat n_l+2-\hat n_i},\label{eq:Spm-defs-1} \\
&\hat{\mathcal{S}}_o^{+-}= \hat \chi_i^+ (\lambda^-)^{1-\hat n_o}\sqrt{\hat n_l+2\hat n_o},\label{eq:Spm-defs-2} \\
&\hat{\mathcal{S}}_o^{-+}= \hat \chi_i^- (\lambda^+)^{1-\hat n_o}\sqrt{\hat n_l+1+\hat n_o},\label{eq:Spm-defs-3}\\
&\hat{\mathcal{S}}_o^{--}= \hat \chi_o^- (\lambda^-)^{\hat n_i}\sqrt{\hat n_l+2(1-\hat n_i)},\label{eq:Spm-defs-4}
\end{align}
\end{subequations}
and
\begin{subequations}
\label{eq:Spm-defs-cont}
\begin{align}
&\hat{\mathcal{S}}_i^{++}= \hat \chi_i^+ (\lambda^+)^{\hat n_o}\sqrt{\hat n_l+2-\hat n_o}.
\label{eq:Spm-defs-5} \\
&\hat{\mathcal{S}}_i^{+-}= \hat \chi_o^- (\lambda^+)^{1-\hat n_i}\sqrt{\hat n_l+1+\hat n_i},\label{eq:Spm-defs-6}  \\
&\hat{\mathcal{S}}_i^{-+}= \hat \chi_o^+ (\lambda^-)^{1-\hat n_i}\sqrt{\hat n_l+2\hat n_i}, \label{eq:Spm-defs-7}\\
&\hat{\mathcal{S}}_i^{--}= \hat \chi_i^- (\lambda^-)^{\hat n_o}\sqrt{\hat n_l+2(1-\hat n_o)}.\label{eq:Spm-defs-8}
\end{align}
\end{subequations}
Besides the operators that appear in the Hamiltonian, we need pure loop operators that change the $n_l$ quantum number. These were used in Sec.~\ref{sec:string-setup} to construct a dynamical string operator in Eq.~\eqref{eq:String_operator_LSH}. These operators are defined as:
\begin{subequations}
\label{eq:L-ops}
\begin{align}
    &\hat{\overline{\mathcal{L}}}^{++}=f(\hat N_L)\hat{\lambda}^+\sqrt{\big(\hat{n}_l+1\big)\big(\hat{n}_l+2+(\hat{n}_i\oplus\hat{n}_o)\big)}f(\hat N_R), \\
    &\hat{\overline{\mathcal{L}}}^{+-}=-f(\hat N_L)\hat{\chi}^\dagger_i\hat{\chi}_of(\hat N_R), \\
    &\hat{\overline{\mathcal{L}}}^{-+}=f(\hat N_L)\hat{\chi}_i\hat{\chi}^\dagger_of(\hat N_R),\\
    &\hat{\overline{\mathcal{L}}}^{--}=f(\hat N_L)\hat{\lambda}^-\sqrt{\hat{n}_l\big(\hat{n}_l+1+(\hat{n}_i\oplus\hat{n}_o)\big)}f(\hat N_R),
\end{align}
\end{subequations}
where $f(\hat O)\coloneq\frac{1}{\sqrt{\hat O+1}}$ and $\oplus$ denotes addition modulo two. The action of the string and loop operators on a few low-flux onsite LSH basis states is shown pictorially in Fig.~\ref{fig:string-action}.

%%%%%%%%%%%%%%%%%%%%%%%%%%%%%%%%%%%%%%%%%%%%%%%%%%%%%%%%%%%%%%%%%%%%%%%%%%%%%%%%%%%%%%%%%%%%%%%%%%%%%%%%%%%%%%%%%%%%%%%%%%%%%%%%%%%%%%%%%%%%%%%%%%%%%%%%%%%%%%%%%%
\subsection{Ground-state energy without static charges}
\label{app:ground-state}
\begin{figure*}[t!]
    \centering
    \includegraphics[width=\textwidth]{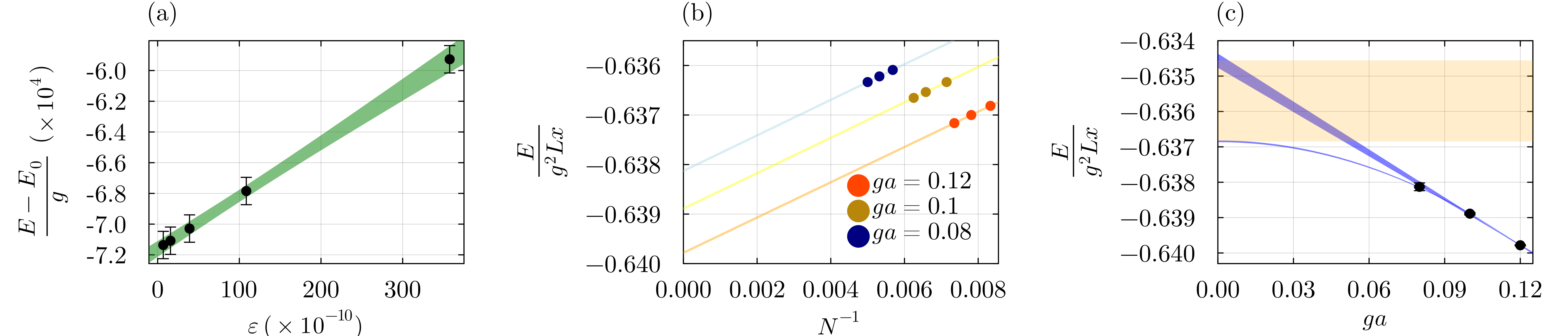}
    \caption{(a) Extrapolation of the no-static-charge ground-state energy (in units of coupling $g$) to zero MPS truncation error, $\varepsilon$, for $(\frac{m}{g},N, ga,j_{\rm max}, gl)=(0.5,140, 0.1,\frac{3}{2}, 0)$. The five truncation-error values correspond to the bond-dimension values $(100, 125, 150, 175, 200)$. The extrapolated ground-state energy is given by -890.868716(5). Using the method described in this appendix, the subsequent extrapolation in $j_{\rm max}$ gives the energy value -890.86873(1) at $(\frac{m}{g},N,ga,gl)=(0.5,140,0.1,0)$. The process is repeated for other values of $N$ and $ga$ (while keeping $\frac{m}{g}$ and $gl$ fixed). (b) Extrapolation of the ground-state energy density $\frac{E}{g^2Lx}$ to the thermodynamic limit using a linear fit of energy values as a function of $N^{-1}$. Three $N$ values are used at each $ga$ value with fixed $\frac{m}{g}=0.5$: $N \in \{120,128,136\}$ for $ga=0.12$, and $N \in \{140,152,160\}$ for $ga=0.1$, $N \in \{176,188,200\}$ for $ga=0.08$. (c) Extrapolation of the ground-state energy to zero lattice spacing $ga$. Two fit forms, linear and quadratic, are tried, with $R$ values of 0.99842 and 0.99992, respectively. The continuum-extrapolated value is obtained by taking the average of the continuum value obtained through these two fits, and turns out to be -0.635(1). The error contains both the statistical fit uncertainties and a systematic uncertainty associated with the difference in the extrapolated values from both fits. The data-point error bars correspond to the input uncertainties rescaled by the fit quality, as explained in the text. The shaded bands indicate the one-standard-error uncertainty of the fitted mean curve.  
    }
    \label{fig:gs}
\end{figure*}
\noindent In this appendix, we present the results for the ground-state energy in the symmetry sector which contains no static charges, to which the strong-coupling vacuum belongs. The first step toward obtaining the ground-state energy is to extrapolate the energy at a fixed value of $(\frac{m}{g},N,x,j_{\rm max},gl)$ to infinite bond dimension. However, it is easier to instead extrapolate to zero MPS truncation error. The reason is that the ground-state energy exhibits a simple linear relationship to the truncation error $\varepsilon$~\cite{wouters2014density}. This linear behavior is evident in the leftmost plot of the energy versus truncation error shown in Fig.~\ref{fig:gs}. 

The relative standard errors on DMRG energies are estimated based on the finite convergence tolerance for DMRG. These relative errors are then used for performing a weighted least squares fit. Next, the fit is then used to compute a reduced chi-squared, $\chi^2_{\rm red}$, which is an estimate for the absolute scale for the variance associated with the energy data points. Thus, when $\chi_{\rm red}^2> 1$, we scale up the relative standard errors by $\sqrt{\chi_{\rm red}^2}$ to estimate the absolute variance on the data points. However, when $\chi_{\rm red}^2<1$, we do not scale down the relative errors. This is the prescription used for determining the error bars associated with the data in all plots. The error bands, on the other hand, describe the uncertainty associated with the fit.Such a weighted fitting scheme based on relative errors, and with estimations of absolute error bars using the reduced chi-squared, are used for all extrapolations (except the extrapolation in flux cutoff) in this appendix and in Sec.~\ref{sec:static}.

In the next step, we extrapolate the energy obtained in the previous step in the flux cutoff $j_{\rm max}$ while holding $(N, x, \frac{m}{g},gl)$ fixed. Computations are performed at three cutoff values $j_{\rm max} \in \{1,\frac{3}{2},2\}$. It is observed that the difference in the energy for the two higher values of cutoff, i.e., $j_{\rm max}=\frac{3}{2}$ and $j_{\rm max}=2$, is less than 10$\%$ of the difference between the energies at the two lower values of cutoff, i.e., $j_{\rm max}=1$ and $j_{\rm max}=\frac{3}{2}$. Thus, the energy at the maximum allowed cutoff $\frac{N}{2}$ is estimated by taking the mean of the energies at the two higher values of flux cutoff. Apart from the statistical error propagated from the previous extrapolation, a systematic error characterized by the difference between these two energy values is also incorporated. This is a fairly conservative error estimate which could be tightened greatly by performing computations at $j_{\rm max}=\frac{5}{2}$ but suffices for our purpose. The MPS truncation and flux-cutoff extrapolations described above are also used in the computations of ground-state energies in the presence of static charges, as described in the main text. \\  

We now proceed to compute the thermodynamic and continuum limits of the energy density.
Fixing $ga$, the energy density, $\frac{E}{g^2Lx}$, is extrapolated to the thermodynamic limit $N\rightarrow \infty$, as shown in the middle plot in Fig.~\ref{fig:gs}. Even though each computation is performed only for three values of lattice volume, it is known that energy depends linearly on $\frac{1}{N}$~\cite{Banuls:2017ena}, and this linear dependence is exploited to perform the extrapolation. The continuum limit, i.e., the value at $ga=0$ (or $x \to \infty$), is estimated by averaging over the results from linear and two-parameter quadratic fits, as shown in the rightmost plot in Fig~\ref{fig:gs}. The continuum-extrapolated value of $\frac{E}{g^2Lx}=-0.635(1)$ turns out to be close to the analytically determined $\frac{m}{g}=0$ value of $-\frac{2}{\pi}$~\cite{Hamer:1981yq}. The model uncertainty in the $ga\rightarrow 0$ extrapolation dominates the error budget.\\ 

%%%%%%%%%%%%%%%%%%%%%%%%%%%%%%%%%%%%%%%%%%%%%%%%%%%%%%%%%%%%%%%%%%%%%%%%%%%%%%%%%%%%%%%%%%%%%%%%%%%%%%%%%%%%%%%%%%%%%%%%%%%%%%%%%%%%%%%%%%%%%%%%%%%%%%%%%%%%%%%%%%
\subsection{On the static-charge sectors
\label{app:static_charges}}
\noindent In this appendix, we describe the notion of external charges used for the computation of string tension in Sec.~\ref{sec:static}. We begin by reviewing the definitions of static charges in the Kogut-Susskind formulation  previously used in the literature~\cite{Zohar:2015hwa, Kuhn:2015zqa, Sala:2018dui}. Then, we describe how this notion can be adapted to the LSH formulation.

The Gauss's law operators $\hat G^{\rm a}(r)$, defined in Eq.~\eqref{eq:Ga}, satisfy the $\mathfrak{su}(2)$ algebra, and hence do not commute. A convenient, maximally commuting subset consists of the quadratic Casimir,  
\begin{align}
    \hat{\mathcal{J}}(r) \coloneq \hat G^{\rm a}(r)\hat G^{\rm a}(r),
\end{align}
together with one Cartan generator, e.g.,
\begin{align}\hat{\mathcal{M}}(r)=\hat G^3(r).
\end{align}
Their joint eigenvalues are denoted by a spin quantum number $\mathcal{J}(r)$ and a magnetic quantum number $\mathcal{M}(r)$. In this way, the local Gauss's law operators decompose the Kogut-Susskind Hilbert space $\mathbb{H}_{\text{KS}}$ into a direct sum of Gauss's law sectors,  
\begin{align}
    \mathbb{H}_{\text{KS}}\;
    =\;\bigoplus_{\{\mathcal{J}(r),\mathcal{M}(r)\}} \mathbb{H}_{\{\mathcal{J}(r),\mathcal{M}(r)\}}.
\end{align}
We refer to these non-Abelian Gauss's law sectors as \emph{non-Abelian static-charge sectors}. Different values of $\{\mathcal{J}(r),\mathcal{M}(r)\}$ specify different external-charge configurations~\cite{Zohar:2015hwa}. For instance, the states that satisfy $\mathcal{J}(r)=0$ at all sites $r$ contain no static charges. A nonzero value for $\mathcal{J}(r)$ at any site $r$, on the other hand, corresponds to the presence of static charges. In particular, we are interested in the sector with a pair of fundamental SU(2) static charges with opposite colors located at the sites $r_1$ and $r_2$ (with $r_1 < r_2$):
\begin{align}
    \mathcal{J}(r_1)=\mathcal{J}(r_2)=\tfrac{1}{2},~~ \mathcal{M}(r_1)=+\tfrac{1}{2},~~ \mathcal{M}(r_2)=-\tfrac{1}{2}.
    \label{eq:static_string_sector}
\end{align}

Another perspective is to enlarge the Hilbert space by introducing explicit static-charge degrees of freedom. For the spin-$\tfrac{1}{2}$ pair, this setting amounts to introducing:
\begin{equation}
\mathbb{H}'_{\text{KS}} \;\coloneq\; \mathbb{H}_{\text{KS}}\otimes \mathbb{C}^2\otimes \mathbb{C}^2,
\end{equation}
where the two $\mathbb{C}^2$ spaces carry the fundamental SU(2) representations at sites $r_1$ and $r_2$. Because the charges are static, the Hamiltonian remains the Kogut-Susskind Hamiltonian in Eq.~\eqref{eq:HKS}, extended trivially to $\hat H' \coloneq \hat H\otimes \hat{\mathds{1}} \otimes \hat{\mathds{1}}$. The only modification is in Gauss’s laws, which now act jointly on the system and static degrees of freedom:  
\begin{align}
    \hat G'^{\rm a}(r)\coloneq \hat G^{\rm a}(r)+\hat \rho^{\text{ext},{\rm a}}(r).
\end{align}
Here,
\begin{align}
    \hat \rho^{\text{ext},{\rm a}}(r)\coloneq\delta_{r,r_1}\frac{\hat \sigma^{\rm a}(r_1)}{2}+\delta_{r,r_2}\frac{\hat \sigma^{\rm a}(r_2)}{2},
\end{align}
where $\hat \sigma^{\rm a}(r_1)$ and $\hat \sigma^{\rm a}(r_2)$ are the spin Pauli operators acting on the Hilbert spaces of the two external charges. Due to this combined Gauss's law constraint, the ground state of $\hat H'$ exhibits entanglement between the dynamical and static degrees of freedom. In some contexts, this entanglement is of physical interest, as discussed in Ref.~\cite{Sala:2018dui}, but here only the dynamical sector is of interest. Thus, we project onto a fixed state in the external sector, and  work with states of the form,
\begin{align}
    \gket{\Phi}=\gket{\phi}\otimes \gket{\chi(r_1)}\otimes \gket{\chi(r_2)}\in \mathbb{H}'_{\rm KS},
\end{align}
where $\gket{\phi}\in \mathbb{H}_{\rm KS}$ and $\gket{\chi(r_1)}, \gket{\chi(r_2)}\in \mathbb{C}^2$. The extended Gauss's law condition on  $\gket{\Phi}$, i.e.,
\begin{align}
    \hat G'^{\rm a}(r)\gket{\Phi}=0,
\end{align}
leads to
\begin{align}
\label{eq:modified-KS-GL}
    \hat G^{\rm a}(r)\gket{\phi}&\otimes \gket{\chi_1(r_1)}\otimes\gket{\chi_2(r_2)}=\nonumber \\
    &-\gket{\phi}\otimes \hat \rho^{\text{ext},{\rm a}}(r)\big(\chi_1(r_1)\otimes\chi_2(r_2)\big).
\end{align}
Thus, if one chooses the state of the external charges to be
\begin{align}
    \gket{\chi_1(r_1)}=\begin{pmatrix}
        0 \\ 1
    \end{pmatrix},~~ \gket{\chi_2(r_2)}=\begin{pmatrix}
        1 \\ 0
    \end{pmatrix},
\end{align}
Eq.~\eqref{eq:modified-KS-GL} essentially constrains the state $\gket{\phi}\in \mathbb{H}_{\rm KS}$ to be in the static charge sector defined in Eq.~\eqref{eq:static_string_sector}.

In practice, this approach is implemented by preparing an initial state in the desired Gauss's law sector, i.e., that specified by Eq.~\eqref{eq:static_string_sector} for a pair of static charges, then evolving the state toward the ground state within that sector. This evolution to the ground state can be implemented numerically using techniques such as DMRG and imaginary-time evolution. The latter was used in Ref.~\cite{Kuhn:2015zqa}: To obtain a state $\gket{\phi'}$ in the desired sector, a Wilson-line operator connecting the two static charges was applied to a vacuum-sector state. Imaginary-time evolution of such a state then yielded the ground state in the chosen static-charge sector.

\medskip

In the LSH formulation, by contrast, the Hilbert space $\mathbb{H}_{\text{LSH}}$ is annihilated by the local non-Abelian Gauss's law operators $\hat G^a(r)$ by construction. Here, static charges can instead be defined as various sectors of the AGL, i.e., Eq.~\eqref{eq:AGL}. Define the AGL operator acting on sites $r$ and $r+1$ as
\begin{align}
    \hat{\mathcal{Q}}(r,r+1)\coloneq \hat N_L(r)-\hat N_R(r+1).
\end{align}
Under the action of $\hat{\mathcal{Q}}(r,r+1)$, the LSH Hilbert space decomposes as
\begin{align}
    \mathbb{H}_{\text{LSH}}
    =\bigoplus_{\{\mathcal{Q}(r,r+1)\}}\mathbb{H}_{\{\mathcal{Q}(r,r+1)\}},
\end{align}
where $\mathcal{Q}(r,r+1)\in\mathbb{Z}$. Thus, a state $\gket{\phi}\in \mathbb{H}_{\{\mathcal{Q}(r,r+1)\}}$ satisfies
\begin{align}
    \hat{\mathcal{Q}}(r,r+1)\gket{\phi}=\mathcal{Q}(r,r+1)\gket{\phi}.
\end{align}
In particular, we can mimic the insertion of a pair of static fundamental SU(2) charges at sites $r_1$ and $r_2$ (with $r_2 > r_1+1$) by choosing the external flux sector
\begin{align}
    \mathcal{Q}(r_1,r_1+1)=-1,~~ \mathcal{Q}(r_2-1,r_2)=+1.
    \label{eq:AGL-charges}
\end{align}
We ignore the special case $r_1=r_2-1$, for which $\mathcal{Q}(r_1,r_2)=0$. With the described prescription, the additional flux sourced by static charges is accounted for along the link, and with a single link connecting two static charges, the original AGL remains intact everywhere.

This notion can also be understood in the extended Hilbert-space picture. 
When introducing static SU(2) matter at some site $r$, the external matter degree of freedom can form an SU(2) singlet on its own, or by combining with the dynamical left and right Schwinger bosons at the same site.\footnote{We discount the hadrons formed by combining dynamical and static fermions.} These possibilities can be described by the number operators $\hat n_{i/o}^{\text{ext}}(r)$ that count the number of static hadrons, as well as of string-in/out excitations with static sources at site $r$ (in analogy with the number operators $\hat n_{i/o}(r)$ for the dynamical fermions). In particular, as discussed above, we insert nonzero static charges at sites $r_1$ and $r_2>r_1+1$. The extended Hilbert space is:
\begin{equation}
\mathbb{H}'_{\text{LSH}} \;\coloneq\; \mathbb{H}_{\text{LSH}}\otimes \mathbb{S}\otimes \mathbb{S},
\end{equation}
where the four-dimensional space $\mathbb{S}$ is spanned by the four different values for the $(n_i^{\rm ext}(r),n_o^{\rm ext}(r))$ eigenvalues of the number operators defined above for $r=r_1,r_2$. Note that $(n_i^{\rm ext}(r),n_o^{\rm ext}(r))=(0,0)$ when $r\neq r_1,r_2$. Since the external fermions can now source dynamical fluxes, the AGL acts jointly on the system and static degrees of freedom:
\begin{align}
   \hat{\mathcal{Q}}'(r,r+1) \coloneq \hat N_L(r)-\hat N_R(r+1)+\hat N^{\text{ext}}(r,r+1).
\end{align}
Here, we have defined
\begin{align}
    \hat N^{\rm ext}(r,r+1)&\coloneq \hat N^{\rm ext}_L(r)- \hat N^{\rm ext}_R(r+1),
\end{align}
with
\begin{align}
    \hat N^{\rm ext}_{L/R}&\coloneq \hat n^{\text{ext}}_{o/i}(r) \big(1-\hat n^{\text{ext}}_{i/o}(r)\big).
\end{align}
We are only interested in the dynamical sector; thus, we project onto a fixed state in the external sector, i.e.,
\begin{align}
    \gket{\Phi}=\gket{\phi}\otimes \gket{\xi(r_1)}\otimes \gket{\xi(r_2)}\in \mathbb{H}'_{\rm LSH}.
\end{align}
The extended AGL on  $\gket{\Phi}$, i.e.,
\begin{align}
    \hat{\mathcal{Q}}'(r,r+1)\gket{\Phi}=0,
\end{align}
leads to
\begin{align}
    \hat{\mathcal{Q}}(r,r+1)&\gket{\phi}\otimes \gket{\xi_1(r_1)}\otimes\gket{\xi_2(r_2)}=\nonumber \\
    &-\gket{\phi}\otimes \hat N^{\text{ext}}(r,r+1)\big(\gket{\xi_1(r_1)}\otimes\gket{\xi_2(r_2)}\big).
\end{align}
We choose the states $\gket{\xi(r_1)}$ and $\gket{\xi(r_2)}$ to be the eigenstates of the external number operators with eigenvalues $(n^{\rm ext}_i(r_1), n^{\rm ext}_o(r_1))=(0,1)$ and $(n^{\rm ext}_i(r_2), n^{\rm ext}_o(r_2))=(1,0)$, respectively. Then, the above equation constrains the state $\gket{\phi}\in \mathbb{H}_{\rm LSH}$ to be in the AGL sector defined in Eq.~\eqref{eq:AGL-charges}. Note that as desired, $Q(r_1,r_2)=0$ if $r_1=r_2-1$.

In the presence of external charge at site $r$, the total number of Schwinger bosons on the link to the right of site $r$ can be counted by the operator $\hat N^{\rm tot}_L(r)\coloneq\hat N_L(r)+\hat N_L^{\rm ext}(r)$ or $\hat N_R^{\rm tot}(r+1)\coloneq \hat N_R(r+1)+\hat N^{\rm ext}_R(r+1)$ (since these two are constrained to have the same eigenvalue by the modified AGL). Thus, the electric Hamiltonian in Eq.~\eqref{eq:HE_LSH} should be replaced by:
\begin{align}
    \hat H_E'=
    \frac{1}{2}\sum_{r=1}^{N-1}
    \Bigg[&\frac{\hat{N}^\text{tot}_L(r)}{2}\left( \frac{\hat{N}^\text{tot}_L(r)}{2}+1 \right)+\nonumber\\
    &\frac{\hat{N}^\text{tot}_R(r+1)}{2}\left( \frac{\hat{N}^\text{tot}_R(r+1)}{2}+1 \right)\Bigg].
\end{align}
However, we continue working with the electric Hamiltonian in Eq.~\eqref{eq:HE_LSH}. 
The difference
\begin{align}
    \hat H'_E-\hat H_E=\frac{\hat N_L(r_1)+\hat N_R(r_2)}{4}+\frac{3}{4}
\end{align}
leads to negligible effects in the continuum and thermodynamic limits. We have checked this explicitly by verifying that the slope of the static potential in our computations is the same for either forms of the electric Hamiltonian.

Inspired by the procedure in Ref.~\cite{Kuhn:2015zqa}, we initialize a state in this AGL sector by acting on the strong-coupling vacuum (which belongs to the $\hat{\mathcal{Q}}=0$ AGL sector) with the operator $\hat{\mc L}^{++}(r_1+1)\cdots \hat{\mc L}^{++}(r_2-1)$. This state is then used as the initial condition for DMRG, with the penalty term in Eq.~\eqref{eq:penalty} used to forbid departure from the AGL sector.

%%%%%%%%%%%%%%%%%%%%%%%%%%%%%%%%%%%%%%%%%%%%%%%%%%%%%%%%%%%%%%%%%%%%%%%%%%%%%%%%%%%%%%%%%%%%%%%%%%%%%%%%%%%%%%%%%%%%%%%%%%%%%%%%%%%%%%%%%%%%%%%%%%%%%%%%%%%%%%%%%%
\subsection{Meson-string operator
\label{app:meson-string}}
\noindent The gauge-invariant string operator in the Kogut-Susskind formulation can be defined as
\begin{align}
    \hat S_{r,\Delta r}=\hat\psi^{\dagger}(r)\hat U(r)\cdots\hat U(r+\Delta r-1)\hat\psi(r+\Delta r).
    \label{eq:string-original-app}
\end{align}
From Eq.~\eqref{eq:ULUR}, each link operator splits as
\begin{align}
    \hat U(s)=\hat U_L(s)\hat U_R(s+1).
\end{align}
One can then combine the leftmost (rightmost) $\hat U_L(r)$ ($\hat U_R(r+\Delta r)$) with the fermion operator $\hat \psi^{\dagger}(r)$ ($\hat \psi(r+\Delta r)$) to obtain local string-in (string-out) operators at the two ends. At the middle sites, one can contract $\hat U_R(s)$ with $\hat U_L(s)$ to get local loop operators. This yields the expression given in Eq.~\eqref{eq:String_operator_LSH}. Thus, the string operator in Eq.~\eqref{eq:string-original-app} breaks into a sum of $2^{2\Delta r}$ gauge-invariant operators in the LSH picture. Note that in the original Kogut-Susskind form, one has to implement $2^{2\Delta r+1}$ suboperators with distinct actions on states.\footnote{Consider expressing the string operator in terms of the $SU(2)$ matrix/vector components of $\hat U$, $\hat \psi$, and $\psi^\dagger$, and recall that each $\hat U$ can raise and lower the angular momentum by $\frac{1}{2}$.}
\begin{figure}[t!]
    \centering
    \includegraphics[scale=0.775]{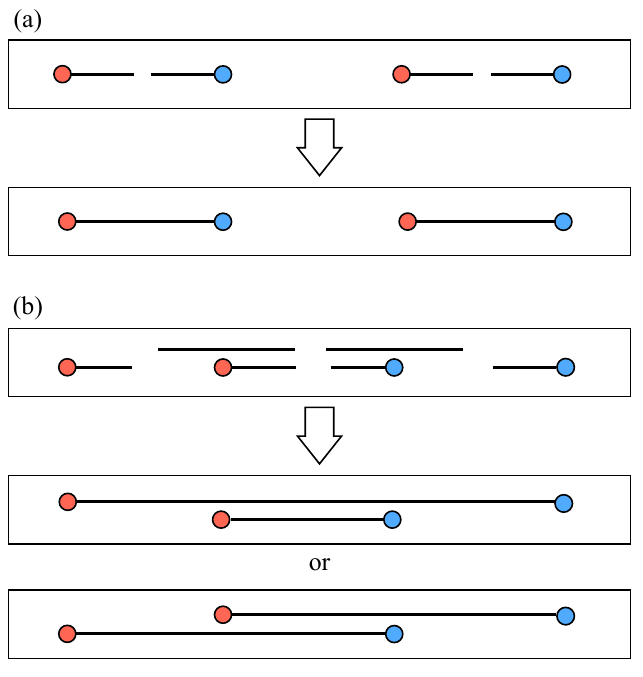}
    \caption{Examples of four-site LSH basis states with the incoming boundary flux set to zero. While there is only one way to connect the flux lines in (a), the LSH basis state in (b) can be graphically represented by two distinct connections between flux lines.}
    \label{fig:flux_conn}
\end{figure}
\begin{figure*}[t!]
    \centering
    \includegraphics[scale=0.505]{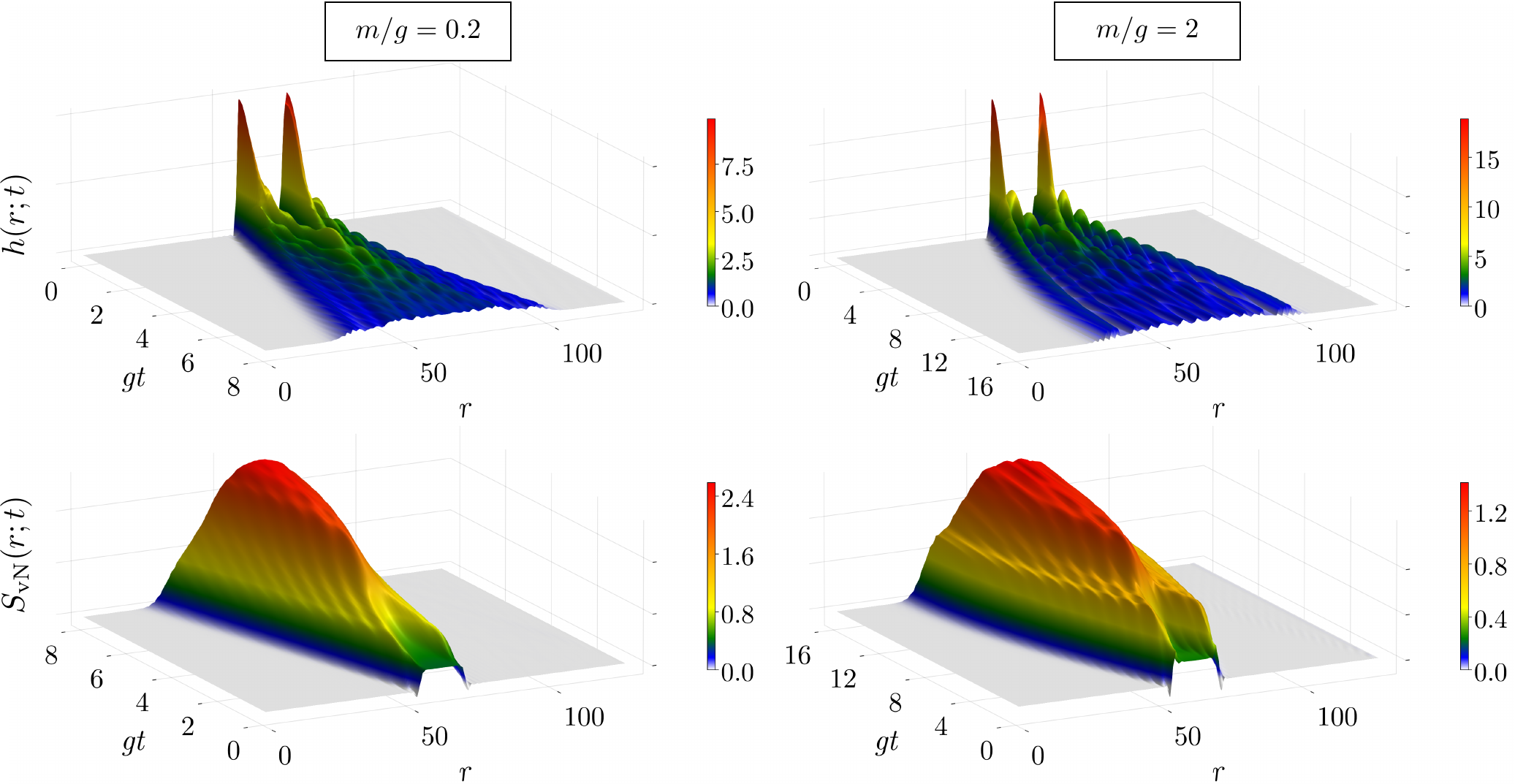}
    \caption{3D plots of evolution of the vacuum-subtracted expectation values of energy density, as well as the vacuum-subtracted bipartite entanglement entropy for a cut at position $r$, for two values of bare fermion mass. The corresponding 2D plots are presented in Fig.~\ref{fig:energy-entropy-2D-plots} of the main text. The other parameters are $N=128$, $x=16$, and $j_{\rm max}=\frac{5}{2}$.
    }
    \label{fig:energy-entropy-3D-plots}
\end{figure*}

Moreover, recall that Dirac fermions $\hat \Psi(r)$ at some position $ra$ in the continuum can be written as
\begin{align}
    \hat \Psi(r)\coloneq 
    \begin{pmatrix}
        \hat \Psi_u(r) \\
        \hat \Psi_l(r).
    \end{pmatrix},
\end{align}
where $\Psi_u(r)$ ($\Psi_l(r)$) denotes the upper (lower) component of the Dirac field in (1+1)D. The Dirac-string operator takes the form
\begin{align}
    \hat{\bar{\Psi}}^\dagger(r) \prod_{s=r}^{r+\Delta r-1} & U(s)\, {\hat \Psi}(r+\Delta r)&
    \nonumber \\
    &= \hat\Psi_u^{\dagger}(r) \prod_{s=r}^{r+\Delta r-1}U(s)\, \hat\Psi_u(r+\Delta r)
    \nonumber\\
    &-\hat\Psi_l^{\dagger}(r) \prod_{s=r}^{r+\Delta r-1}U(s)\, \hat\Psi_l(r+\Delta r),
\end{align}
where we have used the gamma-matrix basis choice
\begin{align}
    \gamma_0=
    \begin{pmatrix}
        1 & 0 \\
        0 & -1
    \end{pmatrix}.
\end{align}

Now consider the staggering prescription in the Kogut-Susskind formulation, which imposes the following identification:
\begin{subequations}
\begin{align}
    &\psi(r)= \Psi_u(r)~\text{if $r$ is even},\\
    &\psi(r)=\Psi_l(r)~\text{if $r$ is odd}.
\end{align}
\end{subequations}
In other words, one keeps only one component of the Dirac spinor at each site and discards the other component.

The above expression indicates that if we pick $r$ to be odd and $\Delta r$ to be even, the equivalent string operator in the staggered formulation becomes:
\begin{align}
    -\hat\psi^{\dagger}(r) \prod_{s=r}^{r+\Delta r-1}U(s)\, \hat\psi(r+\Delta r).
\end{align}
Similarly, if we pick $r$ to be even and $\Delta r$ to be still even, the equivalent string operator in the staggered formulation becomes: 
\begin{align}
   \hat\psi^{\dagger}(r) \prod_{s=r}^{r+\Delta r-1}U(s)\, \hat\psi(r+\Delta r).
\end{align}
To combine the two situations, in Eq.~\eqref{eq:string_op_dirac} of the main text, we define the Dirac-string operator in the staggered formulation to be 
\begin{align}
    -\hat\psi^{\dagger}(r) &\prod_{s=r}^{r+\Delta r-1}U(s)\, \hat\psi(r+\Delta r)
    \nonumber\\
    &+\hat\psi^{\dagger}(r+1) \prod_{s=r+1}^{r+\Delta r}U(s)\, \hat\psi(r+\Delta r),
\end{align}
up to an overall sign and constant. This is merely a choice and, toward the continuum limit, any other choice should produce the same dynamics.

%%%%%%%%%%%%%%%%%%%%%%%%%%%%%%%%%%%%%%%%%%%%%%%%%%%%%%%%%%%%%%%%%%%%%%%%%%%%%%%%%%%%%%%%%%%%%%%%%%%%%%%%%%%%%%%%%%%%%%%%%%%%%%%%%%%%%%%%%%%%%%%%%%%%%%%%%%%%%%%%%%
\subsection{Local versus extended gauge-invariant quantum numbers
\label{app:local-vs-extended}}
\noindent As discussed in Sec.~\ref{sec:microscopic}, interpreting LSH states across multiple sites in terms of extended gauge-invariant meson strings is not possible in a unique way. We provide examples in this appendix to elucidate this statement.

Consider the four-site LSH basis states shown in Fig.~\ref{fig:flux_conn}, where the incoming boundary flux is set to zero. In the upper panel of Fig.~\ref{fig:flux_conn}(a), there is a unique way to connect the flux lines across sites, and thus, this basis state can also be represented with connected flux lines, as in the lower panel of Fig.~\ref{fig:flux_conn}(a). Such a state, consists of two meson strings. These mesonic strings are created by acting on the strong-coupling vacuum by the string operator defined in Eq.~\eqref{eq:string-original-app} or its Hermitian conjugate. Thus, the state represented in Fig.~\ref{fig:flux_conn}(a) is proportional to 
$S_{1,1}^\dagger S_{3,1}^\dagger$ acting on the strong-coupling vacuum.

In contrast, when the maximum flux flowing in the lattice exceeds $n_l = 1$ (equivalently $j=\frac{1}{2}$), i.e., there are overlapping flux lines, there is no unique way to connect the fluxes, and the LSH formulation provides no prescription for connecting them. For instance, the state in the top panel of Fig.~\ref{fig:flux_conn}(b) could be represented with connected flux lines in two different ways, as shown in the middle and bottom panels of Figs.~\ref{fig:flux_conn}. The state in the top panel is, in fact, a superposition of the states created by the action of $\hat{S}_{1,3}^\dagger \hat{S}_{2,1}$ and $ \hat{S}_{1,2} \hat{\mc H}^{++}(1)^\dagger \hat{S}_{2,2}^\dagger \hat{\mc H}^{++}(2)$ acting on the strong-coupling vacuum, where the hadron creation operator $\hat{\mc H}^{++}(r)$ at site $r$ is defined in Eq.~\eqref{eq:Hpp}.

The examples of this appendix make it clear  why it is more meaningful to track the string-in/out states and the loop quantum numbers separately throughout the evolution. LSH does not specify a definite connectivity for extended states, hence making identification of mesonic content of states ambiguous.\\
\begin{figure*}[t!]
    \centering
    \includegraphics[scale=0.51]{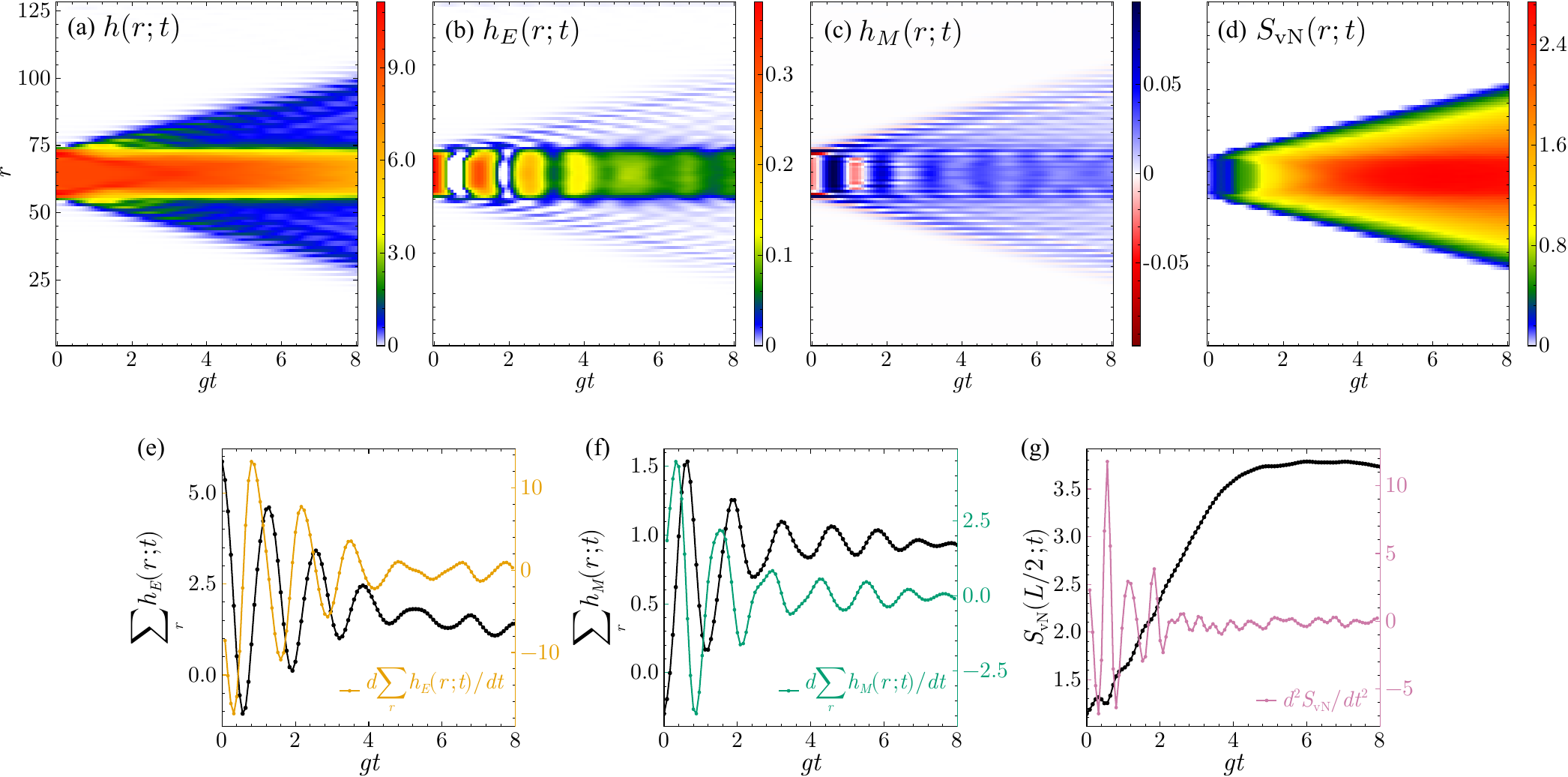}
    \caption{
    The effect of a low flux cutoff $j_\text{max}=\frac{1}{2}$ on the evolution of the vacuum-subtracted expectation values of the energy density (a), electric-energy density (b), mass-energy density (c), vacuum-subtracted bipartite entanglement entropy for a cut at position $r$ (d), the total vacuum-subtracted electric energy and its first derivative (e), the total vacuum-subtracted mass energy and its first derivative (f), and the vacuum-subtracted half-cut entanglement entropy and its second derivative (g). Other parameters are $\frac{m}{g}=0.2$, $N=128$, and $x=16$. These plots should be contrasted with those included in Figs.~\ref{fig:energy-entropy-2D-plots} and \ref{fig:energy-entropy-1D-plots} of the main text where $j_\text{max}=\frac{5}{2}$.
    }
    \label{fig:combined_plot_lowcutoff}
\end{figure*}
\begin{figure*}[t!]
    \centering
    \includegraphics[scale=0.5]{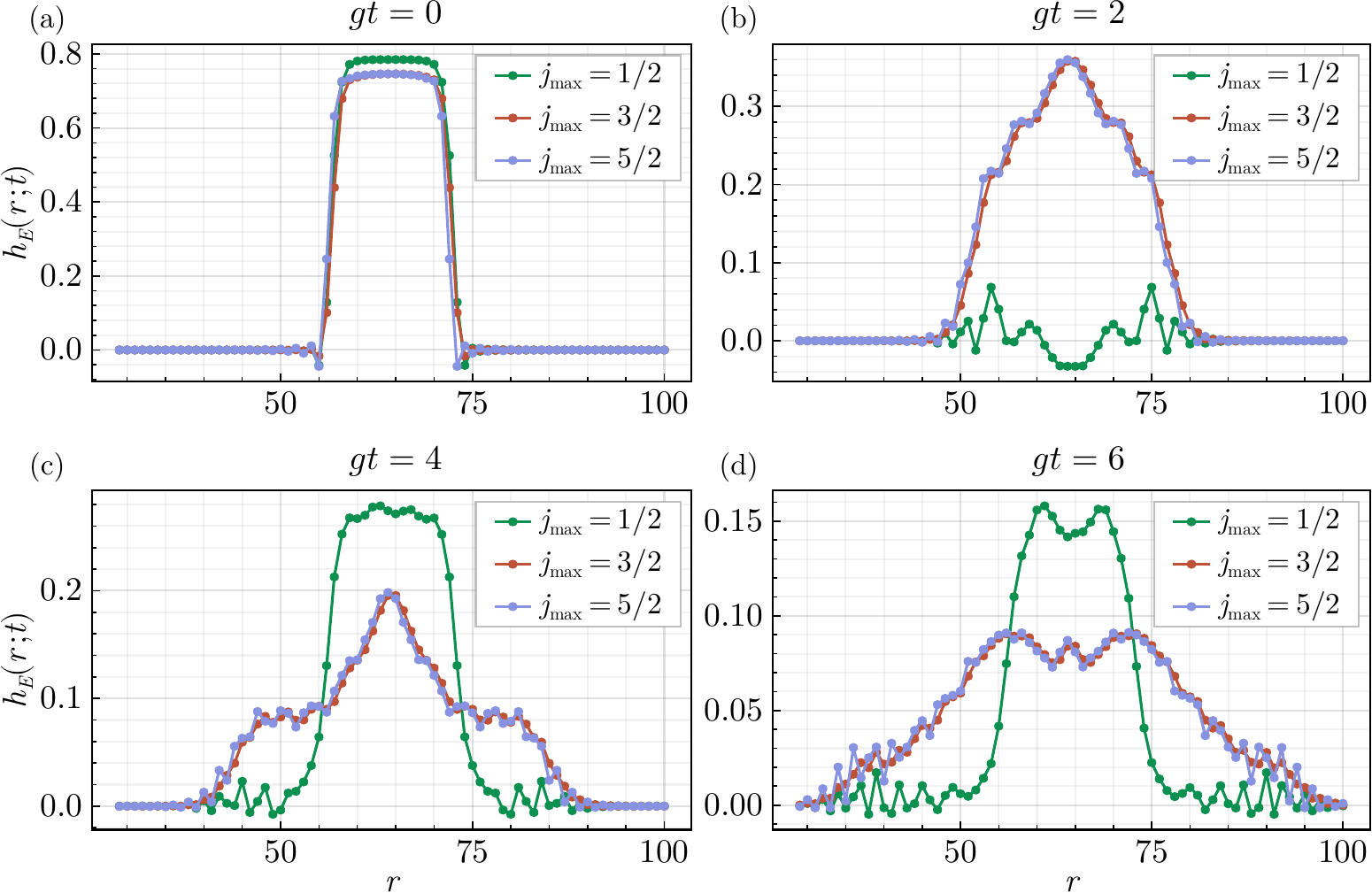}
    \caption{The site-resolved vacuum-subtracted electric energy at the initial time and three subsequent times during the meson-string evolution, for three cutoff values $j_\text{max} \in \{\frac{1}{2},\frac{3}{2},\frac{5}{2} \}$. Other parameters are $\frac{m}{g}=0.2$, $N=128$, $x=16$, and $D_\text{max}=200$.}
    \label{fig:HE_jmax_err}
\end{figure*}
%

%%%%%%%%%%%%%%%%%%%%%%%%%%%%%%%%%%%%%%%%%%%%%%%%%%%%%%%%%%%%%%%%%%%%%%%%%%%%%%%%%%%%%%%%%%%%%%%%%%%%%%%%%%%%%%%%%%%%%%%%%%%%%%%%%%%%%%%%%%%%%%%%%%%%%%%%%%%%%%%%%%
\subsection{Dynamical-string evolution: 3D plots
\label{app:3D-plots}}
\noindent In Sec.~\ref{sec:energy-entropy} of the main text, we presented in Fig.~\ref{fig:energy-entropy-2D-plots} 2D heatmap plots for the vacuum-subtracted energy density $h(r;t)$ and vacuum-subtracted bipartite entanglement entropy $S_\text{vN}(r;t)$. For enhanced visualization, we provide the corresponding 3D plots in this appendix. The decay of lumps of energy density at the meson-string endpoints; the transport of energy along the particle-excitation showers; and the enhancement of local energy density at the primary and secondary scattering events in the bulk are clearly visible in the 3D energy plots. The abundant generation of entanglement entropy toward the center of the lattice, particularly as the primary and secondary scattering events occur; and for the heavier mass, the two-stage depletion of the entanglement entropy away from the center (i.e., the formation of a narrow ledge) is clearly visible from the 3D entanglement-entropy plots.

%%%%%%%%%%%%%%%%%%%%%%%%%%%%%%%%%%%%%%%%%%%%%%%%%%%%%%%%%%%%%%%%%%%%%%%%%%%%%%%%%%%%%%%%%%%%%%%%%%%%%%%%%%%%%%%%%%%%%%%%%%%%%%%%%%%%%%%%%%%%%%%%%%%%%%%%%%%%%%%%%%
\subsection{Effect of Hilbert-space truncation on dynamics
\label{app:truncation-error}}
\begin{figure*}[t!]
    \centering
    \includegraphics[scale=0.65]{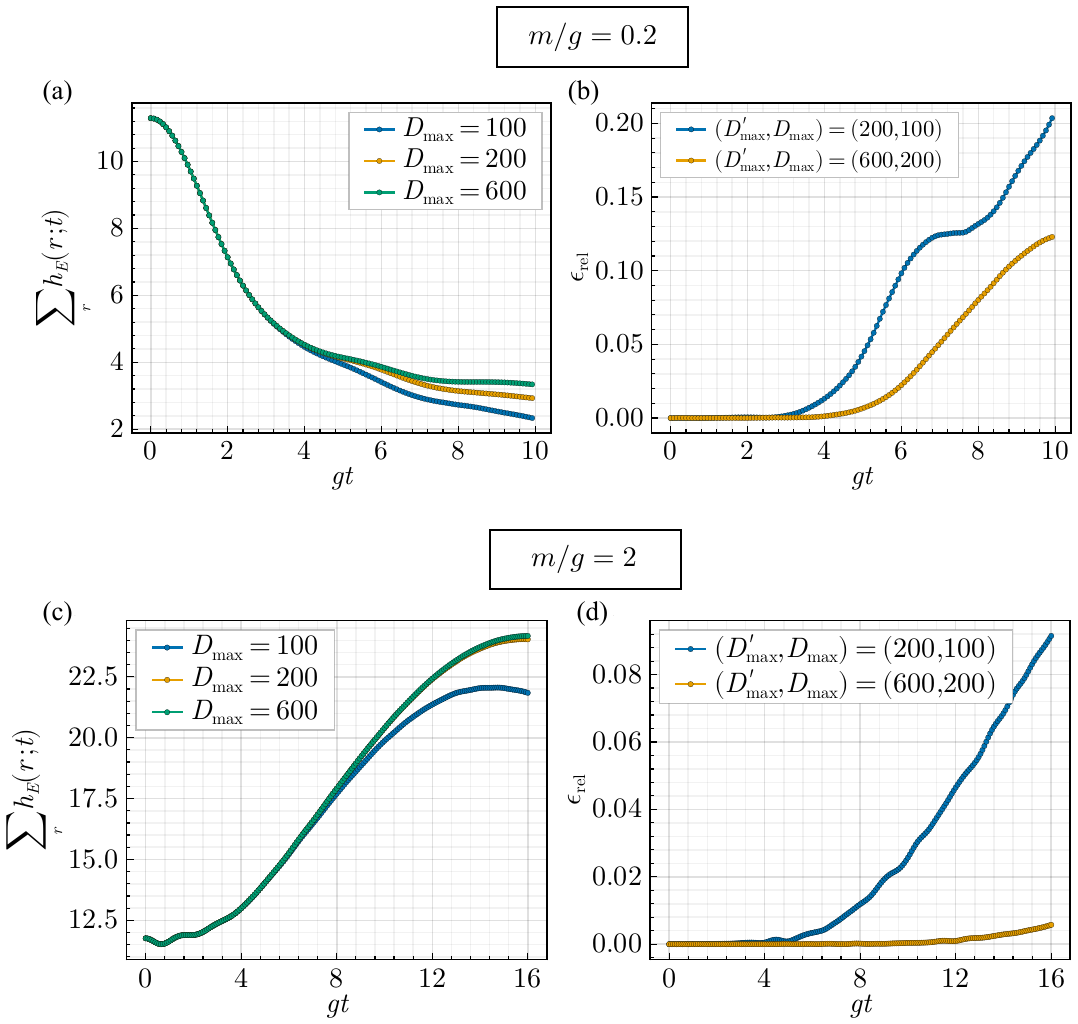}
    \caption{The total vacuum-subtracted electric energy (a) at three bond-dimension values, along with the corresponding relative errors (b), as a function of evolution time for $\frac{m}{g}=0.2$. (c,d) The same quantities for $\frac{m}{g}=2$
    . Other parameters are $j_\text{max}=\frac{5}{2}$, $N=128$, and $x=16$.
    }
    \label{fig:rel_err_heavymass}
\end{figure*}
\noindent In this appendix, we analyze the role of Hilbert-space truncation on observables, and demonstrate the sufficiency of the cutoff used in the main text for the dynamical computations.

First, we repeat the dynamical computations of Sec.~\ref{sec:dynamic} for the harshest nontrivial Hilbert-space truncation of $j_\text{max} = \frac{1}{2}$. We consider only the lighter mass, i.e., $\frac{m}{g}=0.2$; the other parameters are those used the main text except the maximum allowed bond dimension is set to $D_\text{max} = 200$ using a 2-site TDVP algorithm (see the next appendix for a discussion of the bond-dimension truncation effects). Figures~\ref{fig:combined_plot_lowcutoff}(a)-(d) display the exact same set of observables shown in Fig.~\ref{fig:energy-entropy-2D-plots} of the main text, with the only difference being the reduced cutoff. Dramatic differences are observed in the dynamics captured by each energy operator and the bipartite entanglement entropy. The electric- and mass-energy density plots show an oscillatory behavior, and damped conversion between string and broken string, and a confined transport, in stark contrast with the behavior observed for the higher cutoff values. The entanglement entropy displays qualitatively different behavior compared to its counterpart at larger $j_{\text{max}}$. The same features are captured in the total vacuum-subtracted electric and mass energies, and the vacuum-subtracted half-cut entanglement entropy, as a function time, along with their derivatives, shown in Figs.~\ref{fig:energy-entropy-2D-plots}(e)-(g). These discrepancies together suggest that achieving reliable computations requires a sufficiently large flux cutoff so as to ensure convergence in the observables of interest. 

To demonstrate the convergence at the cutoff value $j_\text{max}=\frac{5}{2}$ used in the main text, we plot in Fig.~\ref{fig:HE_jmax_err} the site-resolved vacuum-subtracted electric energy at the initial time and three subsequent times during the meson-string evolution, for the cutoff values $j_\text{max} \in \{\frac{1}{2},\frac{3}{2},\frac{5}{2} \}$, mass $\frac{m}{g}=0.2$, and other parameters as in the main text. The $j_\text{max}=\frac{1}{2}$ results clearly start deviating from the larger $j_\text{max}$ results quickly, while the $j_\text{max}=\frac{3}{2}$ and $j_\text{max}=\frac{5}{2}$ results follow each other pretty closely. We conclude that the difference between the results at $j_\text{max}=\frac{5}{2}$ and the next cutoff is small and convergence is achieved for the evolution times considered. The same behavior is observed in other quantities considered in the main text, and for the heavier-mass case as well.

\begin{figure*}[t!]
    \centering
    \includegraphics[scale=0.725]{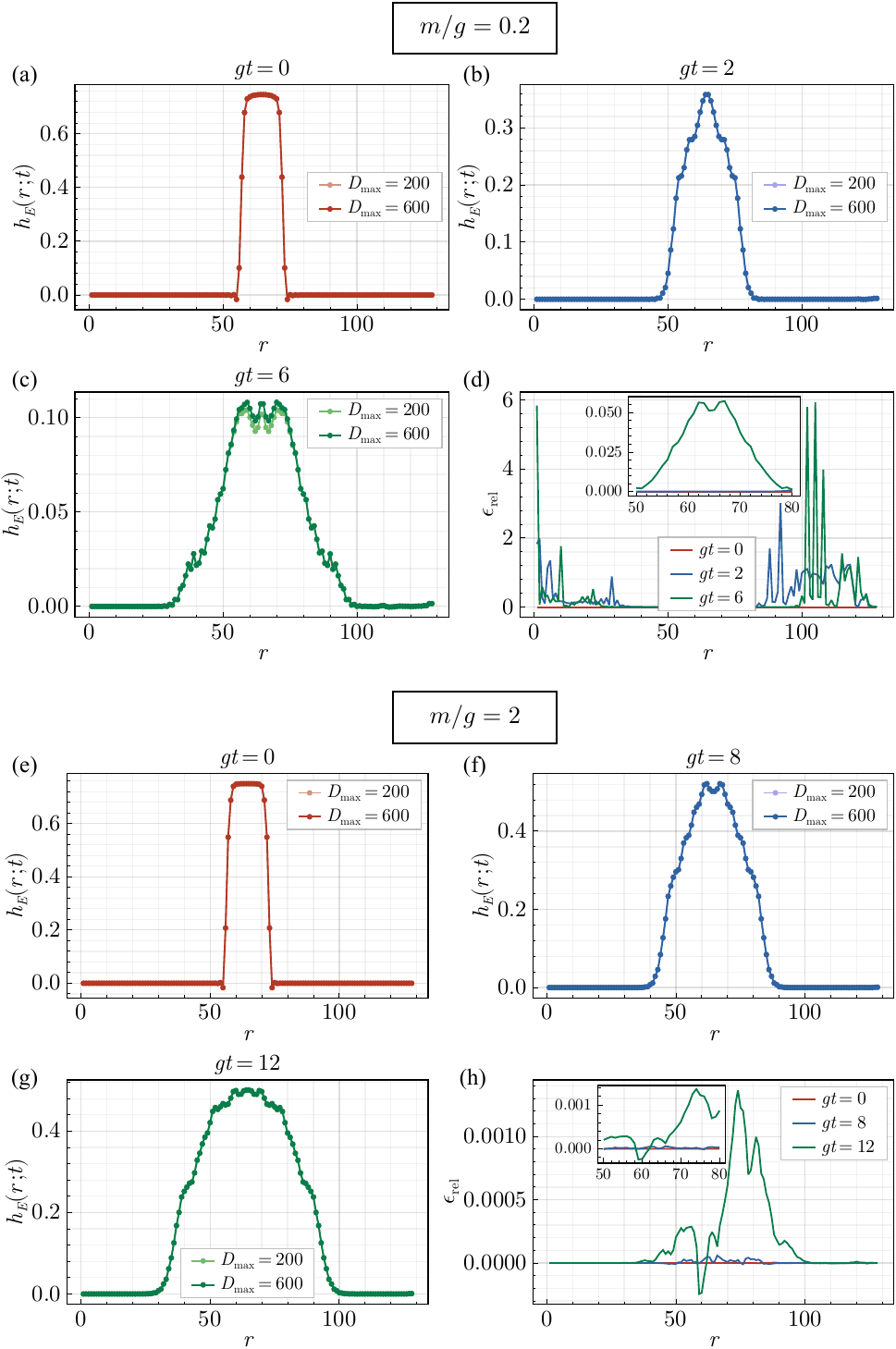}
    \caption{The vacuum-subtracted electric-energy density (a,b,c) at two bond-dimension values and select times, along with the corresponding relative errors (d), as a function of position for $\frac{m}{g}=0.2$. (e,f,g,h) The same quantities for $\frac{m}{g}=2$. Other parameters are $j_\text{max}=\frac{5}{2}$, $N=128$, and $x=16$.
    }
    \label{fig:relative_err_siteresolved}
\end{figure*}
%

%%%%%%%%%%%%%%%%%%%%%%%%%%%%%%%%%%%%%%%%%%%%%%%%%%%%%%%%%%%%%%%%%%%%%%%%%%%%%%%%%%%%%%%%%%%%%%%%%%%%%%%%%%%%%%%%%%%%%%%%%%%%%%%%%%%%%%%%%%%%%%%%%%%%%%%%%%%%%%%%%%
\subsection{Bond-dimension convergence in time evolution
\label{app:bond-dimension}}
\noindent In this appendix, we explain the TDVP algorithm's error-estimation procedure for the time-evolved states. We perform three separate computations with the maximum allowed bond dimensions $D_\text{max} \in \{100,200,600\}$. For the two lowest $D_\text{max}$ values, we use the 2-site TDVP algorithm, while for the highest $D_\text{max}$ value, we use the 1-site TDVP algorithm. The reason is that the 2-site algorithm is computationally more demanding~\cite{Paeckel:2019yjf}; hence the lower bond dimensions. We observe that after a few steps of the TDVP algorithms, the bond dimension saturates. For the largest allowed maximum bond dimension, the saturation occurs at bond dimension $577$ for the lighter mass and at $447$ for the heavier mass.
To assess the effects of bond-dimension truncation, we compute the relative errors for the total electric-energy density of the time-evolved state. The relative error is defined as
\begin{equation}
    \epsilon_\text{rel}(O) \coloneq \frac{O_{D'_\text{max}} - O_{D_\text{max}}}{O_{D'_\text{max}}},
\end{equation}
where $\hat{O}_D$ is the observable of interest computed with a maximum bond dimension $D$, and $D'_\text{max}>D_\text{max}$. In the following analysis, $O$ is chosen to be $\sum_r h_E(r;t)$ and $h_E(r;t)$, with $h_E(r;t)$ defined in Eq.~\eqref{eq:hEr-defs}. 

Figure~\ref{fig:rel_err_heavymass} displays the total vacuum-subtracted electric energy at the three bond-dimension values along with the corresponding relative errors as a function of evolution time for both the lighter and heavier masses. The two largest bond dimensions used for each mass nearly coincide for all evolution times considered for the heavier mass, with a relative error of only subpercent at $gt=16$. In contrast, for the lighter mass, the relative error for the two largest bond dimensions is up to a few percent for $gt=8$. We observe the same trend in other observables and, hence, limit the time evolution to $gt=8$ in the lighter-mass case.

Figure~\ref{fig:relative_err_siteresolved} displays the vacuum-subtracted electric-energy density at the two highest bond-dimension values as a function of position at select evolution times, along with the corresponding relative errors. The convergence is clear from the plots; the errors in the bulk of the lattice are at few percent (subpercent) levels for the lighter (heavier) mass. The large relative errors at the boundaries are an artifact of the small values of the quantities and have no bearing on the results. Similar behavior is observed in other quantities, which gives confidence that the time-evolved state is converged relatively well in the bond dimension.

\bibliography{bibi.bib}

\end{document}